\documentclass[prd,twocolumn,a4paper,superscriptaddress,nofootinbib,preprintnumbers,twoside,raggedbottom,eqsecnum,showpacs]{revtex4}

\usepackage{graphics,bbm,amsmath,amssymb}

\catcode`\à=\active \defà{\`a} \catcode`\À=\active \defÀ{\`A}
\catcode`\á=\active \defá{\'a} \catcode`\Á=\active \defÁ{\'A}
\catcode`\ä=\active \defä{\"a} \catcode`\Ä=\active \defÄ{\"A}
\catcode`\â=\active \defâ{\^a} \catcode`\Â=\active \defÂ{\^A}
\catcode`\å=\active \defå{{\aa}} \catcode`\Å=\active \defÅ{{\AA}}
\catcode`\ç=\active \defç{\c{c}} \catcode`\Ç=\active \defÇ{\c{C}}
\catcode`\è=\active \defè{\`e} \catcode`\È=\active \defÈ{\`E}
\catcode`\é=\active \defé{\'e} \catcode`\É=\active \defÉ{\'E}
\catcode`\ë=\active \defë{\"e} \catcode`\Ë=\active \defË{\"E}
\catcode`\ê=\active \defê{\^e} \catcode`\Ê=\active \defÊ{\^E}
\catcode`\ì=\active \defì{\`{\i}} \catcode`\Ì=\active \defÌ{\`{\I}}
\catcode`\í=\active \defí{\'{\i}} \catcode`\Í=\active \defÍ{\'{\I}}
\catcode`\ï=\active \defï{\"{\i}} \catcode`\Ï=\active \defÏ{\"{\I}}
\catcode`\î=\active \defî{\^{\i}} \catcode`\Î=\active \defÎ{\^{\I}}
\catcode`\ò=\active \defò{\`o} \catcode`\Ò=\active \defÒ{\`O}
\catcode`\ó=\active \defó{\'o} \catcode`\Ó=\active \defÓ{\'O}
\catcode`\ö=\active \defö{\"o} \catcode`\Ö=\active \defÖ{\"O}
\catcode`\ô=\active \defô{\^o} \catcode`\Ô=\active \defÔ{\^O}
\catcode`\ù=\active \defù{\`u} \catcode`\Ù=\active \defÙ{\`U}
\catcode`\ú=\active \defú{\'u} \catcode`\Ú=\active \defÚ{\'U}
\catcode`\ü=\active \defü{\"u} \catcode`\Ü=\active \defÜ{\"U}
\catcode`\û=\active \defû{\^u} \catcode`\Û=\active \defÛ{\^U}
\catcode`\ý=\active \defý{\'y} \catcode`\Ý=\active \defÝ{\'Y}
\catcode`\ÿ=\active \defÿ{\"y} \catcode`\˜=\active \def˜{\"Y}
\catcode`\½=\active \def½{!`}
\catcode`\¾=\active \def¾{?`}
\catcode`\ß=\active \defß{{\ss}}

\newcommand{\tmop}[1]{\ensuremath{\operatorname{#1}}}
\newcommand{\tmem}[1]{{\em #1\/}}
\newcommand{\mathpi}{\pi}
\newcommand{\emdash}{---}
\newcommand{\mathe}{\mathrm{e}}
\newcommand{\mathi}{\mathrm{i}}
\newcommand{\tmfloatcontents}{}
\newlength{\tmfloatwidth}
\newcommand{\tmfloat}[5]{
  \renewcommand{\tmfloatcontents}{#4}
  \setlength{\tmfloatwidth}{\widthof{\tmfloatcontents}+1in}
  \ifthenelse{\equal{#2}{small}}
    {\ifthenelse{\lengthtest{\tmfloatwidth > \linewidth}}
      {\setlength{\tmfloatwidth}{\linewidth}}{}}
    {\setlength{\tmfloatwidth}{\linewidth}}  \begin{minipage}[#1]{\tmfloatwidth}
    \begin{center}
      \tmfloatcontents
      \captionof{#3}{#5}
    \end{center}
  \end{minipage}}
\newcommand{\Iota}{\mathrm{I}}
\newcommand{\mathd}{\mathrm{d}}
\newcommand{\tmmathbf}[1]{\ensuremath{\boldsymbol{#1}}}
\newcommand{\bignone}{}

\begin{document}

\title{Lepton-number violation and right-handed neutrinos in Higgs-less effective theories}

\author{Johannes Hirn}
\affiliation{IFIC, Departament de F\'\i sica Te\`orica, CSIC - Universitat de València\\ Edifici d'Instituts de Paterna, Apt. Correus 22085, 46071 Val\`encia, Spain}
\email{johannes.hirn@ific.uv.es}

\author{Jan Stern}
\affiliation{Groupe Physique Th\'eorique, Unit\'e mixte de recherche 8608 du CNRS\\ IPN Orsay, Universit\'e~Paris-Sud~XI, 91406~Orsay, France}
\email{stern@ipno.in2p3.fr}

\begin{abstract}
Following previous work, we identify a symmetry $S_{\text{nat}}$ that
  generalizes the concept of custodial symmetry, keeping under control
  deviations from the Standard Model (SM). To realize $S_{\text{nat}}$
  linearly, the space of gauge fields has to be extended. Covariant
  constraints formulated in terms of spurions reduce $S_{\text{nat}}$ back to
  $\tmop{SU} \left( 2 \right)_L \times \mathrm{U} \left( 1 \right)_Y$. This
  allows for a covariant introduction of explicit $S_{\text{nat}}$-breaking parameters.
  We assume that $S_{\text{nat}}$ is at play in a theory of electroweak
  symmetry-breaking without a light Higgs particle. We describe some
  consequences of this assumption, using a non-decoupling effective theory in
  which the loop expansion procedure is based on both momentum and spurion
  power counting, as in Chiral Perturbation Theory. A hierarchy of
  lepton-number violating effects follows. Leading corrections to the SM are
  non-oblique. The effective theory includes stable light right-handed
  neutrinos, with an unbroken $\mathbb{Z}_2$ symmetry forbidding neutrino
  Dirac masses. $\nu_R$ contribution to dark matter places bounds on their
  masses.
\end{abstract}

\preprint{IFIC/05-23}
\preprint{FTUV/05-0428}

\maketitle

\section{Introduction}

In this paper, we detail a systematic low-energy expansion procedure applied
to scenarios of electroweak symmetry-breaking (EWSB) without a Higgs particle,
and without any other so far undiscovered particle below the $\tmop{TeV}$
scale (except maybe right-handed neutrinos). The aim is to construct an
effective theory in which the smallness of deviations from the Standard Model
(SM) would be controlled by a symmetry, i.e. {\tmem{technically natural}}
{\footnote{Throughout this paper, we use the words {\tmem{(technically)
natural}} and {\tmem{naturally}} as follows: it is technically natural for a
parameter to be small if there is a symmetry related to the limit in which it
is sent to zero {\cite{'tHooft:1979bh}}. }}. In its minimal version, the
theory only contains $\tmop{SU} \left( 2 \right)_L \times \mathrm{U} \left( 1
\right)_Y$ Yang-Mills fields and chiral fermions coupled to three Goldstone
bosons (GBs) which disappear from the spectrum, resulting in three of the
vector fields acquiring a mass. All these degrees of freedom are light
compared to the scale $\Lambda_{\text{w}}$
\begin{eqnarray}
  \Lambda_{\text{w}} & \simeq & 4 \mathpi v \hspace{0.2em} \simeq \hspace{0.2em} 3
  \tmop{TeV} .  \label{1.1}
\end{eqnarray}
Under these circumstances, the theory is not renormalizable in the usual
sense, i.e. in powers of coupling constants. Instead, it has to be defined,
renormalized (and unitarized) in powers of external momenta, generalizing the
example and techniques of Chiral Perturbation Theory ($\chi$PT)
{\cite{Weinberg:1979kz,Gasser:1984yg,Gasser:1985gg}}. This presumes a
self-consistent infrared power counting which would make it possible to order
operators in the lagrangian $\mathcal{L}_{\text{eff}}$ as well as loops
according to their importance in the low-energy limit. It is then understood
that each operator allowed by the symmetries has to be included into
$\mathcal{L}_{\text{eff}}$ at the order corresponding to its infrared
dimension. The relevant infrared power-counting for operators and Feynman
diagrams is reviewed in Sections \ref{3-WPCF}, \ref{WPC} and in Appendix
\ref{A-WPC}.

The essential new ingredient concerns the symmetry $S_{\text{nat}}$ of the
lagrangian underlying the low-energy effective theory (LEET). We require
$S_{\text{nat}}$ to be sufficiently large as to force the leading $\mathcal{O}
\left( p^2 \right)$ order of the LEET to coincide with the tree-level
Higgs-less vertices of the SM in the limit of vanishing fermion masses. It is
more easy to motivate this requirement than to find its appropriate
mathematical implementation:

i) First, a LEET based on $\tmop{SU} \left( 2 \right)_L \times \mathrm{U}
\left( 1 \right)_Y$ as the maximal symmetry group {\emdash}the case considered
in the past
{\cite{Appelquist:1985rr,Holdom:1990tc,Georgi:1991ci,Espriu:1992vm,Feruglio:1993wf,Wudka:1994ny,Bagan:1998vu,Nyffeler:1999ap,Nyffeler:1999hp}}
{\emdash} does {\tmem{not}} fulfill the above requirement: indeed, there are
several non-standard and unobserved $\tmop{SU} \left( 2 \right)_L \times
\mathrm{U} \left( 1 \right)_Y$-invariant vertices that appear at the leading
order $\mathcal{O} \left( p^2 \right)$ {\cite{Hirn:2004ze}}. They represent a
priori unsuppressed tree-level contributions to the $S$
{\cite{Holdom:1990tc,Peskin:1992sw}} and $T$ {\cite{Longhitano:1980iz}}
parameters, non-standard couplings of left-handed fermions to vector bosons
{\cite{Appelquist:1985rr,Peccei:1990kr}}, or introduce couplings of
right-handed fermions to the $W^{\pm}$. They are discussed in Section
\ref{3.2-irr-H-less} where it is shown that a LEET based on such a small
symmetry group would in addition allow a large Majorana mass for left-handed
neutrinos, and large $\mathcal{O} \left( p^2 \right)$ lepton-number violating
(LNV) vertices in general.

ii) Next, if the absence of non-standard $\mathcal{O} \left( p^2 \right)$
vertices is to be explained by a {\tmem{higher symmetry group}}
$S_{\text{nat}}$
\begin{eqnarray}
  S_{\text{nat}} & \supset & S_{\text{red}} \hspace{0.2em} = \hspace{0.2em}
  \tmop{SU} \left( 2 \right)_L \times \mathrm{U} \left( 1 \right)_Y, 
\end{eqnarray}
the action of $S_{\text{nat}} / S_{\text{red}}$ on the GBs and on the original
SM set of gauge fields, must be non-linear. This complicates the task of
inferring $S_{\text{nat}}$ from the SM lagrangian. The nature of this problem
can be illustrated by the following example, given here for a pedagogical
purpose: suppose that one deduced from $\pi \pi$ scattering experiments the
following effective lagrangian {\cite{Weinberg:1979kz}}
\begin{eqnarray}
  \mathcal{L}_{\pi \pi}^{\text{eff}} & = & \frac{1}{2} \partial_{\mu}
  \overset{\rightarrow}{\pi} \cdot \partial^{\mu} \overset{\rightarrow}{\pi}
+
  \frac{1}{2 f^2}  \left( 1 - \frac{\overset{\rightarrow}{\pi}^2}{f^2}
  \right)^{- 1}   \nonumber\\ &&\times \left( \overset{\rightarrow}{\pi} \cdot \partial_{\mu}
  \overset{\rightarrow}{\pi} \right)  \left( \overset{\rightarrow}{\pi} \cdot
  \partial^{\mu} \overset{\rightarrow}{\pi} \right) +\mathcal{O} \left( p^4
  \right) .  \label{1.3}
\end{eqnarray}
We could then ask why at the leading order $\mathcal{O} \left( p^2 \right)$,
various terms allowed by the linear isospin symmetry $O \left( 3 \right)$ are
actually absent. The task would then be to rediscover the $O \left( 4 \right)
/ O \left( 3 \right)$ non-linearly realized chiral symmetry. It is known that
the problem is greatly simplified by the introduction of an additional field
$\sigma$ such that the action of $O \left( 4 \right)$ on the enlarged manifold
$\left( \sigma, \overset{\rightarrow}{\pi} \right)$ is linear. The field
$\sigma$ is subject to the $O \left( 4 \right)$-invariant constraint
\begin{eqnarray}
  \sigma^2 + \overset{\rightarrow}{\pi}^2 & = & f^2,  \label{1.4}
\end{eqnarray}
and the lagrangian (\ref{1.3}) is equivalently rewritten as
\begin{eqnarray}
  \mathcal{L}_{\pi \pi}^{\text{eff}} & = & \frac{1}{2}  \left( \partial_{\mu}
  \sigma \partial^{\mu} \sigma + \partial_{\mu} \overset{\rightarrow}{\pi}
  \cdot \partial^{\mu} \overset{\rightarrow}{\pi} \right) +\mathcal{O} \left(
  p^4 \right) .  \label{1.5}
\end{eqnarray}
This lagrangian does not describe any interaction unless the constraint
(\ref{1.4}) is applied.

iii) We show in Section \ref{4-spurions} that a similar procedure exists in
the case of Higgs-less vertices of the SM, leading to an explicit description
of its ``hidden symmetry'' $S_{\text{nat}}$. This higher symmetry
$S_{\text{nat}} \supset S_{\text{red}}$ can be linearized by adding a set of
nine auxiliary gauge fields to the original four present in the SM. The
additional gauge fields are no more physical than the $\sigma$ field of the
previous example. At the end, they are eliminated by
$S_{\text{nat}}$-invariant constraints akin to (\ref{1.4}). Before these
constraints are applied, the lagrangian of the theory at $\mathcal{O} \left(
p^2 \right)$ consists of two decoupled sectors, as in equation (\ref{1.5}): a)
the symmetry-breaking sector containing three GBs together with six gauge
connections of the gauged spontaneously-broken $\tmop{SU} \left( 2
\right)_{\Gamma_L} \times \tmop{SU} \left( 2 \right)_{\Gamma_R}$ symmetry
{\tmem{and}} b) the unbroken $\tmop{SU} \left( 2 \right)_{G_L} \times
\tmop{SU} \left( 2 \right)_{G_R} \times \mathrm{U} \left( 1 \right)_{B - L}$
gauge theory with the $L \leftrightarrow R$ symmetric coupling of local left
and right isospin to chiral fermion doublets~{\cite{Senjanovic:1975rk}}.

$S_{\text{nat}}$ may be defined as the maximal local linear symmetry the
theory could have if the symmetry breaking sector and gauge/fermion sector are
decoupled. In the present case, one has
\begin{eqnarray}
  S_{\text{nat}} & = & \left[ \tmop{SU} \left( 2 \right) \times \tmop{SU}
  \left( 2 \right) \right]^2 \times \mathrm{U} \left( 1 \right)_{B - L}, 
  \label{1.6}
\end{eqnarray}
and the unconstrained theory contains thirteen gauge fields: four are
physical, whereas the nine remaining ``$\sigma$-type fields'' extend the
custodial symmetry, originally related to the right-isospin group
{\cite{Sikivie:1980hm,Longhitano:1980iz}}. The $S_{\text{nat}}$-invariant
constraints eliminate the nine redundant ``$\sigma$-type fields'', reduce the
{\tmem{linear symmetry}} $S_{\text{nat}}$ to its electroweak subgroup
$S_{\text{red}}$ and induce couplings between the symmetry-breaking and
gauge/fermion sectors. At this stage, $W^{\pm}$ and $Z^0$ become massive,
whereas all fermions remain massless. In this way one recovers all Higgs-less
vertices of the SM. The symmetry $S_{\text{nat}} / S_{\text{red}}$ is realized
non-linearly and hidden similarly to the $O ( 4 ) / O ( 3 )$ symmetry of the
non-linear $\sigma$-model hidden in the pion lagrangian (\ref{1.3}). The main
effect of $S_{\text{nat}}$ is the elimination of all non-standard $\mathcal{O}
\left( p^2 \right)$ vertices.

iv) The analogy with the non-linear $\sigma$-model is however not complete,
and this fact makes the problem even more interesting. Whereas the constraint
(\ref{1.4}) concerns spin-$0$ fields, the constraints in the electroweak case
operate with gauge fields: a gauge configuration $G_{\mu} \in s_{\text{nat}}$~{\footnote{Small letters denote the algebra of a group.}} satisfies the
constraint if there exists a gauge transformation $\Omega$ such that the image
$H_{\mu}$ of $G_{\mu}$ by $\Omega$ belongs to the sub-algebra $s_{\text{red}}$
\begin{eqnarray}
  G_{\mu} & \overset{\Omega}{\longmapsto} & H_{\mu} \in s_{\text{red}} \subset
  s_{\text{nat}} .  \label{1.7}
\end{eqnarray}
Covariance of (\ref{1.7}) implies that $\Omega$ itself should transform in a
definite (non-linear) way under $S_{\text{nat}}$. The analysis of this
constraint parallels that of the spontaneous symmetry breaking $S_{\text{nat}}
\longrightarrow S_{\text{red}}$, and of the corresponding Goldstone theorem.
There exist $n = \dim s_{\text{nat}} - \dim s_{\text{red}}$ scalar objects
{\emdash}referred to as {\tmem{spurions}} {\emdash} that live in the coset
space $S_{\text{nat}} / S_{\text{red}}$ and transform under $S_{\text{nat}}$
as GBs would do. However, we shall see that the constraint (\ref{1.7}) implies
that spurions have vanishing covariant derivatives and, consequently, in
contrast to GBs, the spurions do not propagate and do not generate mass terms
for vector fields either. There exists a ``standard gauge'' in which spurions
reduce to a set of constants. In the actual case of the group $S_{\text{nat}}$
(\ref{1.6}), the nine spurions reduce to three constants, denoted $\xi, \eta$
and $\zeta$. This reflects the structure of the coset space which, in this
case, is a product of three $\tmop{SU} \left( 2 \right)$ groups.

Spurions allow to keep track of the original symmetry $S_{\text{nat}}$, even
if the latter is explicitly broken. In particular, $\xi, \eta$ and $\zeta$ can
be considered as small expansion parameters describing perturbatively the
explicit breaking of $S_{\text{nat}}$. From this point of view, spurions play
a role similar to quark masses in $\chi$PT: they allow for a classification of
explicit symmetry-breaking operators
under the assumption that such effects are small. As in the case of quark masses, the
smallness of $\xi, \eta$ and $\zeta$ is protected by the symmetry. The
complete LEET invariant under $S_{\text{nat}}$ should be defined as a double
expansion: in powers of momenta and in powers of spurions. The LEET at leading
order coincides with the Higgs-less vertices of the SM, used at tree-level.
The non-standard $\mathcal{O} \left( p^2 \right)$ vertices mentioned above now
reappear as $S_{\text{nat}}$-invariant operators explicitly containing
spurions, i.e. suppressed by powers of the parameters $\xi, \eta$ and $\zeta$.

Such spurions have been introduced in {\cite{Hirn:2004ze,Hirn:2004zz}}, where
their existence and properties have been postulated and justified {\tmem{ad
hoc}}. In the present work, a deeper group-theoretical insight into the origin
and role of spurions is established and presented in a self-contained way in
Section \ref{4-spurions}. Spurions thus appear as unavoidable elements of a
LEET in which the dominance of SM vertices is a consequence of a higher
symmetry $S_{\text{nat}}$ {\footnote{In models based on reduction of
extra-dimensional theories (see for the Higgs-less case
{\cite{Cacciapaglia:2004rb}} and references therein), the suppression of
certain vertices need not result from extra symmetries, but rather from the
locality along the extra dimension. This suppression itself can be improved by
considering a curved extra-dimension.}}.

In Sections \ref{2-SM}, \ref{3-H-less} and \ref{4-spurions}, the origin of
spurions and of the covariant reduction of the symmetry $S_{\text{nat}}$ to
$S_{\text{red}} = \tmop{SU} \left( 2 \right)_L \times \mathrm{U} \left( 1
\right)_Y$ is discussed in details. The emphasis is put from the beginning on
the special status of the lepton-number violating (LNV) sector in a
non-decoupling LEET. The following Sections \ref{5-masses}, \ref{7-list} and
\ref{LNV-NLO} are devoted to observable consequences of the spurion formalism,
starting with the Dirac masses of charged fermions in Section
\ref{quark-masses}. Special attention is paid to the spurion effects which
{\emdash}by power-counting arguments{\emdash} are expected to contribute
before loops. These are genuine effects beyond the SM. Section \ref{7-list}
contains a complete list of such next-to-leading (NLO) effects in the
lepton-number conserving sector. These effects merely involve universal
non-standard couplings of fermions to massive vector bosons. They are
suppressed by the spurions $\xi$ and $\eta$. The oblique corrections only
appear at the NNLO, together with loops, and consequently, they are even more
suppressed. The phenomenological analysis of the NLO is underway
{\cite{Oertel}}.

In Section \ref{4-spurions} it is shown that the existence of $\Delta L = 2$
LNV vertices is a necessary consequence of the reduction $S_{\text{nat}}
\rightarrow S_{\text{red}}$: one of the resulting spurions carries two units
of the $B - L$ charge. All LNV effects, in particular the Majorana masses of
both left-handed and right-handed neutrinos, are suppressed by the
corresponding spurion strength $\zeta^2$. Since the symmetry group
$S_{\text{nat}}$ includes the right isospin group as the origin of the custodial
symmetry, the LEET necessarily contains three species of {\tmem{light}}
right-handed neutrinos. The usual see-saw mechanism
{\cite{Minkowski:1977sc,Mohapatra:1979ia,Gell-Mann:1980vs,Yanagida}} is not efficient. Instead,
we assume in Section \ref{5-masses} that a non-anomalous $\mathbb{Z}_2$
subgroup of the flavor symmetry, which we call $\nu_R$ sign-flip symmetry,
remains unbroken; it forbids neutrino Dirac masses. In this manner, the
right-handed neutrinos decouple from the other fermions. There are three
different possibilities for the introduction of the ($B - L$)-breaking
parameter $\zeta^2$, resulting in different estimates for the ratio of right-
to left-handed neutrinos masses. Two of them seem to be allowed by
cosmological observations constraining the contribution of light and stable
sterile right-handed neutrinos to dark matter (DM), as is discussed in Section
\ref{6-nuR}. Section \ref{LNV-NLO} focuses on the relative importance of
indirect and direct LNV contributions to the process $W^- W^- \longrightarrow
e^- e^-$, a building block for neutrino-less double beta decay ($0 \nu 2
\beta$) and other $\Delta L = 2$ processes. We give our conclusions in Section
\ref{8-concl}.

\section{LNV in theories with elementary scalars} \label{2-SM}

Before we turn to LNV in Higgs-less effective theories, we recall the fate of
this accidental symmetry in the SM.

\subsection{Right isospin and corresponding notations} \label{2-notations}

The complex Higgs doublet of the SM transforms under weak $\tmop{SU} \left( 2
\right)_L$ gauge transformations $G$ and the $\mathrm{U} \left( 1 \right)_Y$
gauge function $\alpha_Y$ as
\begin{eqnarray}
  \varphi & = & \left(\begin{array}{c}
    \varphi_+\\
    \varphi_0
  \end{array}\right) \hspace{0.2em} \longmapsto \hspace{0.2em} G \mathe^{- \mathi
  \frac{\alpha_Y}{2}} \varphi . 
\end{eqnarray}
For further convenience, we display the custodial symmetry {\cite{Sikivie:1980hm}},
using a two-by-two matrix $\Phi$ {\cite{Appelquist:1980vg}}
\begin{eqnarray}
  \Phi & \equiv & \left(\begin{array}{c}
    \varphi_c, \varphi
  \end{array}\right) \nonumber\\ &=& \left(\begin{array}{cc}
    \varphi_0^{\ast} & \varphi_+\\
    - \varphi_+^{\ast} & \varphi_0
  \end{array}\right) \hspace{0.2em} \longmapsto \hspace{0.2em} G \Phi
  \mathe^{\mathi \frac{\tau^3}{2} \alpha_Y},  \label{1.69}
\end{eqnarray}
where the conjugate $\varphi_c \equiv \mathi \tau^2 \varphi^{\ast} \longmapsto
G \mathe^{\mathi \frac{\alpha_Y}{2}} \varphi_c$ of the Higgs doublet $\varphi$
has been introduced. The most general renormalizable lagrangian involving
$\Phi$ and invariant under $\tmop{SU} \left( 2 \right)_L \times \mathrm{U}
\left( 1 \right)_Y$ is
\begin{eqnarray}
  \mathcal{L}_{\text{Higgs}} & = & \frac{1}{2}  \left\langle D_{\mu}
  \Phi^{\dag} D^{\mu} \Phi \right\rangle - \frac{\lambda}{4}  \left(
  \left\langle \Phi^{\dag} \Phi \right\rangle - v^2 \right)^2,  \label{1.64}
\end{eqnarray}
where $\left\langle A \right\rangle \equiv \tmop{Tr} A$ and $D_{\mu} \Phi =
\partial_{\mu} \Phi - \mathi gG_{\mu} \Phi + \mathi g' \Phi b_{\mu} 
\frac{\tau^3}{2}$.

In the limit $g' = 0$, the lagrangian $\mathcal{L}_{\text{Higgs}}$
(\ref{1.64}) is invariant under global $\tmop{SU} \left( 2 \right)_R$
transformations acting from the right on $\Phi$ {\footnote{As pointed out in
the introduction, the $S_{\text{nat}}$ symmetry can be viewed as defining this
custodial symmetry (and its extension to the left-handed non-abelian sector)
without requiring the gauge couplings to be zero.}}. We stress that, when $g'
\neq 0$, only the third component of the $\tmop{SU} \left( 2 \right)_R$ group
is gauged. This $\tmop{SU} \left( 2 \right)_R$ is the group of right isospin
transformations, as can be seen when writing the transformations of the
fermions (depending on their baryon and lepton numbers $B$ and $L$
respectively)
\begin{eqnarray}
  \chi_L & = & \frac{1 - \gamma_5}{2} \chi \hspace{0.2em} \longmapsto
  \hspace{0.2em} G \mathe^{- \mathi \frac{B - L}{2} \alpha_Y} \chi_L, 
  \label{pb}\\
  \chi_R & = & \frac{1 + \gamma_5}{2} \chi \hspace{0.2em} \longmapsto
  \hspace{0.2em} \mathe^{- \mathi \frac{\tau^3}{2} \alpha_Y} \mathe^{- \mathi
  \frac{B - L}{2} \alpha_Y} \chi_R,  \label{1.28}
\end{eqnarray}
where $\chi$ consists of two Dirac spinors arranged in a column, and will be
denoted by $q$ in the case of quarks ($B - L = 1 / 3$) and $\ell$ for the case
of leptons ($B - L = - 1$). Note that this writing does not imply the
existence of a right-handed neutrino, since such a field is not charged under
the $\tmop{SU} \left( 2 \right)_L \times \mathrm{U} \left( 1 \right)_Y$ gauge
symmetry. On the other hand, (\ref{1.69}) shows that the Higgs doublet has a
non-zero value for the third component of the right isospin: $T_R^3 = 1 / 2$,
but vanishing ($B - L$), as should be. In fact, the right isospin group is
explicitly broken not only by gauging the hypercharge, but also by the Yukawa
terms for the fermions: invariance under $\tmop{SU} \left( 2 \right)_L \times
\mathrm{U} \left( 1 \right)_Y$ allows for different masses for the two
components of a fermion doublet
\begin{eqnarray}
  \mathcal{L}_{\text{Yukawa}} & = & - \sum_{i, j = 1}^3 \left(
  \overline{q^i_L} \Phi \left(\begin{array}{cc}
    y^u_{i j} & 0\\
    0 & y^d_{i j}
  \end{array}\right) q^j_R \right) \nonumber\\ && - \sum_{i, j = 1}^3 \left( \overline{\ell^i_L} \Phi \left(\begin{array}{cc}
    0 & 0\\
    0 & y^e_{i j}
  \end{array}\right) \ell^j_R \right) + \text{h.c} .  \label{1.89}
\end{eqnarray}
The sum running over $i, j = 1, \cdots, 3$ in the above equation corresponds
to the three generations. Adding the gauge-invariant kinetic terms for the
gauge fields and the fermions, one recovers the lagrangian for the SM. As an
outcome of the symmetry-breaking mechanism, the generator corresponding to the
unbroken $\mathrm{U} \left( 1 \right)_Q$ subgroup can be expressed as
{\cite{Senjanovic:1975rk}}
\begin{eqnarray}
  Q & = & T_L^3 + T_R^3 + \frac{B - L}{2},  \label{1.31a}
\end{eqnarray}
where the third component of the left isospin $\tmop{SU} \left( 2 \right)_L$
and right isospin $\tmop{SU} \left( 2 \right)_R$, respectively $T_L^3$ and
$T_R^3$, appear. This formula also applies to the Higgs doublet, but it does
not imply the presence in the theory of both left and right isospin gauge
fields.

\subsection{The unique mass-dimension five effective operator}

The renormalizable SM lagrangian is defined to consist of all operators of
mass-dimension $d_M \leqslant 4$ and invariant under $\tmop{SU} \left( 2
\right)_L \times \mathrm{U} \left( 1 \right)_Y$ that can be built with the
fields introduced above. These fields necessarily involve elementary Higgs
scalars, and they transform linearly with respect to the symmetry group
$\tmop{SU} \left( 2 \right)_L \times \mathrm{U} \left( 1 \right)_Y$. Taking
the view that the SM is an effective theory, which has to be completed at
higher energies, one augments the renormalizable SM lagrangian with effective
operators of dimension higher than four, constructed with the same fields, and
respecting the same symmetries
\begin{eqnarray}
  \mathcal{L}_{\text{eff}} & = & \mathcal{L}_{\text{SM}} + \sum_{d_M > 4}
  \frac{1}{\Lambda^{d_M - 4}} \mathcal{O}_{d_M} .  \label{2.x}
\end{eqnarray}
The effective operators $\mathcal{O}_{d_M}$ describe the effects of new
physics beyond the SM. They are ``irrelevant'' at low energies: they are
suppressed by  a dimensionful scale $\Lambda$, related to that at which new
physics appears. Due to the property of renormalizability, the separation of
the lagrangian in two distinct parts {\emdash}the SM lagrangian on one side,
and the effective operators on the other{\emdash} is preserved in loop
calculations. In particular, there is no theoretical inconsistency in assuming
that the dimensionful scale $\Lambda$ is arbitrarily large, in which case the
effects of new physics become vanishingly small {\emdash}hence the
qualification ``decoupling''.

The only mass-dimension five $\tmop{SU} \left( 2 \right)_L \times \mathrm{U}
\left( 1 \right)_Y$-invariant effective operator that can be built with the
fields of the SM, and which is therefore suppressed by the smallest power of
$\Lambda$, is the following lepton number violating operator
{\cite{Weinberg:1979sa}}
\begin{eqnarray}
  \mathcal{L}_{\text{Majorana} L} & = & - \frac{1}{\Lambda}  \sum_{i, j = 1}^3
   c_{i j}  \overline{\ell^i_L} \Phi \tau^+ \Phi^{\dag}  \left( \ell^j_L
  \right)^c + \text{h.c}  .  \label{1.94}
\end{eqnarray}
We have used the following definition for the conjugate of a doublet of
left-handed leptons
\begin{eqnarray}
  \left( \ell_L \right)^c & = & \mathi \tau^2 C \overline{\left( \ell_L
  \right)}^T \hspace{0.2em} = \hspace{0.2em} \left(\begin{array}{c}
    \left( e_L \right)^c\\
    - \left( \nu_L \right)^c
  \end{array}\right),  \label{conj}
\end{eqnarray}
where $C$ is the charge conjugation matrix, defined to satisfy $C^{- 1}
\gamma_{\mu} C = - \gamma_{\mu}^T$.

The effective operator (\ref{1.94}) encodes the low-energy consequences of an
unknown mechanism generating masses for the left-handed neutrinos. The
description by an effective operator is useful if the scale $\Lambda$ is large
enough (with eigenvalues of order unity for the matrix $c_{i j}$) compared to
the vacuum expectation value $v$ of the Higgs field. One
concludes that the left-handed neutrinos have Majorana masses of order $v^2 /
\Lambda$. Provided $\Lambda$ is large enough, this may well be much smaller
than the masses of the charged fermions, which are of order $v$ times a Yukawa
coupling~{\footnote{We will not get into the problem of accounting for the six
orders of magnitude between the electron and top quark masses, which is part
of the flavor problem.}}. The see-saw mechanism
{\cite{Minkowski:1977sc,Mohapatra:1979ia,Gell-Mann:1980vs,Yanagida}} would provide one possible
dynamical origin for the term (\ref{1.94}). In this case, $\Lambda$ is given
in terms of the right-handed Majorana masses.

\section{At which chiral order does LNV appear in the Higgs-less effective
theory?} \label{3-H-less}

In the case of Higgs-less EWSB, we do not have a renormalizable theory to
start with, and the framework of decoupling effective theories (\ref{2.x}) is
not suitable. In the alternative framework of {\tmem{non-decoupling}}
effective theories, all operators respecting the symmetries must be included
in the effective lagrangian, but they are no longer ordered according to their
mass-dimension $d_M$ as in (\ref{2.x}): one must use instead the infrared
(or chiral) dimension $d_{\text{IR}}$ reflecting their behavior in the
low-energy limit $p \rightarrow 0$. The effective lagrangian is expressed as
\begin{eqnarray}
  \mathcal{L}_{\text{eff}} & = & \sum_{d_{\text{IR}} \geqslant 2}
  \mathcal{L}_{d_{\text{IR}}},  \label{3.1}
\end{eqnarray}
where in the limit of small momenta $p$, the operators of
$\mathcal{L}_{d_{\text{IR}}}$ behave as
\begin{eqnarray}
  \mathcal{L}_{d_{\text{IR}}} & = & \mathcal{O} \left( p^{d_{\text{IR}}}
  \right) . 
\end{eqnarray}
The loops are renormalized order by order in the low-energy expansion
(\ref{3.1}). We summarize the rules of the {\tmem{infrared power-counting}}
underlying the expansion (\ref{3.1}) and the corresponding order-by-order
renormalization. Afterwards, we ask at which place of the expansion
(\ref{3.1}) LNV appears for the first time.

\subsection{Infrared power counting} \label{3-WPCF}

In the minimal Higgs-less theory, the low-energy description of the
symmetry-breaking sector comprises only the three GBs that are eaten to give
masses to the gauge bosons $W^{\pm}$ and $Z^0$. In addition, there are chiral
fermions $\chi_L, \chi_R$. As in $\chi$PT, the three GBs
$\overset{\rightarrow}{\pi} \left( x \right)$ are collected in a matrix
$\Sigma \left( x \right) \in \tmop{SU} \left( 2 \right)$, and they transform
non-linearly with respect to $\tmop{SU} \left( 2 \right)_L \times \mathrm{U}
\left( 1 \right)_Y$ (compare (\ref{1.69}))
\begin{eqnarray}
  \Sigma \left( x \right) & \equiv & \mathe^{\frac{\mathi}{f} 
  \overset{\rightarrow}{\pi} \left( x \right) \cdot
  \overset{\rightarrow}{\tau}} \hspace{0.2em} \longmapsto \hspace{0.2em} G \left(
  x \right) \Sigma \left( x \right) \mathe^{\mathi \alpha_Y \left( x \right) 
  \frac{\tau^3}{2}} .  \label{3.3}
\end{eqnarray}
All particles included in the LEET should have their masses protected by a
symmetry; masses then only come in with an explicit power of expansion
parameters. A crucial ingredient for the non-decoupling effective theory is
therefore technical naturalness {\cite{'tHooft:1979bh}}: there should be a
limit in the parameter space, in which all particles become massless, and the
symmetry of $\mathcal{L}_{\text{eff}}$ is increased. The LEET involves a
systematic expansion around that limit. The original discussion of power
counting for the GB sector in $\chi$PT can be found in
{\cite{Weinberg:1979kz,Gasser:1984yg,Gasser:1985gg}}. Generalizations to
include other degrees of freedom have already been considered in
{\cite{Wudka:1994ny,Urech:1995hd}} for the case of gauge fields, and in
{\cite{Wudka:1994ny,Nyffeler:1999ap}} for the case of chiral fermions.

\subsubsection{Goldstone bosons}

Due to the non-linear transformation (\ref{3.3}), the GB fields
$\overset{\rightarrow}{\pi}$ carry no infrared dimension, i.e.
$\overset{\rightarrow}{\pi} =\mathcal{O} \left( 1 \right)$. Their physical
mass-dimension one is compensated by the dimensionful constant $f$ which
represents an intrinsic scale of the theory and is not affected by the
low-energy limit: in $\chi$PT, it coincides with the pion decay constant
$f_{\pi} \simeq 92.4 \tmop{MeV}$, whereas in the case of EWSB, $f$ replaces
the Higgs vacuum expectation value $f \simeq 250 \tmop{GeV}$. Hence, in the
low-energy limit
\begin{eqnarray}
  \Sigma & = & \mathcal{O} \left( 1 \right), \\
  D_{\mu} \Sigma & = & \mathcal{O} \left( p \right),  \label{3.5}
\end{eqnarray}
where $D_{\mu}$ is the covariant derivative. The lowest-order contribution of GBs to
$\mathcal{L}_{\text{eff}}$ takes the form
\begin{eqnarray}
  \mathcal{L}_{\text{GB}} & = & \frac{f^2}{4}  \left\langle D_{\mu}
  \Sigma^{\dag} D^{\mu} \Sigma \right\rangle \hspace{0.2em} = \hspace{0.2em}
  \mathcal{O} \left( p^2 \right) .  \label{3.6}
\end{eqnarray}
Estimating loop contributions, one infers that the LEET should be applicable
for momenta {\cite{Manohar:1984md,Georgi:1985kw}}
\begin{eqnarray}
  p & \ll & 4 \mathpi f \hspace{0.2em} \simeq \hspace{0.2em} 3 \tmop{TeV} . 
\end{eqnarray}

\subsubsection{Gauge fields}

In order to ensure that the vector bosons $W_{\mu}$ and $Z_{\mu}$ be naturally
light, one has to treat them as weakly-coupled gauge fields
\begin{eqnarray}
  gG_{\mu} \hspace{0.2em} \in \hspace{0.2em} \tmop{su} ( 2 ), &  & g' b_{\mu}
  \hspace{0.2em} \in \hspace{0.2em} \mathrm{u} \left( 1 \right), 
\end{eqnarray}
acquiring their masses via the Higgs mechanism (without a Higgs particle). We
stress that this is a necessity once one requires a low-energy power-counting
involving these vector fields
{\cite{Georgi:1990xy,Burgess:1993gx,Wudka:1994ny}}. At leading order, $M_W$
and $M_Z$ can be directly inferred from (\ref{3.6}). The consistency of the
low-energy expansion requires the squared masses to be counted as $\mathcal{O}
\left( p^2 \right)$
\begin{eqnarray}
  M_W^2 & = & \frac{g^2}{4} f^2 \hspace{0.2em} = \hspace{0.2em} \mathcal{O} \left(
  p^2 \right),  \label{A-85}\\
  M_Z^2 & = & \frac{g^2 + g'^2}{4} f^2 \hspace{0.2em} = \hspace{0.2em} \mathcal{O}
  \left( p^2 \right) .  \label{A-93}
\end{eqnarray}
Since $f^2$ is a fixed scale, one must count the gauge couplings as
\begin{eqnarray}
  g^2 & \simeq & \frac{4 M_W^2}{f^2} \hspace{0.2em} = \hspace{0.2em} \mathcal{O}
  \left( p^2 \right),  \label{3.11}\\
  g'^2 & \simeq & \frac{4}{f^2}  \left( M_Z^2 - M_W^2 \right) \hspace{0.2em} =
  \hspace{0.2em} \mathcal{O} \left( p^2 \right) .  \label{3.12}
\end{eqnarray}
Hence, in the low-energy limit, gauge couplings must be considered as
vanishing proportionally to external momenta. Since the covariant derivative
$D_{\mu} \Sigma$ (\ref{3.5}) involves gauge connections $gG_{\mu}$ and $g'
b_{\mu}$, (\ref{3.11}-\ref{3.12}) in turn imply that the infrared dimension of
canonically normalized gauge fields $G_{\mu}$ and $b_{\mu}$ vanishes
\begin{eqnarray}
  G_{\mu} & = & \mathcal{O} \left( 1 \right), \\
  b_{\mu} & = & \mathcal{O} \left( 1 \right) . 
\end{eqnarray}
In the field strength $G_{\mu \nu}$, both the derivative and the non-linear
terms thus count as $\mathcal{O} \left( p \right)$
\begin{eqnarray}
  G_{\mu \nu} & = & \partial_{\mu} G_{\nu} - \partial_{\nu} G_{\mu} - \mathi g
  \left[ G_{\mu}, G_{\nu} \right] \hspace{0.2em} = \hspace{0.2em} \mathcal{O}
  \left( p \right) . 
\end{eqnarray}
Consequently, in the Yang-Mills lagrangian
\begin{eqnarray}
  \mathcal{L}_{\text{YM}} \hspace{0.2em} = \hspace{0.2em} - \frac{1}{2} 
  \left\langle G_{\mu \nu} G^{\mu \nu} \right\rangle & = & \mathcal{O} \left(
  p^2 \right),  \label{3.16}
\end{eqnarray}
the kinetic term {\tmem{and}} both trilinear and quartic couplings are counted
as $\mathcal{O} \left( p^2 \right)$. Hence the low-energy expansion preserves
gauge invariance order by order, in contrast with the more familiar expansion
in powers of coupling constants {\tmem{only}}, used in renormalizable
theories. Finally, since $M_W, M_Z =\mathcal{O} \left( p \right)$, the massive
gauge boson propagators exhibit the homogeneous low-energy behavior
$\mathcal{O} \left( p^{- 2} \right)$.

\subsubsection{Chiral fermions} \label{3.1.3}

As was the case for GBs and gauge fields, the infrared dimension of chiral
fermion fields is one unit less than their physical dimension
\begin{eqnarray}
  \chi_{R, L} & = & \mathcal{O} \left( p^{1 / 2} \right) . 
\end{eqnarray}
This corresponds to the expected low-energy behavior of fermion bilinears
$\overline{\chi} \Gamma \chi =\mathcal{O} \left( p \right)$. In particular,
the fermion kinetic term contributes to the lagrangian as
\begin{eqnarray}
  \mathcal{L}_f & = & \mathi \overline{\chi} \gamma^{\mu} D_{\mu} \chi
  \hspace{0.2em} = \hspace{0.2em} \mathcal{O} \left( p^2 \right), 
\end{eqnarray}
similarly to GBs (\ref{3.6}) and to gauge fields (\ref{3.16}). On the other
hand, the fermion mass term is counted as
\begin{eqnarray}
  \mathcal{L}_{\text{mass}} & = & - m \overline{\chi} \chi \hspace{0.2em} =
  \hspace{0.2em} \mathcal{O} \left( mp \right) . 
\end{eqnarray}
Unless the fermion mass $m$ is suppressed {\tmem{at least}} as $\mathcal{O}
\left( p \right)$, i.e.
\begin{eqnarray}
  m & = & \mathcal{O} \left( p^n \right), \hspace{0.2em} n \geqslant 1, 
\end{eqnarray}
the low-energy behavior $\mathcal{O} \left( p^{- 1} \right)$ of the fermion
propagator is destroyed. We shall return to this potential problem shortly.

The above discussion can be summarized as follows. A local
operator/interaction vertex $\mathcal{O}$ built from GBs $\Sigma$, gauge
fields $G_{\mu}$ and $b_{\mu}$, and from fermions $\chi_{L, R}$ carries the
infrared dimension
\begin{eqnarray}
  d_{\text{IR}} \left[ \mathcal{O} \right] & = & n_{\partial} \left[
  \mathcal{O} \right] + n_g \left[ \mathcal{O} \right] + \frac{1}{2} n_f
  \left[ \mathcal{O} \right],  \label{3.16a}
\end{eqnarray}
where $n_{\partial} \left[ \mathcal{O} \right]$ is the number of derivatives
entering the operator $\mathcal{O}$, $n_g \left[ \mathcal{O} \right]$ the
number of gauge coupling constants and $n_f \left[ \mathcal{O} \right]$ the
number of fermion fields. This infrared counting rule provides the basis for
the ordering of contributions to the effective lagrangian (\ref{3.1}).

\subsection{Generalized Weinberg power-counting formula} \label{WPC}

We consider a connected Feynman diagram $\Gamma$ built from vertices
$\mathcal{O}_v$ of the lagrangian (\ref{3.1}) labelled by $v = 1, \cdots, V$.
Replacing external lines by the corresponding fields, the diagram $\Gamma$ can
be compared with an operator of the effective lagrangian (\ref{3.1}). We call
its infrared dimension $d_{\text{IR}} \left[ \Gamma \right]$. The low-energy
limit implies a rescaling of gauge couplings and external and internal momenta
\begin{eqnarray}
  p_i & \longmapsto & tp_i,  \label{20}\\
  g & \longmapsto & tg .  \label{21}
\end{eqnarray}
The masses must be rescaled as well
\begin{eqnarray}
  M_{W, Z} & \longmapsto & tM_{W, Z}, \\
  m_{\chi} & \longmapsto & t^n m_{\chi}, \hspace{0.2em} n \geqslant 1, 
\end{eqnarray}
and all external fermion fields
\begin{eqnarray}
  \chi^{\text{ext}} & \longmapsto & t^{1 / 2} \chi^{\text{ext}} .  \label{22}
\end{eqnarray}
$d_{\text{IR}} \left[ \Gamma \right]$ appears as the homogeneous degree of the
diagram $\Gamma$ in the low-energy limit $t \longrightarrow 0$
\begin{eqnarray}
  \Gamma & \longmapsto & t^{d_{\text{IR}} \left[ \Gamma \right]} \Gamma . 
\end{eqnarray}
Hence, $d_{\text{IR}} \left[ \Gamma \right]$ measures the degree of
suppression of a Feynman diagram in the low-energy expansion, i.e. for $t
\longrightarrow 0$. This is true if one uses dimensional regularization and a
mass-independent renormalization scheme, in which case there are no powers of
a cut-off involved, and the naive power-counting is valid
{\cite{Georgi:1993qn,Pich:1998xt}}. As re-derived in Appendix \ref{A-WPC},
$d_{\text{IR}} \left[ \Gamma \right]$ is given by 
{\cite{Weinberg:1979kz,Wudka:1994ny}} (see also \cite{He:1996rb})
\begin{eqnarray}
  d_{\text{IR}} \left[ \Gamma \right] & = & 2 + 2 L + \sum^V_{v = 1} \left(
  d_{\text{IR}} \left[ \mathcal{O}_v \right] - 2 \right),  \label{3.28}
\end{eqnarray}
where $L$ is the number of loops and $d_{\text{IR}} \left[ \mathcal{O}_v
\right]$ is the infrared dimension (\ref{3.16a}) of the vertex
$\mathcal{O}_v$. This formula is formally analogous to Weinberg's
power-counting formula, originally introduced in an effective theory involving
GBs only {\cite{Weinberg:1979kz}}. The point is that it can also include
vector fields (when introduced as gauge fields) and chiral fermions (if they
remain massless, or if their masses can consistently be counted as
$\mathcal{O} \left( p^1 \right)$).

The result (\ref{3.28}) calls for two comments. The first is that, for a given
precision (a given chiral order $d_{\text{IR}}$), only a finite number of
diagrams contributes, since all terms on the right-hand side are positive: we
have $L \geqslant 0$, and, if the effective lagrangian only contains terms of
order $\mathcal{O} \left( p^2 \right)$ or higher, $d_{\text{IR}} \left[
\mathcal{O} \right] \geqslant 2$. This last point is crucial to ensure that
the low-energy limit is weakly interacting. In particular, in a LEET
containing scalar fields other than GBs (or superpartners of chiral fermions),
the requirement $d_{\text{IR}} \left[ \mathcal{O} \right] \geqslant 2$ is
difficult to satisfy.

The second comment is that the chiral expansion is intimately related to a
loop expansion: the order $d_{\text{IR}}$ of a diagram increases with the
number of loops $L$. More precision can systematically be achieved by going to
the next order in the expansion, involving Feynman diagrams with one
additional loop. The divergences in loops are absorbed by the (finite number
of) new operators appearing at the corresponding order $d_{\text{IR}}$ in the
lagrangian, yielding finite and renormalization-scale independent results
{\footnote{This has been checked explicitly at one loop for the bosonic sector
of an effective theory of EWSB without a Higgs in {\cite{Nyffeler:1999ap}}.}}.
See also Appendix \ref{Unitarity} in this respect.

\subsection{ What becomes of the irrelevant operators of the SM in the
Higgs-less effective theory?} \label{3.2-irr-H-less}

In the simplest effective theory approach to Higgs-less EWSB, one constructs
the most general lagrangian with the same fields as in Section \ref{2-SM},
except that the Higgs fields (the matrix $\Phi$ with mass-dimension one) is
replaced by the $2 \times 2$ unitary and unimodular matrix $\Sigma$,
describing the three GBs. The most important difference is that the rule for
ordering the operators is changed from (\ref{2.x}) to (\ref{3.1}). On the
other hand, the assumed symmetry is the same as in the SM, i.e. $\tmop{SU}
\left( 2 \right)_L \times \mathrm{U} \left( 1 \right)_Y$. As shown below,
basing the construction on this sole symmetry leads to difficulties both in
the comparison with experiment, and for the consistency of the expansion
itself. This will be remedied in Section \ref{4-spurions}.

\subsubsection{The $\mathcal{O} \left( p^1 \right)$ lepton-number violating
operator} \label{3.3.1}

Using a left-handed lepton doublet $\ell_L$ transforming as in (\ref{pb}) with
$B - L = - 1$, one can construct the following $\tmop{SU} \left( 2 \right)_L
\times \mathrm{U} \left( 1 \right)_Y$ invariant, which breaks custodial
symmetry as well as lepton number, compare (\ref{1.94})
\begin{eqnarray}
  \Lambda \overline{\ell_L} \Sigma \tau^+ \Sigma^{\dag}  \left( \ell_L
  \right)^c & = & \mathcal{O} \left( p^1 \right) .  \label{5.26}
\end{eqnarray}
According to the power-counting rules given in \ref{3-WPCF}, it appears at
$\mathcal{O} \left( p^1 \right)$, without any suppression factor
{\footnote{The operator in equation (\ref{5.26}) appears multiplied by the
appropriate power of $\Lambda$ (or $f$) in order to match the dimension of a
term in the lagrangian. One can always argue as to which one is more
appropriate, the difference being a factor of $4 \mathpi$. This does not upset
the argument regarding the formal ordering of operators in the low-energy
expansion.\label{foot-Lambda-or-f}}}. Since this operator (\ref{5.26}) has
chiral dimension less than two, it violates the requirement of $d_{\text{IR}}
\left[ \mathcal{O} \right] \geqslant 2$, which was pointed out at the end of
Section \ref{3-WPCF} to be crucial for the low-energy expansion to make sense.
We conclude that the presence, at this order, of the lepton-number violating
effective operator (\ref{5.26}) would not only ruin any chance of
phenomenological success of such a LEET, but also endanger its very
consistency.

\subsubsection{A whole class of $\mathcal{O} \left( p^2 \right)$ operators}
\label{3-other}

Among all $\tmop{SU} \left( 2 \right)_L \times \mathrm{U} \left( 1
\right)_Y$-invariant operators of lowest order in the Higgs-less theory, one
finds $\mathcal{O} \left( p^2 \right)$ operators which have no equivalent in
the renormalizable lagrangian of the SM. On the other hand, one could build
the corresponding invariants using the fields of the SM: they would have
mass-dimension six, and their effects are therefore suppressed by a
two powers of a dimensionful scale in the SM case. In absence of the Higgs particle, due the
concomitant ``replacement'' of $\Phi$ by $\Sigma$, this suppression by a
dimensionful scale no longer holds: in addition to their appearing at
$\mathcal{O} \left( p^2 \right)$, these operators are not dimensionally
suppressed anymore. The first operator can be described as ``giving a
tree-level contribution to the $S$ parameter''
{\cite{Holdom:1990tc,Peskin:1992sw}}
\begin{eqnarray}
  b_{\mu \nu}  \left\langle \Sigma \frac{\tau^3}{2} \Sigma^{\dag} G^{\mu \nu}
  \right\rangle & = & \mathcal{O} \left( p^2 \right),  \label{E.17}
\end{eqnarray}
and the second one as ``contributing to the $T$ parameter''
{\cite{Longhitano:1980iz}} {\footnote{Once again, footnote
\ref{foot-Lambda-or-f} applies.}}
\begin{eqnarray}
  f^2  \left\langle \frac{\tau^3}{2} \Sigma^{\dag} D_{\mu} \Sigma
  \right\rangle^2 & = & \mathcal{O} \left( p^2 \right) .  \label{E.16}
\end{eqnarray}
If such operators appear at $\mathcal{O} \left( p^2 \right)$, then they modify
the two-point functions of vector fields already at that level. If they appear
at tree-level, these two operators can be directly interpreted as oblique
corrections: with our normalization of (\ref{E.16}), the constants appearing
in front of them are then, respectively identified as $- gg' S / \left( 16
\mathpi \right)$ and $e^2 T / \left( 64 \mathpi \right)$.  Here again, it
seems that a direct application of LEETs to Higgs-less EWSB over-predicts
deviations from the SM: compared to the default estimate of $1$, one would
need a suppression by more than a factor of $4 \mathpi$ in order to agree with
the current limits from {\cite{LEP,PDG}}, of order $10^{- 3}$. In addition,
rather than imposing a suppression by hand on these two operators, we would
like it to be systematic, based on a symmetry that protects it.

Other ``unwanted'' operators involve fermions: the following non-universal
couplings {\cite{Appelquist:1985rr,Peccei:1990kr}} to massive vector bosons
appear at $\mathcal{O} \left( p^2 \right)$
\begin{eqnarray}
  \mathi \overline{\chi^i_L} \gamma^{\mu}  \left( \Sigma D_{\mu} \Sigma^{\dag}
  \right) \chi^j_L & = & \mathcal{O} \left( p^2 \right),  \label{E.23}\\
  \mathi \overline{\chi^i_R} \gamma^{\mu}  \left( \Sigma^{\dag} D_{\mu} \Sigma
  \right) \chi^j_R & = & \mathcal{O} \left( p^2 \right) .  \label{E.24}
\end{eqnarray}
In addition, (\ref{E.24}) introduces couplings of the right-handed fermions to
the $W^{\pm}$. Both types of operators would also be a new source of
flavor-changing currents, requiring a redefinition of the CKM matrix, which
would not be unitary anymore. These operators would also introduce
flavor-changing neutral currents (FCNCs) at this level.

The magnetic moment operators such as
\begin{eqnarray}
  \frac{1}{\Lambda}  \overline{\chi_L} \Sigma \sigma^{\mu \nu} \chi_R b^{\mu
  \nu} & = & \mathcal{O} \left( p^2 \right),  \label{blast}\\
  \frac{1}{\Lambda}  \overline{\chi_L} G_{\mu \nu} \Sigma \sigma^{\mu \nu}
  \chi_R & = & \mathcal{O} \left( p^2 \right),  \label{last}
\end{eqnarray}
also have chiral dimension two, but mass-dimension five, and can therefore be
suppressed thanks to a dimensional scale. Nonetheless, this suppression is not
as strong as in the linear case: these operators would have mass-dimension six
if constructed with SM fields. In fact, if the scale appearing in front of
these operators was indeed $\Lambda_{\text{w}} \sim 3 \tmop{TeV}$, the
anomalous magnetic moment of the muon generated would still be larger than is
measured {\cite{Arzt:1992wz,Einhorn:2001mf}}.

The operators (\ref{E.17}-\ref{E.24}) have already been mentioned in
{\cite{Hirn:2004ze}}, but the case of the lepton-number violating one
(\ref{5.26}) was not fully analyzed, and the common interpretation of this
class of operators as corresponding to mass-dimension six operators in the SM
was not given.

There are other $\mathcal{O} \left( p^2 \right)$ operators
which correspond to irrelevant operators in the SM (four-fermion
interactions), but they are also suppressed by a scale in the Higgs-less
effective theory. Such operators are not related to symmetry
breaking (they do not involve the Higgs doublet in the SM or the GB matrix in
the Higgs-less case), the scale involved need not be the same as
$\Lambda_{\text{w}} \sim 3 \tmop{TeV}$, but may be larger. If so, the
difficulty would not be more acute that in the SM. Therefore, we disregard
these operators in this paper, since we have no control over the values of
these dimensionful scales: we simply assume that they are large enough so that
we can neglect the corresponding operators.

\subsubsection{Dirac mass terms} \label{3.2.3-Yuk}

Although they were not irrelevant in the SM, we also mention Yukawa terms,
which have mass-dimension three in the Higgs-less case. One finds that they
can be written down at $\mathcal{O} \left( p^1 \right)$
\begin{eqnarray}
  \Lambda \overline{\chi_L} \Sigma \chi_R & = & \mathcal{O} \left( p^1
  \right),  \label{B.20}\\
  \Lambda \overline{\chi_L} \Sigma \tau^3 \chi_R & = & \mathcal{O} \left( p^1
  \right) .  \label{E.21}
\end{eqnarray}
This is dangerous for the consistency of the expansion: the Weinberg
power-counting formula (\ref{3.28}) then shows that the loop expansion does
not make sense, since there are operators with degree $d_{\text{IR}} \left[
\mathcal{O} \right] < 2$.

\section{Higher symmetry $S_{\text{nat}}$, its reduction and spurions} \label{4-spurions}

We ask whether an expansion procedure exists that is consistent with the
principles of a LEET, and in which the unwanted operators of the previous
section are relegated to higher orders. This is the motivation behind the
$S_{\text{nat}}$ symmetry and the spurion formalism.

\subsection{The symmetry group $S_{\text{nat}}$}

To achieve the aforementioned goal, we require the lagrangian to be invariant
under a symmetry group $S_{\text{nat}}$, which contains as a subgroup the
electroweak one $S_{\text{red}} = \tmop{SU} \left( 2 \right)_L \times
\mathrm{U} \left( 1 \right)_Y$. The symmetry group $S_{\text{nat}}$ is a
product of groups acting separately on the composite sector (which produces
the three GBs as its only massless bound states) and separately on the
elementary sector (quarks, leptons and Yang-Mills fields). The two sectors are
coupled via constraints from which the existence of spurions follows. In
reference {\cite{Hirn:2004ze}}, the group $S_{\text{nat}} = \left[ \tmop{SU}
\left( 2 \right) \times \tmop{SU} \left( 2 \right) \right]^2 \times \mathrm{U}
\left( 1 \right)$ was shown to be large enough to introduce a suppression of
the operators described in (\ref{E.17}-\ref{E.24}). Since we are interested in
introducing the minimal number of particles, we will stick to this group. In
order to clarify the discussion, the formalism is rephrased from the onset,
implying some modifications with respect to the generic notation of
{\cite{Hirn:2004ze}}.

We assume an underlying theory responsible for the spontaneous symmetry
breaking of $\tmop{SU} \left( 2 \right)_{\Gamma_L} \times \tmop{SU} \left( 2
\right)_{\Gamma_R} \subset S_{\text{nat}}$ down to its vector subgroup. This
produces a triplet of GBs, which we parametrize by a unitary unimodular matrix
$\Sigma$ transforming according to
\begin{eqnarray}
  \Sigma & \longmapsto & \Gamma_L \Sigma \Gamma_R^{\dag},  \label{00000}
\end{eqnarray}
where $\Gamma_L \in \tmop{SU} \left( 2 \right)_{\Gamma_L}$ and $\Gamma_R \in
\tmop{SU} \left( 2 \right)_{\Gamma_R}$. The global $\tmop{SU} \left( 2
\right)_{\Gamma_L} \times \tmop{SU} \left( 2 \right)_{\Gamma_R}$ symmetry is
then promoted to a local one, in order to define a generating functional for
its Noether currents. The corresponding $\Gamma_{L \mu}$ and $\Gamma_{R \mu}$
sources transform according to
\begin{eqnarray}
  \Gamma_{L \mu} & \longmapsto & \Gamma_L \Gamma_{L \mu} \Gamma_L^{\dag} +
  \mathi \Gamma_L \partial_{\mu} \Gamma_L^{\dag}, \\
  \Gamma_{R \mu} & \longmapsto & \Gamma_R \Gamma_{R \mu} \Gamma_R^{\dag} +
  \mathi \Gamma_R \partial_{\mu} \Gamma_R^{\dag} . 
\end{eqnarray}
This composite sector will be at the origin of the electroweak symmetry
breaking once gauge fields are appropriately coupled to its conserved
currents. Note that the $\tmop{SU} \left( 2 \right)_{\Gamma_L} \times
\tmop{SU} \left( 2 \right)_{\Gamma_R}$ structure parallels that of the Higgs
sector in the renormalizable SM, see Section \ref{2-notations}, and is
necessary for the custodial symmetry to play its role.

Turning now to the elementary sector, we recall the formula (\ref{1.31a}), which
indicates that we have to introduce an $\tmop{SU} \left( 2 \right)_{G_L}
\times \tmop{SU} \left( 2 \right)_{G_R} \times \mathrm{U} \left( 1 \right)_{B
- L}$ structure acting on elementary fermions {\footnote{This is indeed the
gauge group in Higgs-less models {\cite{Csaki:2003zu}}, and in left-right symmetric
models {\cite{Senjanovic:1975rk}}. Here, although we will introduce in total
thirteen vector fields, they will subsequently be restricted to take values in the algebra
of $\tmop{SU} \left( 2 \right)_L \times \mathrm{U} \left( 1 \right)_Y$ in
Section \ref{4.2-spurions}.}}. The elementary $G_{L \mu}$ and $G_{R \mu}$
gauge fields transform under $G_L \in \tmop{SU} \left( 2 \right)_{G_L}$ and
$G_R \in \tmop{SU} \left( 2 \right)_{G_R}$ as
\begin{eqnarray}
  g_L G_{L \mu} & \longmapsto & G_L g_L G_{L \mu} G_L^{\dag} + \mathi G_L
  \partial_{\mu} G_L^{\dag}, \\
  g_R G_{R \mu} & \longmapsto & G_R g_R G_{R \mu} G_R^{\dag} + \mathi G_R
  \partial_{\mu} G_R^{\dag} . 
\end{eqnarray}
The elementary fermion doublets with $B - L = - 1$ will be denoted by $\ell$
and those with $B - L = 1 / 3$ by $q$. Their transformations represent an
extension of those of (\ref{pb}-\ref{1.28})
\begin{eqnarray}
  \chi_L & \longmapsto & G_L \mathe^{- \mathi \frac{B - L}{2} \alpha} \chi_L, 
  \label{cp}\\
  \chi_R & \longmapsto & G_R \mathe^{- \mathi \frac{B - L}{2} \alpha} \chi_R .
  \label{cq}
\end{eqnarray}
The corresponding $\mathrm{U} \left( 1 \right)_{B - L}$ gauge field
transforms as
\begin{eqnarray}
  g_B G_{B \mu} & \longmapsto & g_B G_{B \mu} - \partial_{\mu} \alpha, 
\end{eqnarray}
which we rewrite, embedding $\mathrm{U} \left( 1 \right)$ into an $\tmop{SU}
\left( 2 \right)$ group, as
\begin{eqnarray}
  g_B G_{B \mu}  \frac{\tau^3}{2} & \longmapsto & \mathe^{- \mathi \alpha
  \frac{\tau^3}{2}} g_B G_{B \mu}  \frac{\tau^3}{2} \mathe^{\mathi \alpha
  \frac{\tau^3}{2}} \nonumber\\
&&+ \mathi \mathe^{- \mathi \alpha \frac{\tau^3}{2}}
  \partial_{\mu} \mathe^{\mathi \alpha \frac{\tau^3}{2}} .  \label{4.9}
\end{eqnarray}
Note that, in this formalism, the $\chi_R$ are $\tmop{SU} \left( 2 \right)_{G_R}$ doublets. This implies, in
particular, that we have introduced $\nu_R$ degrees of freedom.

The symmetry group of the theory is
\begin{eqnarray}
  S_{\text{nat}} & = & \tmop{SU} \left( 2 \right)_{G_L} \times \tmop{SU}
  \left( 2 \right)_{\Gamma_L} \times \tmop{SU} \left( 2 \right)_{\Gamma_R}
  \times \tmop{SU} \left( 2 \right)_{G_R} \nonumber\\
&&\times \mathrm{U} \left( 1
  \right)_{B - L},  \label{Snat}
\end{eqnarray}
hence the effective lagrangian should contain all invariants under this group,
organized according to their chiral dimensions $d_{\text{IR}}$. The operators with
chiral dimension $\mathcal{O} \left( p^2 \right)$ involving the fields
introduced at this stage are collected in $\mathcal{L} \left( p^2 \right)$
\begin{eqnarray}
  \mathcal{L} \left( p^2 \right) & = & \frac{f^2}{4}  \left\langle D_{\mu}
  \Sigma^{\dag} D^{\mu} \Sigma \right\rangle + \mathi \overline{\chi_L}
  \gamma^{\mu} D_{\mu} \chi_L + \mathi \overline{\chi_R} \gamma^{\mu} D_{\mu}
  \chi_R \nonumber\\
  & - & \frac{1}{2}  \left\langle G_{L \mu \nu} G_L^{\mu \nu} + G_{R \mu \nu}
  G^{\mu \nu}_R \right\rangle - \frac{1}{4} G_{B \mu \nu} G_B^{\mu \nu} . 
  \label{lag}
\end{eqnarray}
The covariant derivatives are defined in the standard manner in terms of the
connections $\Gamma_{L \mu}, \Gamma_{R \mu}, g_L G_{L \mu}, g_R G_{R \mu}, g_B
G_{B \mu}$, following the transformation properties (\ref{00000}) and
(\ref{cp}-\ref{cq}).

Equation (\ref{lag}) contains all $S_{\text{nat}}$-invariant operators that
are $\mathcal{O} \left( p^2 \right)$ except four-fermion operators without
derivatives. The latter are dimensionally suppressed by the inverse squared of
a scale $\Lambda_{4 \text{f}}$ which we assume to be much higher than the
scale $\Lambda_{\text{w}}$ (\ref{1.1}) of the LEET. Notice that such $\mathcal{O} \left( p^2
\right)$ four-fermion operators cannot be generated by loops.

\subsection{Covariant reduction of the symmetry $S_{\text{nat}} \rightarrow
\tmop{SU} \left( 2 \right)_L \times \mathrm{U} \left( 1 \right)_Y$}
\label{4.2-spurions}

The symmetry $S_{\text{nat}}$ is large enough to eliminate all the unwanted
couplings discussed in Section \ref{3.2-irr-H-less} at the leading chiral
order $\mathcal{O} \left( p^2 \right)$ described by the lagrangian
(\ref{lag}). On the other hand, $S_{\text{nat}}$ is too large: the lagrangian
(\ref{lag}) contains thirteen gauge connections $g_L G_{L \mu}, g_R G_{R \mu},
g_B G_{B \mu}, \Gamma_{L \mu}, \Gamma_{R \mu}$, as compared to four in the SM.
Due to the GB term in (\ref{lag}) (first term on the right-hand side), the
three combinations $\Gamma_{R \mu}^a - \Gamma_{L \mu}^a$ acquire a mass term
by the Higgs mechanism, whereas all ten remaining vector fields as well as
fermions remain massless. Furthermore, the lagrangian (\ref{lag}) does not
contain any coupling that would transmit the symmetry breaking from the
composite sector ($\Sigma, \Gamma_{L \mu}, \Gamma_{R \mu}$) to the elementary
sector ($G_{L \mu}, G_{R \mu}, G_{B \mu}, \chi$). Following
{\cite{Hirn:2004ze}}, such a coupling will be introduced {\emdash}and the
number of degrees of freedom reduced{\emdash} via constraints identifying
certain connections of $S_{\text{nat}}$ up to a gauge transformation. This
will provide the reduction $S_{\text{nat}} \longrightarrow \tmop{SU} \left( 2
\right)_L \times \mathrm{U} \left( 1 \right)_Y$. In order not to loose the
benefit of the large symmetry $S_{\text{nat}}$, the constraints must be
invariant under $S_{\text{nat}}$. We show that such a procedure is tantamount
to introducing a set of unitary fields with definite transformations under
$S_{\text{nat}}$ and vanishing covariant derivatives. Introducing
multiplication factors into these matrices, one obtains the spurions of
{\cite{Hirn:2004ze}}. The reciprocal {\emdash}i.e. postulating the existence
of covariantly constant spurions and thereby obtaining the reduction{\emdash}
was already discussed in {\cite{Hirn:2004ze,Hirn:2004zz}}.

\subsubsection{Origin of spurions} \label{origin-spurions}

We want to identify $\Gamma_{L \mu}$ to $g_L G_{L \mu}$, up to a gauge
transformation $\Omega_L \in \tmop{SU} \left( 2 \right)$, i.e.
\begin{eqnarray}
  \Gamma_{L \mu} & = & \Omega_L \left( x \right) g_L G_{L \mu} \Omega_L^{- 1}
  \left( x \right) \nonumber\\ && + \mathi \Omega_L \left( x \right) \partial_{\mu}
  \Omega_L^{- 1} \left( x \right) .  \label{39}
\end{eqnarray}
This will reduce the group $\tmop{SU} \left( 2 \right)_{G_L} \times \tmop{SU}
\left( 2 \right)_{\Gamma_L}$ to its vector subgroup, which will be recognized
as the $\tmop{SU} \left( 2 \right)_L$ introduced in the SM {\emdash}see
Section \ref{2-notations}. Requiring the invariance of the constraint
(\ref{39}) with respect to the whole symmetry $\tmop{SU} \left( 2
\right)_{G_L} \times \tmop{SU} \left( 2 \right)_{\Gamma_L}$ amounts to
promoting the gauge function $\Omega_L$ to a field variable transforming
according to
\begin{eqnarray}
  \Omega_L & \longmapsto & \Gamma_L \Omega_L G_L^{\dag} .  \label{40}
\end{eqnarray}
Note that equation (\ref{39}) now replaces the original set of gauge
connections $\{ G_{L \mu}, \Gamma_{L \mu} \}$ by a smaller set $G_{L \mu}$
{\tmem{and}} $\Omega_L$, maintaining a well-defined action of the $\tmop{SU}
\left( 2 \right)_{G_L} \times \tmop{SU} \left( 2 \right)_{\Gamma_L}$symmetry
group on this smaller manifold $\{ G_{L \mu}, \Omega_L \}$. The field
$\Omega_L$ is {\tmem{not}} a GB, but rather a non-propagating
{\tmem{spurion}}: this follows from the constraint (\ref{39}) itself, since the
latter can be equivalently rewritten as the condition of covariant constancy
of $\Omega_L$, reflecting the transformation (\ref{40})
\begin{eqnarray}
  D_{\mu} \Omega_L & \equiv & \partial_{\mu} \Omega_L - \mathi \Gamma_{L \mu}
  \Omega_L + \mathi g_L \Omega_L G_{L \mu} \hspace{0.2em} = \hspace{0.2em} 0 . 
  \label{41}
\end{eqnarray}

Next, we proceed to a similar reduction in the right-handed sector. The
connections $\Gamma_{R \mu}$ and $G_{R \mu}$ are identified up to a gauge
through the constraint
\begin{eqnarray}
  \Gamma_{R \mu}  & = & \Omega_R \left( x \right) g_R G_{R \mu} \Omega_R^{- 1}
  \left( x \right)  \nonumber\\ && + \mathi \Omega_R \left( x \right) \partial_{\mu}
  \Omega_R^{- 1} \left( x \right) .  \label{44}
\end{eqnarray}
This is done by  introducing of the spurion $\Omega_R \in \tmop{SU} \left(
2 \right)$ transforming as
\begin{eqnarray}
  \Omega_R & \longmapsto & \Gamma_R \Omega_R G_R^{\dag}, 
\end{eqnarray}
and enforcing the covariant constancy of $\Omega_R$
\begin{eqnarray}
  D_{\mu} \Omega_R & = & \partial_{\mu} \Omega_R - \mathi \Gamma_{R \mu}
  \Omega_R + \mathi g_R \Omega_R G_{R \mu} \hspace{0.2em} = \hspace{0.2em} 0 . 
  \label{4.17}
\end{eqnarray}
The action of the whole symmetry $\tmop{SU} \left( 2 \right)_{\Gamma_R} \times
\tmop{SU} \left( 2 \right)_{G_R}$ on the reduced manifold $\{ G_{R \mu},
\Omega_R \}$ is still at work, but only the diagonal subgroup
{\emdash}identified with the right-handed isospin $\tmop{SU} \left( 2
\right)_R$ {\emdash} is linearly realized in the space of propagating fields.

\begin{figure}
\includegraphics{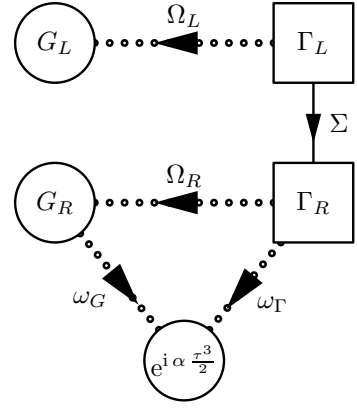}
\caption{Transformation properties of {\tmem{unitary}} spurions (dotted lines) and GBs (continuous line).\label{T1}}
\end{figure}

So far, the symmetry $S_{\text{nat}}$ has been reduced to $\tmop{SU} \left( 2
\right)_L \times \tmop{SU} \left( 2 \right)_R \times \mathrm{U} \left( 1
\right)_{B - L}$, involving seven gauge connections and two spurions fields
$\Omega_L, \Omega_R$. In order to end up with the physical degrees of freedom
of the electroweak sector, with $\tmop{SU} \left( 2 \right)_L \times
\mathrm{U} \left( 1 \right)_Y$ as the maximal linearly realized symmetry, it
remains to further reduce $\tmop{SU} \left( 2 \right)_R \times \mathrm{U}
\left( 1 \right)_{B - L}$ to $\mathrm{U} \left( 1 \right)_Y$. This can be
achieved identifying $\tmop{SU} \left( 2 \right)_R$ with $\mathrm{U} \left( 1
\right)_{B - L}$ embedded into $\tmop{SU} \left( 2 \right)$ according to
equation (\ref{4.9}). Such an identification amounts to the constraint
\begin{eqnarray}
  \Gamma_{R \mu} & = & \omega_{\Gamma} g_B G_{B \mu}  \frac{\tau^3}{2}
  \omega_{\Gamma}^{- 1} + \mathi \omega_{\Gamma} \partial_{\mu}
  \omega_{\Gamma}^{- 1},  \label{42}
\end{eqnarray}
where the spurion $\omega_{\Gamma} \in \tmop{SU} \left( 2 \right)$ transforms
as
\begin{eqnarray}
  \omega_{\Gamma} & \longmapsto & \Gamma_R \omega_{\Gamma} \mathe^{\mathi
  \alpha \frac{\tau^3}{2}} .  \label{45}
\end{eqnarray}
Again, the constraint (\ref{42}) is equivalent to the vanishing of the
corresponding covariant derivative
\begin{eqnarray}
  D_{\mu} \omega_{\Gamma} & \equiv & \partial_{\mu} \omega_{\Gamma} - \mathi
  \Gamma_{R \mu} \omega_{\Gamma} + \mathi g_B \omega_{\Gamma} G_{B \mu} 
  \frac{\tau^3}{2} \hspace{0.2em} = \hspace{0.2em} 0 .  \label{49}
\end{eqnarray}
We note in passing, that the constraints (\ref{44}) and (\ref{42}) imply, by
transitivity, the relation
\begin{eqnarray}
  g_R G_{R \mu} & = & \omega_G g_B G_{B \mu}  \frac{\tau^3}{2} \omega_G^{- 1}
  + \mathi \omega_G \partial_{\mu} \omega_G^{- 1},  \label{43}
\end{eqnarray}
where $\omega_G$ is defined as
\begin{eqnarray}
  \omega_G & \equiv & \Omega_R^{\dag} \omega_{\Gamma} \hspace{0.2em} \longmapsto
  \hspace{0.2em} G_R \omega_{\Gamma} \mathe^{\mathi \alpha \frac{\tau^3}{2}} . 
  \label{46}
\end{eqnarray}
The constraint (\ref{42}) implies the orientation of the right-isospin in the
third direction together with the reduction of the remaining symmetry
$\mathrm{U} \left( 1 \right)_{T^3_R} \times \mathrm{U} \left( 1 \right)_{B -
L}$ to its diagonal subgroup $\mathrm{U} \left( 1 \right)_Y$ where
\begin{eqnarray}
  \frac{Y}{2} & = & T_R^3 + \frac{B - L}{2} . 
\end{eqnarray}
We thus end up with the reduced symmetry
\begin{eqnarray}
  S_{\text{red}} & = & \tmop{SU} \left( 2 \right)_L \times \mathrm{U} \left( 1
  \right)_Y .  \label{51}
\end{eqnarray}
The reduction of the symmetry from $S_{\text{nat}}$ to $S_{\text{red}}$ is
schematically represented in FIG.
\ref{T1}.
The symmetry $S_{\text{red}}$ (\ref{51}) acts linearly on
physical fields, exactly as in the SM. In particular, the right-handed
neutrino is inert under $S_{\text{red}}$ (\ref{51}) and in the lagrangian (\ref{lag}), it
decouples once the constraints (\ref{39}), (\ref{44}) and (\ref{42}) are
applied. Yet, its presence in the theory is enforced by the symmetry
$S_{\text{nat}}$. The latter did not disappear: the transformations belonging
to
\begin{eqnarray}
  S_{\text{hidden}} & = & \frac{S_{\text{nat}}}{S_{\text{red}}}, 
\end{eqnarray}
merely act on the spurion fields $\Omega_L, \Omega_R, \omega_G$ and
consequently they become apparent only when terms explicitly involving
spurions are included.

\subsubsection{Magnitude of spurions} \label{strength-spurions}

The covariant constraints (\ref{41}), (\ref{49}) and (\ref{43}) can now be
plugged into the $\mathcal{O} \left( p^2 \right)$ lagrangian (\ref{lag}),
eliminating nine of the connections in favor of four $\tmop{SU} \left( 2
\right)_L \times \mathrm{U} \left( 1 \right)_Y$ gauge fields, and the unitary
fields $\omega_{\Gamma}, \omega_G$ and $\Omega_L$. The latter keep track of
the original symmetry $S_{\text{nat}}$, they do not propagate (their covariant
derivatives vanish, c.f. (\ref{41}), (\ref{49}) and (\ref{43})) and they can
be transformed away by redefinition of the remaining fields. The lagrangian
then involves the four $\tmop{SU} \left( 2 \right)_L \times \mathrm{U} \left(
1 \right)_Y$ gauge fields with all vertices, mixing and vector boson masses
identical to the SM without the following: left-over Higgs particle, Yukawa
couplings and fermion mass terms.

The problem of unwanted terms reappears as long as the spurions are described
by the unitary variables $\omega_G, \omega_{\Gamma}, \Omega_L$, introducing no
new parameters that cannot be eliminated by a gauge choice: the spurions
$\omega_G, \omega_{\Gamma}, \Omega_L$ can be used to construct other
$S_{\text{nat}}$-invariants that are still $\mathcal{O} \left( p^2 \right)$
but are not contained in (\ref{lag}). These additional terms are exactly all
the terms of the type (\ref{5.26}-\ref{last}), which still have to be
suppressed.

This can be achieved by adding a new ingredient: we now admit multiplication
of the unitary spurions by small constants (expansion parameters), which does
not modify (\ref{39}) and (\ref{44}). For the first identifications (\ref{41})
and (\ref{4.17}), this is implemented via the requirement that only the
objects $\mathcal{X},\mathcal{Y}$
\begin{eqnarray}
  \mathcal{X} & \equiv & \xi \Omega_L,  \label{X}\\
  \mathcal{Y} & \equiv & \eta \Omega_R,  \label{1st-Y}
\end{eqnarray}
and positive powers of them or of their hermitian conjugate may be inserted in
the operators of Section \ref{3.2-irr-H-less} in order to make them invariant
under $S_{\text{nat}}$. The order of magnitude of $\xi$ and $\eta$ should
later be estimated from experiments, but they will be considered as expansion
parameters. Indeed, the $S_{\text{nat}}$ symmetry guarantees that the
condition of small $\xi, \eta$ will not be upset by the loop expansion.

It is convenient to decompose $\mathcal{Y}$ as
\begin{eqnarray}
  \mathcal{Y} & = & \mathcal{Y}_u +\mathcal{Y}_d, 
\end{eqnarray}
where
\begin{eqnarray}
  \mathcal{Y}_u & \equiv & \eta \omega_{\Gamma}  \left(\begin{array}{cc}
    1 & 0\\
    0 & 0
  \end{array}\right) \omega_G^{\dag} \hspace{0.2em} \longmapsto \hspace{0.2em}
  \Gamma_R \mathcal{Y}_u G_R^{\dag},  \label{Y}\\
  \mathcal{Y}_d & \equiv & \mathcal{Y}_{u c} \hspace{0.2em} \equiv \hspace{0.2em}
  \tau^2 \mathcal{Y}_u^{\ast} \tau^2  \nonumber\\ &=&  \eta
  \omega_{\Gamma}  \left(\begin{array}{cc}
    0 & 0\\
    0 & 1
  \end{array}\right) \omega_G^{\dag} \hspace{0.2em} \longmapsto \hspace{0.2em}
  \Gamma_R \mathcal{Y}_d G_R^{\dag} .  \label{58a}
\end{eqnarray}
Note also the following relations, useful for the future construction of the
lagrangian
\begin{eqnarray}
  \mathcal{Y}_u \mathcal{Y}_d^{\dag} & = & 0,  \label{55}\\
  \mathcal{Y}_u \mathcal{Y}_u^{\dag} +\mathcal{Y}_d \mathcal{Y}_d^{\dag} & = &
  \eta^2  \mathbbm{1} .  \label{63}
\end{eqnarray}
As for $\mathcal{X}$, its transformation law is immediate from the definition
(\ref{X})
\begin{eqnarray}
  \mathcal{X} & \longmapsto & \Gamma_L \mathcal{X}G_L^{\dag} .  \label{trsf-X}
\end{eqnarray}
For the triangular identification (\ref{44}), (\ref{42}) and (\ref{43}) in the
right-handed sector, one must specify two independent combinations of unitary
matrices $\Omega_R$ , $\omega_{\Gamma}$ and $\omega_G$ in front of which small
expansion parameters will be introduced. We have already performed a choice
when writing (\ref{1st-Y}) and/or (\ref{Y}-\ref{58a}). This choice is
motivated by the fact that $\mathcal{Y}_u$ and $\mathcal{Y}_d$ are neutral
under $\mathrm{U} \left( 1 \right)_{B - L}$, whereas both $\omega_{\Gamma}$
and $\omega_G$ carry a non-zero $B - L$ charge. Furthermore, it is an
empirical fact that the right-handed isospin violation is less suppressed than
the violation of $B - L$. It remains to specify how the remaining $\mathrm{U}
\left( 1 \right)_{B - L}$ breaking spurion strength will be introduced. For
the moment, we focus on the simplest case, which is most symmetric between the
left- and right-handed sectors {\emdash}but is not the one presented in
{\cite{Hirn:2004ze}}. The general case will be discussed in section
\ref{alternatives}. We define
\begin{eqnarray}
  \mathcal{Z} & \equiv & \zeta_{\Iota}^2 \omega_{\Gamma} \tau^+
  \omega_{\Gamma}^{\dag} .  \label{Z}
\end{eqnarray}
From (\ref{45}), we find that the spurion $\mathcal{Z}$ so defined transform
as
\begin{eqnarray}
  \mathcal{Z} & \longmapsto & \mathe^{+ \mathi \alpha} \Gamma_R \mathcal{Z}
  \Gamma_R^{\dag} .  \label{spuf}
\end{eqnarray}
All the other objects that can be constructed from products of
$\mathcal{X},\mathcal{Y}_{u, d}$ and $\mathcal{Z}$ automatically carry
definite powers of $\xi, \eta$ or $\zeta_{\Iota}$. Note that $\mathcal{Z}$ and
$\mathcal{Z}^{\dag}$satisfy a Grassman algebra
\begin{eqnarray}
  \left\{ \mathcal{Z},\mathcal{Z}^{\dag} \right\} & = & \zeta_{\Iota}^4 
  \mathbbm{1},  \label{Grassman}\\
  \mathcal{Z}^2 & = & 0 .  \label{4.37}
\end{eqnarray}
Other related properties of the spurions that can be deduced from the above
are
\begin{eqnarray}
  \mathcal{Y}_d^{\dag} \mathcal{Z} \hspace{0.2em} = \hspace{0.2em} \mathcal{Z}
  \mathcal{Y}_u & = & 0,  \label{4.52}\\
  \mathcal{Z}^{\dag} & = & -\mathcal{Z}_c, \\
  \tmop{Tr} \mathcal{Z} & = & 0 .  \label{58}
\end{eqnarray}

\subsubsection{Reciprocal}

The reciprocal of the statement (Section \ref{origin-spurions}) that spurions
are a necessary consequence of the covariant identification of subgroups of
$S_{\text{nat}}$ also holds. To show this, one considers a set of three
spurions $\mathcal{X},\mathcal{Y}_u$ and $\mathcal{Z}$ transforming as in
(\ref{trsf-X}), (\ref{Y}) and (\ref{spuf}) and satisfying (\ref{55}) and
(\ref{4.37}), and imposes the following constraints on them
\begin{eqnarray}
  D_{\mu} \mathcal{X} & \equiv & \partial_{\mu} \mathcal{X}- \mathi \Gamma_{L
  \mu} \mathcal{X}+ \mathi g_L \mathcal{X}G_{L \mu} \hspace{0.2em} =
  \hspace{0.2em} 0,  \label{c1}\\
  D_{\mu} \mathcal{Y}_u & \equiv & \partial_{\mu} \mathcal{Y}_u - \mathi
  \Gamma_{R \mu} \mathcal{Y}_u + \mathi g_R \mathcal{Y}_u G_{R \mu}
  \hspace{0.2em} = \hspace{0.2em} 0,  \label{6.28}\\
  D_{\mu} \mathcal{Z} & \equiv & \partial_{\mu} \mathcal{Z}- \mathi g_R 
  \left[ \Gamma_{R \mu},\mathcal{Z} \right] + \mathi g_B G_{B \mu} \mathcal{Z}
  \hspace{0.2em} = \hspace{0.2em} 0 .  \label{c2}
\end{eqnarray}
One then finds that equations (\ref{c1}-\ref{c2}) have no non-trivial solution
($\mathcal{X},\mathcal{Y}_u,\mathcal{Z} \neq 0$) unless the thirteen gauge
connections of $S_{\text{nat}}$ are aligned in a definite way. As shown in
{\cite{Hirn:2004ze}}, the integrability condition of the system
(\ref{c1}-\ref{c2}) amounts to the existence of a gauge, specified by
$\Omega_L = \omega_G = \omega_{\Gamma} = 1$. In this gauge, called the
``standard gauge'', we automatically have $\Omega_R \equiv \omega_G
\omega_{\Gamma}^{\dag} = 1$ and
\begin{eqnarray}
  \Gamma_{L \mu} & \overset{\text{s.g.}}{=} & g_L G_{L \mu},  \label{id1}\\
  \Gamma^{1, 2}_{R \mu} & \overset{\text{s.g.}}{=} & g_R G_{R \mu}^{1, 2}
  \quad \overset{\text{s.g.}}{=} \quad 0,  \label{id2}\\
  \Gamma^3_{R \mu} & \overset{\text{s.g.}}{=} & g_R G_{R \mu}^3 \quad
  \overset{\text{s.g.}}{=} \quad g_B G^3_{B \mu} .  \label{id3}
\end{eqnarray}
In the aforementioned gauge, the spurions reduce to three real functions $\xi
\left( x \right)$, $\eta \left( x \right)$ and $\zeta_{\Iota} \left( x
\right)$ according to
\begin{eqnarray}
  \mathcal{X} & \overset{\text{s.g.}}{=} & \xi \left(\begin{array}{cc}
    1 & 0\\
    0 & 1
  \end{array}\right),  \label{s1}\\
  \mathcal{Y}_u & \overset{\text{s.g.}}{=} & \eta \left(\begin{array}{cc}
    1 & 0\\
    0 & 0
  \end{array}\right), \hspace{2em} \mathcal{Y}_d \hspace{0.2em}
  \overset{\text{s.g.}}{=} \hspace{0.2em} \eta \left(\begin{array}{cc}
    0 & 0\\
    0 & 1
  \end{array}\right), \\
  \mathcal{Z} & \overset{\text{s.g.}}{=} & \zeta_{\Iota}^2 
  \left(\begin{array}{cc}
    0 & 1\\
    0 & 0
  \end{array}\right) .  \label{s2}
\end{eqnarray}
The constraints (\ref{c1}-\ref{c2}) also yield
\begin{eqnarray}
  \partial_{\mu} \xi & = & \partial_{\mu} \eta \hspace{0.2em} = \hspace{0.2em}
  \partial_{\mu} \zeta_{\Iota} \hspace{0.2em} = \hspace{0.2em} 0,  \label{consts}
\end{eqnarray}
stating the space-time independence of $\xi, \eta$ and $\zeta_{\Iota}$. This
is the reciprocal of Section \ref{origin-spurions}.

The solution (\ref{s1}-\ref{s2}) and the conditions (\ref{id1}-\ref{id3}) are
invariant under the electroweak group, which is a subgroup of $S_{\text{nat}}$
\begin{eqnarray}
  S_{\text{red}} & = & \tmop{SU} \left( 2 \right)_L \times \mathrm{U} \left( 1
  \right)_Y .  \label{6.41}
\end{eqnarray}
$S_{\text{red}}$ involves the four SM gauge fields $G^a_{\mu}$  and $b_{\mu}$ defined here through the values of
the other fields in the gauge where (\ref{id1}-\ref{id3}) hold
\begin{eqnarray}
  gG^a_{\mu} & \equiv & g_L G^a_{L \mu} |_{\text{s.g.}} \hspace{0.2em} \left( =
  \Gamma^a_{L \mu} |_{\text{s.g.}} \right), \\
  g' b_{\mu} & \equiv & g_R G^3_{R \mu} |_{\text{s.g.}} \hspace{0.2em} \left( =
  g_B G_{B \mu} |_{\text{s.g.}} = \Gamma^3_{R \mu} |_{\text{s.g.}} \right) . 
\end{eqnarray}
The kinetic $\mathrm{U} \left( 1 \right)_Y$ term is normalized if $g'$ is
defined as
\begin{eqnarray}
  \frac{1}{g'^2} & \equiv & \frac{1}{g_R^2} + \frac{1}{g_B^2} . 
\end{eqnarray}
It has been shown {\cite{Hirn:2004ze}} that, upon inserting the constraints
(\ref{c1}-\ref{c2}) into the leading-order lagrangian (\ref{lag}), the SM
couplings are obtained, between massive $W^{\pm}$, $Z^0$ and photon and
{\tmem{massless}} fermions. Following the Higgs mechanism, the three GBs
contained in the matrix $\Sigma$ are absorbed by the $W^{\pm}$ and $Z^0$, and
there is no scalar particle left in the spectrum. The definitions of the gauge
fields diagonalizing the quadratic terms in the lagrangian are
\begin{eqnarray}
  W_{\mu}^{\pm} & \equiv & \frac{\mathi \sqrt{2}}{g}  \left\langle \tau^{\mp}
  \Sigma^{\dag} D_{\mu} \Sigma \right\rangle |_{\text{s.g.}}, \\
  Z_{\mu} & \equiv & \frac{\mathi c}{g}  \left\langle \tau^3 \Sigma^{\dag}
  D_{\mu} \Sigma \right\rangle |_{\text{s.g.}}, \\
  A_{\mu} & \equiv & \frac{\mathi s}{g}  \left\langle \tau^3 \Sigma^{\dag}
  D_{\mu} \Sigma \right\rangle |_{\text{s.g.}} + \frac{1}{c} G_{B \mu}
  |_{\text{s.g.}}, 
\end{eqnarray}
where $s$ and $c$ are the sinus and cosinus of the Weinberg angle $s =
\frac{g'}{\sqrt{g^2 + g'^2}}$, $c = \frac{g}{\sqrt{g^2 + g'^2}}$.

In all of the sequel, the constants $\xi$, $\eta$ and $\zeta_{\Iota}$ will be
treated as small expansion parameters. This is consistent since, in the
absence of spurions, one recovers the larger symmetry $S_{\text{nat}}$. This
is one role of the spurions: to introduce in a covariant manner these tunable
parameters. Covariance is necessary in order to classify various operators
according to their symmetry-breaking properties. The other role of spurions is
to select the vacuum alignment, i.e. the embedding of the electroweak group
$S_{\text{red}}$ in $S_{\text{nat}}$: this is the identification of
connections given in (\ref{id1}-\ref{id3}).

The next step in the formulation of the LEET is the construction of the
effective lagrangian: one writes down all terms invariant under
$S_{\text{nat}}$ that can be constructed out of the GBs $\Sigma$, the
connections $\Gamma_{L \mu}, \Gamma_{R \mu}$, the gauge fields $G_{L \mu},
G_{R \mu}, G_{B \mu}$, the spurions $\mathcal{X},\mathcal{Y},\mathcal{Z}$, and
fermions. The operators should be ordered according to their chiral
power-counting and, in addition, according to the powers of spurions involved.
To exhibit the physical content of each operator, one then injects the
solution (\ref{id1}-\ref{consts}) of the constraints. This yields a lagrangian
depending on the fermions and on the $S_{\text{red}}$ gauge fields $G_{\mu}$
and $b_{\mu}$, which should be used as dynamical variables to compute loops.
In addition, this lagrangian depends on the three constants $\xi$, $\eta$ and
$\zeta_{\Iota}$. At the leading order $\mathcal{O} \left( p^2 \right)$ without
explicit powers of spurions, the lagrangian describes exactly the SM couplings
but without the Higgs boson, and with all fermions left massless. The origin
of fermion masses, which will come with explicit powers of spurions, thus
appears different from that of vector bosons. Other terms involving explicit
powers of spurions will also bring other interactions. This is the subject of
Sections \ref{5-masses}, \ref{7-list} and \ref{LNV-NLO}.

\subsubsection{Magnitude of ($B - L$)-breaking} \label{alternatives}

Before we describe remaining possibilities of introducing the ($B -
L$)-breaking spurion strength and discuss their physical content, it is worth
stressing that the number of spurions to be introduced is entirely fixed once
we have identified the higher $S_{\text{nat}}$ symmetry, and once we ask to
recover the electroweak group $S_{\text{red}}$ imposing constraints. As should
be clear from the discussion of Section (\ref{origin-spurions}), the
introduction of spurions is not a choice, but follows from the requirement
that the formalism be covariant under $S_{\text{nat}}$.

Given $S_{\text{nat}} = \left[ \tmop{SU} \left( 2 \right) \times \tmop{SU}
\left( 2 \right) \right]^2 \times \mathrm{U} \left( 1 \right)_{B - L}$,
exactly three expansion parameters can be introduced, that suppress the
unwanted operators mentioned in Section \ref{3.2-irr-H-less}. Of these three,
one ($\mathcal{X}$) pertains to the left-handed sector: there is no choice up
to here. In the right-handed sector, due to the triangular identification of
connections (see FIG. \ref{T1}), there would a priori be various possibilities for the
introduction of the expansion parameters. In order that ($B - L$)-breaking
remains a small effect, we required that one of the expansion parameters
($\mathcal{Y}$) does not break ($B - L$). Otherwise, non-($B - L$)-breaking
effects would appear only at quadratic level in the ($B - L$)-breaking
parameter. Once this physical motivation is spelled out, we see that there was
no choice at this stage either.

It turns out that we remain with three inequivalent possibilities for the ($B
- L$)-breaking building block $\mathcal{Z}$, i.e. the one that carries a
non-zero ($B - L$) charge. They correspond to the possibility of introducing
the expansion parameters $\zeta$ in factor of different combinations of the
unitary spurions $\omega_G, \omega_{\Gamma}$. Other cases with fundamental
building blocks bilinear in $\omega_G, \omega_{\Gamma}$, can always be brought
back to one of these three. We already mentioned the simplest case in Section
\ref{strength-spurions}. Instead of $\mathcal{Z}$, we could have introduced
the spurion $\mathcal{Z}_{\tmop{II}}$
\begin{eqnarray}
  \mathcal{Z}_{\tmop{II}} & \equiv & \zeta^2_{\tmop{II}} \omega_{\Gamma}
  \tau^+ \omega_G^{\dag} \hspace{0.2em} \longmapsto \hspace{0.2em} \mathe^{\mathi
  \alpha} \Gamma_R \mathcal{Z}_{\tmop{II}} G_R^{\dag},  \label{ZII}
\end{eqnarray}
to be taken as a fundamental building block rather than $\mathcal{Z}$
{\emdash}but still keeping $\mathcal{Y}$ as an elementary spurion. Another
independent case is with the spurion $\mathcal{Z}_{\tmop{III}}$ instead of
$\mathcal{Z}$
\begin{eqnarray}
  \mathcal{Z}_{\tmop{III}} & \equiv & \zeta^2_{\tmop{III}} \omega_G \tau^+
  \omega_G^{\dag} \hspace{0.2em} \longmapsto \hspace{0.2em} \mathe^{\mathi \alpha}
  G_R \mathcal{Z}_{\tmop{III}} G_R^{\dag} .  \label{ZIII}
\end{eqnarray}
In the three cases, all other combinations must then be constructed out of a
basis of two spurions, using hermitian conjugation, but without using the
inverse of a matrix.

The cases $\tmop{II}$ and $\tmop{III}$ could be recast as the one of Section
\ref{strength-spurions} if one could freely use the following rewritings
\begin{eqnarray}
  \mathcal{Z} & \equiv & \frac{\zeta_{\Iota}^2}{\zeta_{\tmop{II}}^2 \eta}
  \mathcal{Z}_{\tmop{II}} \mathcal{Y}_d^{\dag},  \label{132}\\
  \mathcal{Z} & \equiv & \frac{\zeta_{\Iota}^2}{\zeta_{\tmop{III}}^2 \eta^2}
  \mathcal{Y}_u \mathcal{Z}_{\tmop{III}} \mathcal{Y}^{\dag}_d,  \label{133}
\end{eqnarray}
as can be deduced from a comparison of the definition (\ref{Z}) with those of
(\ref{ZII}-\ref{ZIII}). However, the relations (\ref{132}-\ref{133}) are
singular in the limit of vanishing spurions. Note that the possibility to
build operators that produce an $S_{\text{nat}}$-invariant operator with a
given physical content {\emdash}i.e. are equal in the standard unitary
gauge{\emdash} does not depend on the assumed case $\Iota, \tmop{II}$ or
$\tmop{III}$, but only on the symmetries. What is modified is the  lowest
number of spurion powers necessary to construct such an operator. The three
situations labelled by $\Iota, \tmop{II}, \tmop{III}$ are therefore not
equivalent from the point of view of magnitude estimates, and of the
expansion. As we shall see in Section \ref{alternatives-2}, a difference
appears in the ratio of left- to right-handed neutrino masses. This
distinction in turn implies different cosmological consequences which we
explore in Section \ref{6-nuR}, where we will even see that the physical
hypothesis labeled by $\tmop{III}$ seems a priori excluded.

We can now show that the case mentioned in {\cite{Hirn:2004ze}} corresponds to
the physical situation denoted here by $\tmop{III}$. We first note that the
definition of the spurion $\phi$ used in {\cite{Hirn:2004ze}} would in the
present formalism require the introduction of the following complex doublet
\begin{eqnarray}
  \phi_G & \equiv & \omega_G  \left(\begin{array}{c}
    1\\
    0
  \end{array}\right) \hspace{0.2em} \longmapsto \hspace{0.2em} G_R \mathe^{\mathi
  \frac{\alpha}{2}} \phi_G,  \label{phiG}
\end{eqnarray}
and its conjugate
\begin{eqnarray}
  \left( \phi_G \right)_c & \equiv & \mathi \tau^2 \phi_G^{\ast} \hspace{0.2em}
  \longmapsto \hspace{0.2em} G_R \mathe^{- \mathi \frac{\alpha}{2}}  \left(
  \phi_G \right)_c . 
\end{eqnarray}
With $\phi_G$, one can construct the spurion $\phi$ of {\cite{Hirn:2004ze}}
\begin{eqnarray}
  \phi & \equiv & \zeta_{\tmop{III}} \phi_G . 
\end{eqnarray}
providing the connection between the writing used in {\cite{Hirn:2004ze}} and
the present paper through the further equality
\begin{eqnarray}
  \mathcal{Z}_{\tmop{III}} & = & \phi \phi_c^{\dag} . 
\end{eqnarray}
This also explains the normalization between $\mathcal{Z}_{\tmop{III}}$ and
$\zeta_{\tmop{III}}$ {\emdash}which corresponds to the $\zeta$ of
{\cite{Hirn:2004ze}}.

\subsection{Accidental flavor symmetry} \label{flavor}

In the absence of spurions, the large symmetry $S_{\text{nat}}$ forbids all
non-standard couplings of the type mentioned in Section \ref{3-other}.
Consequently, at the leading order, the couplings of all fermion doublets
(with a given $B - L$) are identical. Furthermore, $S_{\text{nat}}$ forbids
both Dirac and Majorana mass terms {\footnote{Recall that, in Sections
\ref{3.2.3-Yuk} and \ref{3.3.1}, Dirac and Majorana mass terms were identified
as a source of possible difficulties for the LEET.}}.

If each quark and lepton doublet appears in three copies (generations or
families), the lagrangian at the leading order $\mathcal{O} \left( p^2
\right)$ without explicit powers of spurions necessarily enjoys a global
flavor (chiral) symmetry
\begin{eqnarray}
  S_{\text{flavor}} & = & \mathrm{U} \left( 3 \right)_L \times \mathrm{U}
  \left( 3 \right)_R, 
\end{eqnarray}
separately for quarks and for leptons. Denoting by $\chi_{L, R}^i$, $i = 1, 2,
3$ a generic left (right)-handed fermion doublet, the generators of
$S_{\text{flavor}}$ are simply
\begin{eqnarray}
  Q^{i j}_{L, R} & = & \int \mathd^3 \tmmathbf{x} \overline{\chi^i_{L, R}}
  \gamma^0 \chi_{L, R}^j, \hspace{2em} i, j = 1, 2, 3 . \bignone  \label{79}
\end{eqnarray}
Classically, they are invariant under all transformations in $S_{\text{nat}}$,
and, at the leading order, they are conserved. At the quantum level, there
will be anomalies. The symmetry $S_{\text{flavor}}$ is accidental, since its
presence at this level merely follows from the local symmetry $S_{\text{nat}}$
and from the particle content.

The spurion $\mathcal{Y}$ allows one to extend the flavor symmetry $\mathrm{U}
\left( 3 \right)_R$ that acts on the whole doublet $\chi_R^i$ into two
independent transformations of its up and down components. Indeed, the charges
$Q_{R u}^{i j}$ and $Q_{R d}^{i j}$ defined though
\begin{eqnarray}
  \eta^2 Q^{i j}_{R u} & = & \int \mathd^3 \tmmathbf{x} \overline{\chi^i_R}
  \gamma^0 \mathcal{Y}_u^{\dag} \mathcal{Y}_u \chi_R^j, \bignone  \label{80}\\
  \eta^2 Q^{i j}_{R d} & = & \int \mathd^3 \tmmathbf{x} \overline{\chi^i_R}
  \gamma^0 \mathcal{Y}^{\dag}_d \mathcal{Y}_d \chi_R^j, \bignone  \label{81}
\end{eqnarray}
are (classically) invariant with respect to $S_{\text{nat}}$. They obey
\begin{eqnarray}
  Q^{i j}_{R u} + Q^{i j}_{R d} & = & Q^{i j}_R, 
\end{eqnarray}
and commute with each other
\begin{eqnarray}
  \left[ Q_{R u}, Q_{R d} \right] & = & 0 . 
\end{eqnarray}
At the leading order $\mathcal{O} \left( p^2 \right)$ without explicit powers
of spurions, both $Q_{R u}$ and $Q_{R d}$ are conserved, provided the
constraints (\ref{c1}-\ref{c2}) hold. This is in direct connection with the
absence at this order of right-handed charged current interaction. We thus
have an extended flavor horizontal symmetry
\begin{eqnarray}
  S^{\text{extended}}_{\text{flavor}} & = & \mathrm{U} \left( 3 \right)_L
  \times \left[ \mathrm{U} \left( 3 \right)_R^u \times \mathrm{U} \left( 3
  \right)_R^d \right] .  \label{Sfe}
\end{eqnarray}
In the absence of fermion masses, the same extended symmetry also exists in
the SM, reflecting the singlet character of right-handed fermions.

A comment about the left-handed part of $S_{\text{flavor}}$ is in order.
Adding two more powers of the spurion $\mathcal{X}$, it is again possible to
split $\mathrm{U} \left( 3 \right)_L$ into separate $\mathrm{U} \left( 3
\right)$ transformations of up and down components of $\chi^i_L$. The
corresponding charges invariant with respect to $S_{\text{nat}}$ read
\begin{eqnarray}
  \xi^2 \eta^2 Q_{L u}^{i j} & = & \int \mathd^3 \tmmathbf{x}
  \overline{\chi^i_L} \gamma^0 \mathcal{X}^{\dag} \Sigma \mathcal{Y}_u
  \mathcal{Y}^{\dag}_u \Sigma^{\dag} \mathcal{X} \chi_L^j, \\
  \xi^2 \eta^2 Q_{L d}^{i j} & = & \int \mathd^3 \tmmathbf{x}
  \overline{\chi^i_L} \gamma^0 \mathcal{X}^{\dag} \Sigma \mathcal{Y}_d
  \mathcal{Y}^{\dag}_d \Sigma^{\dag} \mathcal{X} \chi_L^j . 
\end{eqnarray}
However, unlike in the right-handed case, the charges $Q_{L u}$ and $Q_{L d}$
are not separately conserved. The obstruction is due to the coupling between
$u_L$ and $d_L$ via charged left-handed current interactions.

What was shown in this section, independently of the flavor symmetry, is that
it is always possible to perform separate $\mathrm{U} \left( 3 \right)$
rotations of the $u$ and $d$ components of the doublet, in a way which is
covariant under the whole symmetry $S_{\text{nat}}$. Such rotations are
generated by the charges (\ref{79}-\ref{81}), no matter whether the latter are
conserved or not.

\section{Leading contributions to fermion masses} \label{5-masses}

In the following three sections, we consider the main spurion effects
associated with leading chiral orders. Dirac masses and $\left( B - L
\right)$-conserving non-standard vertices merely concern the spurions $\xi$
and $\eta$. On the other hand, Majorana masses and $\Delta L = 2$ lepton
couplings are intimately related with the spurion $\zeta$. In general, terms
in the lagrangian that contain spurions also break the horizontal flavor
symmetry $S_{\text{flavor}}^{\text{extended}}$ (\ref{Sfe}) introduced in
Section \ref{flavor}. It is conceivable that different powers of spurions
contributing to the same quantity could describe a hierarchical structure of
$S_{\text{flavor}}^{\text{extended}}$ symmetry breaking. Hereafter we merely
concentrate on the {\tmem{complete list}} of leading spurion effects,
postponing a more detailed discussion of flavor symmetry breaking to a later
stage.

\subsection{Quark masses} \label{quark-masses}

Let us stress once more that the standard Yukawa couplings
(\ref{B.20}-\ref{E.21}) are now forbidden, since they are not invariant under
$S_{\text{nat}}$: the (composite) GBs $\Sigma$ and the (elementary) fermions
$\chi_{L, R}$ transform under different groups {\emdash}cf. equations
(\ref{00000}) and (\ref{cp}-\ref{cq}){\emdash}, reflecting their different
physical origin. The lowest order Dirac mass requires one insertion of the
spurion $\mathcal{X}$ and one of the spurion $\mathcal{Y}$ {\emdash}either
$\mathcal{Y}_u$ or $\mathcal{Y}_d$. Hence, the leading quark mass term in the
lagrangian is of order $\mathcal{O} \left( p^1 \xi^1 \eta^1 \right)$ and reads
\begin{eqnarray}
  \mathcal{L} \left( p^1 \xi^1 \eta^1 \right)_{\text{quarks}} & = & - 
  \mu^u_{i j}  \overline{q^i_L} \mathcal{X}^{\dag} \Sigma \mathcal{Y}_u q^j_R
   \nonumber\\ && - \mu^d_{i j}  \overline{q^i_L} \mathcal{X}^{\dag} \Sigma \mathcal{Y}_d
  q^j_R  + \text{h.c},  \label{eq:minimal-fermion-mass}
\end{eqnarray}
where $q_{L, R}^i$ denotes quark doublets. We next define doublets $q^{u,
d}_{L, R}$ transforming under $\tmop{SU} \left( 2 \right)_{\Gamma_R}$ through
\begin{eqnarray}
  \eta q_R^u & \equiv & \mathcal{Y}_u q_R,  \label{88}\\
  \eta q_R^d & \equiv & \mathcal{Y}_d q_R, \\
  \xi \left( \mathcal{Y}_u \mathcal{Y}_u^{\dag} q^u_L +\mathcal{Y}_d
  \mathcal{Y}_d^{\dag} q^d_L \right) & \equiv & \eta^2 \Sigma^{\dag}
  \mathcal{X}q_L,  \label{89}
\end{eqnarray}
and drop the generation indices on the fields $u$, $d$, since we will always
collect the three families into one column. The newly-defined fields take the
following form in the standard gauge
\begin{eqnarray}
  q^u_{L, R} & \overset{\text{s.g.}}{=} & \left(\begin{array}{c}
    u_{L, R}\\
    0
  \end{array}\right),  \label{90}\\
  q^d_{L, R} & \overset{\text{s.g.}}{=} & \left(\begin{array}{c}
    0\\
    d_{L, R}
  \end{array}\right) .  \label{91}
\end{eqnarray}
With the definition of two three-by-three matrices $M_u$ and $M_d$ through
\begin{eqnarray}
  \left( M_{u, d} \right)_{i j} & \equiv & \eta \xi \mu^{u, d}_{i j}, 
  \label{Mud}
\end{eqnarray}
and using the standard gauge, (\ref{eq:minimal-fermion-mass}) assumes the form
\begin{eqnarray}
  \mathcal{L} \left( p^1 \xi^1 \eta^1 \right)_{\text{quarks}} &
  \overset{\text{s.g.}}{=} & - \overline{u_L} M_u u_R  \nonumber\\ && - \overline{d_L}
  M_d d_R  + \text{h.c} .  \label{eq:quark-masses-SM}
\end{eqnarray}
This shows that, at this level, the freedom for the mass matrix is the same as
in the SM. Therefore, using unitary transformations on the quarks, one will
obtain a CKM matrix with the same number of parameters as in the SM. Equation
(\ref{Mud}) states that quark masses are suppressed {\tmem{at least}} by the
spurion factor $\xi \eta$, but in fact, they can be smaller. Higher-order mass
terms compatible with the symmetry $S_{\text{nat}}$ are also possible,
starting at the order $\mathcal{O} \left( p^1 \xi^1 \eta^3 \right)$.
Accordingly, equation (\ref{Mud}) should actually be understood as
\begin{eqnarray}
  M_{u, d} & = & \xi \eta \left( \mu_0^{u, d} + \mu^{u, d}_1 \eta^2 + \ldots
  \right),  \label{Mudex}
\end{eqnarray}
where $\mu_0^{u, d}$ and $\mu_1^{u, d}$ are three-by-three matrices reflecting
the pattern of flavor symmetry breaking in quark masses. In particular, the
top mass should arise from the leading term in (\ref{Mudex}), i.e. it should
be expected of the order $\mathcal{O} \left( \xi \eta \right)$
\begin{eqnarray}
  m_t & = & \xi \eta \Lambda_Q, \label {topmass} 
\end{eqnarray}
where $\Lambda_Q$ is an a priori unspecified large scale related to the
characteristic scale of our LEET, $\Lambda_{\text{w}} \sim 4 \mathpi f \sim 3
\tmop{TeV}$. Lighter quark masses (including $m_b$) might well be of a higher
order, starting by $\xi \eta^3$, and they can also receive radiative
contributions from loops involving the top quark {\emdash}see also Section
\ref{7-list}.

As mentioned in Section \ref{3.1.3}, consistency of the low-energy power
counting for a fermion propagator inside loops requires fermion masses to
count as $\mathcal{O} \left( p^1 \right)$ or smaller. The above discussion of
quark masses therefore suggests a relation between spurion and momentum
expansion, specified by the counting rule
\begin{eqnarray}
  \xi \eta & = & \frac{m_t}{\Lambda_Q} \hspace{0.2em} = \hspace{0.2em} \mathcal{O}
  \left( p^1 \right) .  \label{corres}
\end{eqnarray}
\subsection{Lepton masses and lepton-number violation} \label{leptons}

\subsubsection{Dirac mass terms}

The Dirac mass terms for leptons can be written in full analogy with the quark
mass term (\ref{eq:minimal-fermion-mass}). We give it here to set up the
notation
\begin{eqnarray}
  \mathcal{L} \left( p^1 \xi^1 \eta^1 \right)_{\text{leptons}} & = & -
  \mu^{\nu}_{i j}  \overline{\ell^i_L} \mathcal{X}^{\dag} \Sigma \mathcal{Y}_u
  \ell^j_R  \nonumber\\ && - \mu^e_{i j}  \overline{\ell^i_L} \mathcal{X}^{\dag} \Sigma
  \mathcal{Y}_d \ell^j_R  + \text{h.c} .  \label{5.122}
\end{eqnarray}
Here, $\ell_L^i$ and $\ell_R^i$ denote left and right-handed lepton doublets,
respectively. Upon defining the unitary gauge components of a doublet as for
quarks in Section \ref{quark-masses}, one sees that the Dirac mass term
(\ref{5.122}) can be re-expressed as
\begin{eqnarray}
  \mathcal{L} \left( p^1 \xi^1 \eta^1 \right)_{\text{leptons}} & = & - 
  \overline{\nu_L} M_{\nu} \nu_R  \nonumber\\ &&  -\overline{e_L} M_e e_R  +
  \text{h.c},  \label{d}
\end{eqnarray}
where the matrices $M_{\nu}$ and $M_e$ are of a form analogous to (\ref{Mud}),
i.e. they are $\mathcal{O} \left( \xi \eta \right)$ or smaller.

\subsubsection{Majorana mass terms} \label{Maj-masses}

In this section, and in the following ones, we work using the spurion
$\mathcal{Z}$ defined in (\ref{Z}), as opposed to $\mathcal{Z}_{\tmop{II}}$
(\ref{ZII}) or $\mathcal{Z}_{\tmop{III}}$ (\ref{ZIII}). The two latter cases
will be considered in Section \ref{alternatives-2}.

The Dirac mass term (\ref{5.122}) is not the only Lorentz-invariant bilinear
form of leptons (which are characterized by $B - L = - 1$). The unsuppressed
$\mathcal{O} \left( p^1 \right)$ operator (\ref{5.26}) with $\Delta L = 2$ is
forbidden by the symmetry $S_{\text{nat}}$: in particular, it is custodial
symmetry breaking. On the other hand, making use of the spurion $\mathcal{Z}$,
one can construct $\Delta L = 2$ operators that are
$S_{\text{nat}}$-invariant. These will in turn allow us to construct ($B -
L$)-violating operators: this is inevitable in the LEET. In the
{\tmem{right-handed sector}}, one can write the $\mathcal{O} \left( p^2 \eta^2
\zeta_{\Iota}^2 \right)$ operator
\begin{eqnarray}
  \mathcal{L} \left( p^1 \eta^2 \zeta_{\Iota}^2 \right)_{\text{leptons}} & = &
  - \mu_{i j}^R  \overline{\ell^i_R} \mathcal{Y}_u^{\dag} \mathcal{Z}
  \mathcal{Y}_d  \left( \ell^j_R \right)^c \nonumber\\
&&+ \text{h.c} .  \label{Majright}
\end{eqnarray}
Conjugation for fermion doublets was defined in equation (\ref{conj}). In the
standard gauge, equation (\ref{Majright}) reduces to the right-handed neutrino
Majorana mass
\begin{eqnarray}
  \mathcal{L} \left( p^1 \eta^2 \zeta_{\Iota}^2 \right)_{\text{leptons}} &
  \overset{\text{s.g.}}{=} & - \overline{\nu_R} M_R  \left( \nu_R \right)^c +
  \text{h.c},  \label{Majright2}
\end{eqnarray}
where
\begin{eqnarray}
  M_R & = & \zeta_{\Iota}^2 \eta^2 \mu^R .  \label{h}
\end{eqnarray}
In the left-handed sector, the minimal power of spurions needed to construct
$S_{\text{nat}}$-invariant Majorana masses involves two spurions $\mathcal{X}$
in addition to $\mathcal{Z}$
\begin{eqnarray}
  \mathcal{L} \left( p^1 \xi^2 \zeta_{\Iota}^2 \right)_{\text{leptons}} & = &
  - \mu_{i j}^L  \left( \overline{\ell^i_L} \mathcal{X}^{\dag} \Sigma
  \mathcal{Z} \Sigma^{\dag} \mathcal{X} \left( \ell^j_L \right)^c \right) \nonumber\\ &&  +
  \text{h.c} .  \label{i}
\end{eqnarray}
Hence, in the standard gauge, the left-handed neutrino Majorana mass term
\begin{eqnarray}
  \mathcal{L} \left( p^1 \xi^2 \zeta_{\Iota}^2 \right)_{\text{leptons}} & = &
  - \overline{\nu_L} M_L  \left( \nu_L \right)^c + \text{h.c},  \label{j}
\end{eqnarray}
is of the order $\mathcal{O} \left( p^1 \xi^2 \zeta_{\Iota}^2 \right)$ with
\begin{eqnarray}
  M_L & = & \zeta_{\Iota}^2 \xi^2 \mu^L .  \label{k}
\end{eqnarray}
At his stage it is worth stressing that the existence of the $\Delta L = 2$
operators (\ref{Majright}) and (\ref{i}) and the hierarchy of spurion powers
in the corresponding Majorana neutrino masses (cf. (\ref{h}) and (\ref{k}))
are a necessary consequence of the symmetry reduction $S_{\text{nat}}
\rightarrow S_{\text{red}} = \tmop{SU} \left( 2 \right)_L \times \mathrm{U}
\left( 1 \right)_Y$ by means of spurions. In particular, the spurion
$\mathcal{Z}$ responsible for the selection of the $\mathrm{U} \left( 1
\right)_Y$ subgroup (see FIG. \ref{T1}), controls {\emdash}via the parameter
$\zeta_{\Iota}$ {\emdash} the strength of all $\Delta L = 2$ operators. This
leads us to the expectation that $\zeta_{\Iota}$ may be much smaller than
$\xi$ and $\eta$, which are themselves smaller than $1$.

\subsection{Smallness of neutrino masses} \label{n-masses}

The Majorana masses for the neutrinos are naturally smaller than the (Dirac)
masses of charged fermions, since they involve powers of $\zeta_{\Iota}$ in
addition to being quadratic in $\xi, \eta$. This is not true of the neutrino
Dirac masses stemming from (\ref{5.122}), which would {\tmem{a priori}} be of
the order of those of the quarks or charged leptons. On the other hand, we
cannot invoke here any see-saw mechanism, since the factor $\zeta_{\Iota}^2$
appearing in the right-handed neutrino mass term (\ref{h}) cannot be assumed
large, without ruining our LEET approach.

Note that, if the LEET has to involve an approximate custodial symmetry
originating from the right-handed isospin, it should contain a doublet partner
$\nu_R$ of $e_R$ with a mass $m_{\nu_R} \ll \Lambda_{\text{w}} \sim 3
\tmop{TeV}$. Hence, we are facing the problem of a natural suppression of the
{\tmem{neutrino Dirac mass matrix}} $M_D$ in equation (\ref{d})
{\footnote{Reference {\cite{Fujikawa:2004jy}} deals with a related situation
in which the $\nu_R$ are light. Though also justified by naturalness, the
solution proposed is different.}}. It turns out that the spurion framework
offers a simple solution to this problem, which we are now going to describe.

\subsubsection{Specificity of $\nu_R$}

The particularity of the three right-handed lepton doublets $\ell_R^i$ is that
they transform under $S_{\text{nat}}$ exactly as the spurion doublet $\phi_G$
defined in (\ref{phiG}), implying that
\begin{eqnarray}
  N_R^i & \equiv & \phi_G^{\dag} \ell^i_R,  \label{m}
\end{eqnarray}
is invariant under $S_{\text{nat}}$. Furthermore, if one sticks to the leading
order $\mathcal{O} \left( p^2 \right)$ without explicit powers of spurions
described by the lagrangian (\ref{lag}), the $N_R^i$ satisfy the free massless
Dirac equation
\begin{eqnarray}
  \gamma^{\mu} \partial_{\mu} N_R^i & = & \gamma^{\mu}  \left\{ \left( D_{\mu}
  \phi_G^{\dag} \right) \ell_R^i + \phi_G^{\dag} D_{\mu} \ell_R^i \right\}
 \nonumber\\ &=& 0,  \label{n}
\end{eqnarray}
as long as the constraint $D_{\mu} \phi_G = 0$ stemming from (\ref{43}) and
the definition (\ref{phiG}) holds. This in turn implies that the $\mathrm{U}
\left( 3 \right)_R^{\nu}$ subgroup of the classical flavor horizontal symmetry
$S_{\text{flavor}}^{\text{extended}}$ (\ref{Sfe}) generated by
\begin{eqnarray}
  \eta^2 Q^{i j}_{R \nu} & \equiv & \int \mathd^3 \tmmathbf{x}
  \overline{\ell^i_R} \bignone \gamma^0 \mathcal{Y}^{\dag}_u \mathcal{Y}_u
  \ell_R^j,  \label{o}
\end{eqnarray}
is free of anomalies for all gauge field configurations for which the
constraints (\ref{c1}-\ref{c2}) hold. Indeed, the charges $Q^{i j}_{R \nu}$
(\ref{o}) can be equivalently written in terms of the free (massless)
right-handed Weyl spinors $N_R^i$ (\ref{m})
\begin{eqnarray}
  Q^{i j}_{R \nu} & = & \int \mathd^3 \tmmathbf{x} \overline{N_R^i} \gamma^0
  N_R^j .  \label{p}
\end{eqnarray}
Notice that this reasoning does not apply in the left-handed case. Even though
one can again define a $S_{\text{nat}}$ invariant left-handed spinor
\begin{eqnarray}
  N_L^i & \equiv & \phi_G^{\dag} \mathcal{Y}^{\dag}_u \Sigma^{\dag}
  \mathcal{X} \ell_L^i \hspace{0.2em} \overset{\text{s.g.}}{=} \hspace{0.2em} \xi
  \eta \nu_L^i,  \label{q}
\end{eqnarray}
it does not satisfy a free Dirac equation, due to the presence of the GB
matrix $\Sigma$ in the definition (\ref{q}).

\subsubsection{$\nu_R$ sign-flip symmetry} \label{flip}

At the leading order $\mathcal{O} \left( p^2 \right)$ without explicit powers
of spurions, we thus have a true symmetry $\mathrm{U} \left( 3
\right)_R^{\nu}$. We should thus ask how the higher-order terms transform with
respect to it. The right-handed Majorana mass $M_R = \zeta_{\Iota}^2 \eta^2
\mu^R$ (\ref{h}) and the neutrino Dirac mass $M_{\nu} = \xi \eta \mu^{\nu}$
both break the $\mathrm{U} \left( 3 \right)_R^{\nu}$ symmetry. However, we can
consider a discrete reflection symmetry
\begin{eqnarray}
  \mathbb{Z}_2 & \subset & \mathrm{U} \left( 3 \right)_R^{\nu},  \label{r}
\end{eqnarray}
acting on doublets $\ell_R^i$
\begin{eqnarray}
  \ell^i_R & \longmapsto & \left( 1 - 2 \Pi \right) \ell^i_R,  \label{5.148}
\end{eqnarray}
where $\Pi$ appearing in (\ref{5.148}) is the projector on the neutrino
component of the right-handed doublet
\begin{eqnarray}
  \eta^2 \Pi & \equiv & \mathcal{Y}^{\dag}_u \mathcal{Y}_u . 
\end{eqnarray}
The transformation (\ref{5.148}) implies, using the properties of the spurions
$\mathcal{Y}_u,\mathcal{Y}_d$ (\ref{63})
\begin{eqnarray}
  \mathcal{Y}_u \ell_R & \longmapsto & -\mathcal{Y}_u \ell_R, \\
  \mathcal{Y}_d \ell_R & \longmapsto & \mathcal{Y}_d \ell_R . 
\end{eqnarray}
In the standard gauge, the reflection (\ref{5.148}) simply amounts to
{\footnote{Note that the $\nu_R$ sign-flip symmetry (\ref{4.76}) can also be
imposed in the case of the SM augmented by $\nu_R$ degrees of freedom; it is
not specific to the Higgs-less case.}}
\begin{eqnarray}
  \nu_R & \longmapsto & \nu_R' \hspace{0.2em} = \hspace{0.2em} - \nu_R . 
  \label{4.76}
\end{eqnarray}
It allows for a non-zero right-handed Majorana mass, but forbids the Dirac
mass matrix $M_{\nu}$ in (\ref{d}).

The only mass terms for neutrinos are then those of the Majorana type, as
given in (\ref{Majright2}) and (\ref{j}). In our case, the smallness of
$\nu_L$ masses compared to those of charged fermions is directly related to
\begin{eqnarray}
  \zeta_{\Iota}^2 \xi^2 & \ll & \xi \eta . 
\end{eqnarray}
This suppression is efficient if $\zeta_{\Iota}$, which appeared first in
these lepton-number violating operators, is small with respect to $\xi$ and
$\eta$.

As for the right-handed neutrinos, their only mass term (\ref{Majright2}) is
proportional to $\zeta_{\Iota}^2 \eta^2$. We will try to get information on
allowed $\nu_R$ masses in Section \ref{6-nuR}, using cosmological
observations, after we discuss their couplings in Section \ref{nuR-couplings}.

\subsubsection{Strength of ($B - L$)-violating spurions and neutrino masses}
\label{alternatives-2}

We briefly come back to the two other alternatives mentioned in Section
\ref{alternatives}. In all three cases the $\nu_R$ sign-flip symmetry of
Section \ref{flip} has to be used, and we are left with Majorana masses
involving a $\zeta_{\Iota}^2, \zeta_{\tmop{II}}^2$ or $\zeta_{\tmop{III}}^2$
factor that is absent in the charged fermion Dirac masses. Hence, in all
cases, the parameter $\zeta^2$ is at the origin of suppression of neutrino
masses relative to charged fermion masses. One can check that this would not be
possible if we had taken two spurions carrying the $B - L$ charge as
independent.

The difference between the three remaining alternatives lies in the
accompanying factors of $\xi$ and $\eta$, as can be seen by writing down the
right-handed and left-handed neutrino Majorana terms.
\begin{itemize}
  \item Case $\mathrm{I}$ was defined in (\ref{Z}) without the explicit label
  $\Iota$. The respective mass terms for right and left-handed neutrinos are
  written in (\ref{Majright}) and (\ref{i}).
  
  \item In case $\tmop{II}$ defined in (\ref{ZII}), we have the following
  neutrino Majorana mass terms, using $\mathcal{Y}_d^{\dag}
  \mathcal{Z}_{\tmop{II}} =\mathcal{Z}_{\tmop{II}} \mathcal{Y}_u^{\dag} = 0$
\end{itemize}
\begin{eqnarray}
  \mathcal{L} \left( p^1 \eta^1 \zeta_{\tmop{II}}^2 \right)_{\text{leptons}} &
  = & - \overline{\ell^i_R} \mathcal{Y}_u^{\dag} \mathcal{Z}_{\tmop{II}} 
  \left( \ell^j_R \right)^c + \text{h.c}, \\
  \mathcal{L} \left( p^1 \xi^2 \eta^1 \zeta_{\tmop{II}}^2
  \right)_{\text{leptons}} & = & - \overline{\ell^i_L} \mathcal{X}^{\dag}
  \Sigma \mathcal{Z}_{\tmop{II}} \mathcal{Y}_d^{\dag} \Sigma^{\dag}
  \mathcal{X} \left( \ell^j_L \right)^c  \nonumber\\ && + \text{h.c} . 
\end{eqnarray}
\begin{itemize}
  \item In case $\tmop{III}$ defined by (\ref{ZIII}), and using $\mathcal{Y}_d
  \mathcal{Z}_{\tmop{III}} =\mathcal{Z}_{\tmop{III}} \mathcal{Y}^{\dag}_u = 0$
\end{itemize}
\begin{eqnarray}
  \mathcal{L} \left( p^1 \zeta_{\tmop{III}}^2 \right)_{\text{leptons}} & = & -
  \overline{\ell^i_R} \mathcal{Z}_{\tmop{III}}  \left( \ell^j_R \right)^c +
  \text{h.c}, \\
  \mathcal{L} \left( p^1 \xi^2 \eta^2 \zeta_{\tmop{III}}^2
  \right)_{\text{leptons}} & = & - \overline{\ell^i_L} \mathcal{X}^{\dag}
  \Sigma \mathcal{Y}_u \mathcal{Z}_{\tmop{III}} \mathcal{Y}^{\dag}_d
  \Sigma^{\dag} \mathcal{X} \left( \ell^j_L \right)^c  \nonumber\\ && + \text{h.c} . 
\end{eqnarray}

We see that the consequence of the different choices of Section
\ref{alternatives} on the ratio of left-handed to right-handed neutrino masses
are physical, since we have in the various cases the following estimates
\begin{eqnarray}
  m^{\mathrm{I}}_{\nu_R} & \sim & \left( \frac{\eta}{\xi} \right)^2 m_{\nu_L},
  \label{134}\\
  m^{\mathrm{\tmop{II}}}_{\nu_R} & \sim & \left( \frac{1}{\xi} \right)^2
  m_{\nu_L}, \\
  m^{\mathrm{\tmop{III}}}_{\nu_R} & \sim & \left( \frac{1}{\xi \eta} \right)^2
  m_{\nu_L} .  \label{136}
\end{eqnarray}
Note that the heaviest $\nu_L$ must be heavier than $\sqrt{\Delta
m^2_{\text{atm}}} \sim 0.05 \tmop{eV}$, but is also constrained to lie below
one $\tmop{eV}$ by cosmological observations (see Sections \ref{6-nuR} and
particularly \ref{Lyman}), i.e. for our purposes, we can consider it to be
experimentally known {\emdash}up to one order of magnitude. The unknown in
relations (\ref{134}-\ref{136}) is then $m_{\nu_R}$: it depends on which of
the scenarios $\Iota, \tmop{II}$ or $\tmop{III}$ is realized, whereas
$m_{\nu_L}$ is assumed to be the physical one. This is the reason why we have
written $m_{\nu_L}$ without an index $\Iota, \tmop{II}, \tmop{III}$ in the
right hand-side of (\ref{134}-\ref{136}). In Section \ref{6-nuR}, we will turn
to cosmological observations in order to constrain $m_{\nu_R}$.

\section{Non-standard couplings induced by spurions} \label{7-list}

In this section, we describe what could be the next-to-leading order (NLO) of
our low-energy expansion. According to the power-counting formula
(\ref{3.28}), loops calculated with lowest order $\mathcal{O} \left( p^2
\right)$ vertices, as described by the lagrangian (\ref{lag}), start to
contribute at chiral order $\mathcal{O} \left( p^4 \right)$. Here, we classify
and explicitly construct all vertices of chiral order $\mathcal{O} \left( p^2
\right)$ or $\mathcal{O} \left( p^3 \right)$ which involve (a minimal number
of) spurions. Such non-standard vertices will be suppressed by a power of
spurions with respect to tree SM contributions. On the other hand, they could
be more important than loop effects of the leading order (LO) $\mathcal{O}
\left( p^2 \right)$ lagrangian, which appear at order $\mathcal{O} \left( p^4
\right)$ and hence are {\emdash}at least formally{\emdash} more suppressed by
the chiral counting rules. A quantitative study of available data, motivated
by this classification of NLO contributions, will be presented separately
{\cite{Oertel}}.

To organize the discussion, we gather operators of a given order $\mathcal{O}
\left( p^{\alpha} \xi^{\beta} \eta^{\gamma} \right)$ according to the value of
\begin{eqnarray}
  \kappa & = & \alpha + \frac{\beta + \gamma}{2},  \label{6.2}
\end{eqnarray}
reflecting the possible relation (\ref{corres}) between momentum and spurion
power-counting. Equation (\ref{6.2}) presumes that the spurions $\xi$ and
$\eta$ are of a comparable size. As for the spurion $\zeta_{\Iota}$, its
presence signals lepton-number violation $\Delta L = 2$. With this in mind,
the same counting rule (\ref{6.2}) can be separately used both in lepton
number conserving and LNV sectors.

\subsection{Lepton-number conserving fermion couplings} \label{non-univ}

The first non-standard vertices between fermions and vector bosons appear in
the LEET at the level $\kappa = 3$. They are $\mathcal{O} \left( p^2 \xi^2
\right)$ and $\mathcal{O} \left( p^2 \eta^2 \right)$. This suggests that, in
the Higgs-less LEET, certain vertex corrections could be more important than
oblique ones. The latter are discussed in Section \ref{oblique}, where it will
be shown that oblique corrections require $\kappa \geqslant 4$, and receive
contributions from loops of the $\mathcal{O} \left( p^2 \right)$ lagrangian
(\ref{lag}).

Let us first concentrate on vector and axial vertices for quarks. We adopt the
following generic parametrization of effective left-handed and right-handed
couplings
\begin{widetext}
\begin{eqnarray}
  \mathcal{L}_{\text{quarks}} & = & eJ^{\mu} A_{\mu}  +\frac{e}{2 cs}  \left\{ \overline{u_L} \mathcal{F}_L \left( u, u
  \right) \gamma^{\mu} u_L + \overline{d_L} \mathcal{F}_L \left( d, d \right)
  \gamma^{\mu} d_L \right\} Z_{\mu} 
  +\frac{e}{2 cs}  \left\{ \overline{u_R} \mathcal{F}_R \left( u, u
  \right) \gamma^{\mu} u_R + \overline{d_R} \mathcal{F}_R \left( d, d \right)
  \gamma^{\mu} d_R \right\} Z_{\mu} \nonumber\\
  & &+ \frac{e}{\sqrt{2} s}  \left( \left\{ \overline{u_L} \mathcal{F}_L \left( u,
  d \right) \gamma^{\mu} d_L + \overline{u_R} \mathcal{F}_R \left( u, d
  \right) \gamma^{\mu} d_R \right\} W_{\mu}^+ + \text{h.c} \right).  \label{6.3}
\end{eqnarray}
\end{widetext}
In the above, $J^{\mu}$ is unchanged with respect to the SM, since $\mathrm{U}
\left( 1 \right)_Q$ is unbroken. The couplings $\mathcal{F}_{L, R}$ are
three-by-three matrices in generation space. They describe the breaking
pattern of the flavor symmetry (\ref{Sfe}), and departure from universality.
For leptons, we use the same conventions as in (\ref{6.3}) after a systematic
substitution
\begin{eqnarray}
  u \hspace{0.2em} \longmapsto \hspace{0.2em} \nu, &  & d \hspace{0.2em} \longmapsto
  \hspace{0.2em} e . 
\end{eqnarray}
\subsubsection{Flavor structure at the NLO} \label{SMFV}

As already pointed out at the beginning of Section \ref{5-masses}, successive
spurion orders may introduce a hierarchical structure of $S_{\text{flavor}}$
symmetry breaking. Within the SM, all direct and indirect effects of flavor
symmetry breaking originate in the fermion mass matrix, and they are
transmitted via loops involving internal fermion lines: at tree level, the six
effective coupling matrices $\mathcal{F}$ are proportional to the unit matrix
in an appropriate flavor basis.

Loop-induced flavor symmetry breaking effects can be parametrized by effective
$\mathcal{F}$ matrices, which are at least quadratic in fermion masses, since
two mass insertions are needed to preserve chirality. According to Section
\ref{quark-masses}, this means that the flavor asymmetry and the
non-universality in the couplings (\ref{6.3}) induced by loops would
contribute by terms of order at least $\mathcal{O} \left( \xi^2 \eta^2
\right)$, i.e. by operators of order $\mathcal{O} \left( p^2 \xi^2 \eta^2
\right)$ or $\kappa = 4$.

In the LEET, a genuine flavor symmetry breaking could already appear at the
tree $\kappa = 3$ level, and FCNCs could then be generated already at the NLO.
In order to eliminate the latter, it would be sufficient that the mass
matrices $M_u, M_d, M_e$ and the neutral current effective couplings
$\mathcal{F}_{L, R} \left( f, f \right)$ satisfy the conditions $( f = u, d, e
)$
\begin{eqnarray}
  \left[ \mathcal{F}_L \left( f, f \right), M_f M_f^{\dag} \right] & = & 0 ,\label{6.5-1}\\
  \left[ \mathcal{F}_R \left( f, f \right), M_f^{\dag} M_f \right]
  &=&0 .  \label{6.5}
\end{eqnarray}
$M_f$ and $\mathcal{F}_{L, R} \left( f, f \right)$ would then be diagonalized
in the same flavor basis, as it happens e.g. within the minimal flavor
violation scheme {\cite{D'Ambrosio:2002ex}}. However, it is not
straightforward to interpret equations (\ref{6.5-1}-\ref{6.5}) as a consequence of a
symmetry. For this reason, we adopt a stronger but simpler assumption than
(\ref{6.5-1}-\ref{6.5}) or  minimal flavor violation: we assume that, including the LO and
NLO orders, all flavor symmetry breaking can be transformed from vertices to
the fermion mass matrix. In particular, there exists a ``flavor-symmetric
basis'' in which the flavor symmetry (\ref{Sfe}) generated by the (conserved)
charges (\ref{79}-\ref{81}) is manifest: in the flavor-symmetric basis, all
six matrices $\mathcal{F}$ of equation (\ref{6.3}) are multiple of the unit
matrix. We refer to this hypothesis as ``soft flavor violation'' (SFV), by
analogy with the standard terminology used in the renormalizable case. At the
NLO, SFV excludes both the FCNCs and a violation of universality. In the basis
in which mass matrices are diagonal, SFV amounts to $( f = u, d )$
\begin{eqnarray}
  \mathcal{F}_{L, R}^{i j} ( f, f ) & = & \delta^{i j} F_{L, R} \left( f, f
  \right) + \text{NNLO},  \label{6.6}
\end{eqnarray}
and, for the charged currents
\begin{eqnarray}
  \mathcal{F}_{L, R}^{i j} ( u, d ) & = & \left( V^{L, R}_{\text{CKM}}
  \right)^{i j} F_{L, R} ( u, d ) + \text{NNLO} .  \label{130}
\end{eqnarray}
Here, $V_{\text{CKM}}^L$ is the CKM matrix and $V_{\text{CKM}}^R$ its
right-handed analogue: denoting by $U_L, U_R, D_L$ and $D_R$ the $\mathrm{U}
\left( 3 \right)$ transformations of $u_L^i, u^i_R, d_L^i$ and $d_R^i$
respectively, to the basis in which $M_u$ and $M_d$ are diagonal, one has
\begin{eqnarray}
  V_{\text{CKM}}^L & = & U_L D_L^{\dag}, \\
  V_{\text{CKM}}^R & = & U_R D_R^{\dag} . 
\end{eqnarray}
By construction, both $V_{\text{CKM}}^L$ and $V_{\text{CKM}}^R$ are unitary.

Consequences of the hypothesis of SFV in the lepton sector are similar to the
case of quarks. They will be discussed shortly.

\subsubsection{The left-handed quark sector}

We now return to the formalism of LEET in order to show how NLO vertices of
the type (\ref{6.3}) can be explicitly constructed with spurions when we
require the full $S_{\text{nat}}$ symmetry. Since the result is different in
the left and right-handed sectors, the two cases will be presented separately.

Let $\chi_L$ denote a generic left-handed fermion doublet ($\chi = q$ or
$\ell$). For each pair of family indices $( i, j )$, there is a single
lepton-number conserving invariant bilinear in $\chi_L$ that contains one
covariant derivative and is independent of the $\mathcal{O} \left( p^2
\right)$ SM coupling $\mathi \overline{\chi_L} \gamma^{\mu} D_{\mu} \chi_L$
\begin{eqnarray}
  L_{\text{NS}}^{i j} & = & \mathi \overline{\chi^i_L} \gamma^{\mu}
  \mathcal{X}^{\dag}  \left( \Sigma D_{\mu} \Sigma^{\dag} \right) \mathcal{X}
  \chi^j_L .  \label{6.11}
\end{eqnarray}
Indeed, (\ref{6.11}) represents the minimal spurion insertion into equation
(\ref{E.23}) that is necessary to restore the $S_{\text{nat}}$ symmetry. The
operator (\ref{6.11}) is $\mathcal{O} \left( p^2 \xi^2 \right)$ and therefore
has $\kappa = 3$ as defined in (\ref{6.2}). Consequently, (\ref{6.11})
contributes before loops and represents the unique non-standard NLO vertex of
left-handed quarks. Using the constraints (\ref{c1}-\ref{c2}) and the
variables introduced in (\ref{88}-\ref{91}) in the standard gauge,
(\ref{6.11}) can be written as
\begin{eqnarray}
  L_{\text{NS}}^{i j} & \overset{\text{s.g.}}{=} & - \xi^2  \frac{e}{2 cs} 
  \left\{ \overline{u^i_L} \gamma^{\mu} Z_{\mu} u^j_L - \overline{d^i_L}
  \gamma^{\mu} Z_{\mu} d^j_L \right. \nonumber\\
&&+ \left. \sqrt{2} c \left( \overline{u^i_L}
  \gamma^{\mu} W^+_{\mu} d^j_L + \text{h.c.} \right) \right\} .  \label{an-1}
\end{eqnarray}
One observes that the NLO modifications of the left-handed vertex still remain
predictive: it is expressed by a single three-by-three matrix in generation
space. Comparing with the general phenomenological representation (\ref{6.3}),
it is seen then the NLO contributions $\delta \mathcal{F}$ to the
phenomenological matrices $\mathcal{F}$ are related as follows
\begin{eqnarray}
  \delta \mathcal{F}_L \left( u, u \right) & = & - \delta \mathcal{F}_L \left(
  d, d \right) \hspace{0.2em} = \hspace{0.2em} \delta \mathcal{F}_L \left( u, d
  \right) .  \label{6.13}
\end{eqnarray}
Incorporating the hypothesis of SFV at NLO (\ref{6.6}-\ref{130}), one can
summarize left-handed quark couplings including LO and NLO as
\begin{eqnarray}
  F_L \left( u, u \right) & = & \left( 1 - \frac{4}{3} s^2 - \xi^2 \lambda_L
  \right) + \text{NNLO}, \\
  F_L \left( d, d \right) & = & \left( - 1 + \frac{2}{3} s^2 + \xi^2 \lambda_L
  \right) + \text{NNLO}, \\
  F_L \left( u, d \right) & = & \left( 1 - \xi^2 \lambda_L \right) +
  \text{NNLO} . 
\end{eqnarray}
At NLO, the left-handed couplings of quarks are described by a single
additional constant $\xi^2 \lambda_L$, compared to the tree-level SM. There is
no modification of the vector electromagnetic coupling at NLO.

\subsubsection{The right-handed quark sector}

In order to make an invariant with respect to $S_{\text{nat}}$ out of
right-handed doublets $\chi_R^i$, instead of (\ref{6.11}), one needs the
spurion $\mathcal{Y}_u$, or its conjugate $\mathcal{Y}_d$. As a result, a
single NLO $\mathcal{O} \left( p^2 \eta^2 \right)$ {\emdash}i.e. $\kappa = 3$
{\emdash} left-handed invariant (\ref{6.11}) now becomes three independent
invariants $\left( a, b = u, d \right)$
\begin{eqnarray}
  R_{\text{NS}}^{i j} \left( a, b \right) & = & \mathi \overline{\chi^i_R}
  \gamma^{\mu} \mathcal{Y}_a^{\dag} \Sigma^{\dag}  \left( D_{\mu} \Sigma
  \right) \mathcal{Y}_b \chi^j_R .  \label{6.15}
\end{eqnarray}
Hence, NLO modifications of SM couplings of right-handed quarks is
characterized by three parameters $\eta^2 \lambda_R^u, \eta^2 \lambda_R^d$ and
$\eta^2 \kappa_R$. Explicitly, using the notation of Section \ref{SMFV}, one
has in the mass-diagonal flavor basis
\begin{eqnarray}
  F_R \left( u, u \right) & = & - \frac{4}{3} s^2 + \eta^2 \lambda_R^u, 
  \label{an-3}\\
  F_R \left( d, d \right) & = & \frac{2}{3} s^2 - \eta^2 \lambda_R^d, 
  \label{an-4}\\
  F_R \left( u, d \right) & = & \eta^2 \kappa_R,  \label{an-2}
\end{eqnarray}
collecting the LO (SM) and NLO contributions. The last equation (\ref{an-2})
represents the charged right-handed quark currents, which is predicted to
appear at the NLO order $\mathcal{O} \left( p^2 \eta^2 \right)$. Notice that
(\ref{an-2}) arises multiplied by the right-handed CKM matrix
$V_{\text{CKM}}^R$ {\emdash}c.f. equations (\ref{6.3}) and
(\ref{130}){\emdash} which remains completely unknown. Informations on $\eta^2
\kappa_R V_{\text{CKM}}^R$ can be expected to arise from $\Delta S = 2$ and
$\Delta S = 1$ FCNC processes induced by loops. An important issue to be
clarified is the influence of $\eta^2 \kappa_R V_{\text{CKM}}^R$ on the
standard determination of the CKM matrix elements $\left( V_{\text{CKM}}^L
\right)^{i j}$.

\subsection{Lepton sector and the interactions of quasi-sterile $\nu_R$}
\label{nuR-couplings}

It is straightforward to extend the previous analysis to leptons and extract
the lepton-number conserving couplings up to and including the NLO. Here we
assume that the hypothesis of SFV applies both for quarks and for leptons. For
left-handed couplings, one gets from equation (\ref{6.11}), in the basis in
which both electron and $\nu_L$ masses are diagonal
\begin{eqnarray}
  \mathcal{F}^{i j}_L \left( \nu, \nu \right) & = & \delta^{i j}  \left( 1 -
  \xi^2 \rho_L \right) + \tmop{NNLO}, \\
  \mathcal{F}^{i j}_L \left( e, e \right) & = & \delta^{i j}  \left( - 1 + 2
  s^2 + \xi^2 \rho_L \right) + \tmop{NNLO}, \\
  \mathcal{F}^{i j}_L \left( \nu, e \right) & = & \left( V^L_{\text{MNS}}
  \right)^{i j}  \left( 1 - \xi^2 \rho_L \right) + \tmop{NNLO}, 
\end{eqnarray}
where $V^L_{\text{MNS}}$ is the MNS matrix describing the transformation
between flavor and mass basis for left-handed neutrinos and electrons.

An important distinction between quark and lepton couplings arises in the
right-handed sector: whereas they were allowed for quarks  (\ref{an-2}),
charged right-handed currents are forbidden for leptons, due to the $\nu_R$
sign-flip symmetry (\ref{5.148}) introduced in the previous Section. Neither
are such couplings generated by loops, as long as the discrete symmetry is
assumed to be exact {\footnote{Though natural and self-consistent, the
assumption that this symmetry is exact is merely dictated by simplicity, and
could therefore be relaxed. It could be easily falsified by a detection of
charged leptonic right-handed currents.}}. Since it also forbids Dirac mass
term for neutrinos, the consequence of the $\nu_R$ sign-flip symmetry is that
the right-handed neutrinos do not interact (or mix) with their left-handed
counterparts. Therefore, there can be no oscillations between the two, and the
so-called LSND anomaly {\cite{Aguilar:2001ty}} cannot be explained in this
way. The right-handed lepton couplings read off from equation (\ref{6.15}) can
be written as
\begin{eqnarray}
  \mathcal{F}^{i j}_R \left( \nu, \nu \right) & = & \delta^{i j} \eta^2
  \rho^{\nu}_R + \tmop{NNLO},  \label{147}\\
  \mathcal{F}^{i j}_R \left( e, e \right) & = & \delta^{i j}  \left( 2 s^2 -
  \eta^2 \rho^e_R \right) + \tmop{NNLO}, \\
  \mathcal{F}^{i j}_R \left( \nu, e \right) & = & 0 . 
\end{eqnarray}

Note that the three right-handed neutrinos are expected to be lighter than the
scale $\Lambda_{\text{w}} \sim 3 \tmop{TeV}$, otherwise they would not belong
to the LEET. This is emphasized in (\ref{h}) by the explicit power of
$\zeta_{\Iota}^2$ appearing in their (Majorana) masses. Therefore, we would
expect them to be light enough in order to be pair-produced in $Z^0$ decays.
Since the $\nu_R$ carry no $\tmop{SU} \left( 2 \right)_L \times \mathrm{U}
\left( 1 \right)_Y$ charge, however, their couplings to electroweak vector
bosons only come with explicit powers of spurions: the right-handed neutrinos
decouple in the limit where the spurions are sent to zero. Their couplings are
therefore suppressed with respect to weak interactions: they are
``quasi-sterile'', but do have interactions with $Z^0$ due to the $\mathcal{O}
\left( p^2 \eta^2 \right)$ operator yielding (\ref{147}), which we rewrite in
full here
\begin{eqnarray}
  \mathcal{L} \left( p^2 \eta^2 \right) & = & \rho^{\nu}_R \eta^2  \frac{e}{2
  cs}  \overline{\nu_R} \gamma^{\mu} \nu_R Z_{\mu} .  \label{6.82}
\end{eqnarray}
If the suppression factor $\eta$ is of order $0.1$ or smaller, we find that
the contribution to the $Z^0$ invisible width is smaller than the errors on
this observable {\cite{LEP:2003}}. We will assume this to be true throughout
the sequel. If this were not the case, then we would conclude that the
Higgs-less scenarios described by the present effective theory including light
quasi-sterile right-handed neutrinos are excluded. One would then have to find
an alternative way of ensuring an approximate custodial symmetry without
introducing right-handed neutrinos at all. As far as we know, this is not
excluded: we have only considered the simplest possibility of interpreting the
right isospin as the origin of custodial symmetry {\cite{Sikivie:1980hm}}.
Since such quasi-sterile $\nu_R$ are not excluded by particle physics
experiments, we will turn to cosmological observations in Section \ref{6-nuR}.

\subsection{Comment on oblique corrections} \label{oblique}

We have considered above the LO and NLO of the Higgs-less LEET which should
both be treated at tree level: they involve operators with respectively
$\kappa = 2$ and $\kappa = 3$. The set of operators usually collected under
the name of ``oblique corrections'' has $\kappa = 4$ in this framework and is
therefore {\tmem{not}} the main ``deviation from the SM'': these operators
appear only at NNLO, i.e. at the same level as loops. This is the reason why
they should be considered after the non-universal couplings of Sections
\ref{non-univ} and \ref{nuR-couplings}. In this case, we have to give up the
formalism of oblique corrections, which was specifically designed to take into
account cases where the main effect from new physics would occur in two-point
function of gauge bosons  {\footnote{We
note that, for the special case of 5D Higgs-less models, it has been stressed
in {\cite{Chivukula:2004af}} that non-oblique corrections were relevant, even
though they were of the type called ``universal'' in {\cite{Barbieri:2004qk}}.
Such corrections may in fact even be required in 5D Higgs-less models, to
compensate in some observables for the non-zero value of the $S$ parameter
{\cite{Cacciapaglia:2004rb,Foadi:2004ps}}.}}. The outcome is that one must resort to observables,
as advocated for instance in {\cite{Sanchez-Colon:1998xg}}: in brief, oblique
corrections are not the first corrections to be looked for in Higgs-less
scenarios, and they cannot be discussed independently of loops.

Having stressed that, we mention the following $\kappa = 4$ operators which
all reduce, in the standard unitary gauge, to the same operator usually
associated with the oblique $S$ parameter
\begin{widetext}
\begin{eqnarray}
  \left\langle G_{L \mu \nu} \mathcal{X}^{\dag} \Sigma \left( \mathcal{Y}_u
  +\mathcal{Y}_d \right) G_R^{\mu \nu}  \left( \mathcal{Y}_u^{\dag}
  +\mathcal{Y}_d^{\dag} \right) \Sigma^{\dag} \mathcal{X} \right\rangle &
  \overset{\text{s.g.}}{=} & \xi^2 \eta^2 b_{\mu \nu}  \left( sA^{\mu \nu} +
  cZ^{\mu \nu} \right),  \label{S1}\\
  \left\langle G_{L \mu \nu} \mathcal{X}^{\dag} \Sigma \left( \mathcal{Y}_u
  -\mathcal{Y}_d \right) G_R^{\mu \nu}  \left( \mathcal{Y}_u^{\dag}
  -\mathcal{Y}_d^{\dag} \right) \Sigma^{\dag} \mathcal{X} \right\rangle &
  \overset{\text{s.g.}}{=} & \xi^2 \eta^2 b_{\mu \nu}  \left( sA^{\mu \nu} +
  cZ^{\mu \nu} \right), \label{S2}\\
  \left\langle \Gamma_{L \mu \nu} \Sigma \left( \mathcal{Y}_u +\mathcal{Y}_d
  \right) G_R^{\mu \nu} \left( \mathcal{Y}_u^{\dag} +\mathcal{Y}_d^{\dag}
  \right) \Sigma^{\dag} \right\rangle & \overset{\text{s.g.}}{=} & g \eta^2
  b_{\mu \nu}  \left( sA^{\mu \nu} + cZ^{\mu \nu} \right), \label{S3}\\
  \left\langle \Gamma_{L \mu \nu} \Sigma \left( \mathcal{Y}_u -\mathcal{Y}_d
  \right) G_R^{\mu \nu} \left( \mathcal{Y}_u^{\dag} -\mathcal{Y}_d^{\dag}
  \right) \Sigma^{\dag} \right\rangle & \overset{\text{s.g.}}{=} & g \eta^2
  b_{\mu \nu}  \left( sA^{\mu \nu} + cZ^{\mu \nu} \right), \label{S4}\\
  \left\langle G_{L \mu \nu} \mathcal{X}^{\dag} \Sigma \Gamma_R^{\mu \nu}
  \Sigma^{\dag} \mathcal{X} \right\rangle & \overset{\text{s.g.}}{=} & g'
  \xi^2 b_{\mu \nu}  \left( sA^{\mu \nu} + cZ^{\mu \nu} \right), 
  \label{Sl-1}\\
  \left\langle \Gamma_{L \mu \nu} \Sigma \Gamma_R^{\mu \nu} \Sigma^{\dag}
  \right\rangle & \overset{\text{s.g.}}{=} & gg' b_{\mu \nu}  \left( sA^{\mu
  \nu} + cZ^{\mu \nu} \right) . \label{Sl}
\end{eqnarray}
\end{widetext}
These operators should be compared with loops generated by the lagrangian
(\ref{lag}), which is also different from the SM, since it does not contain a
Higgs particle. Also, universal contributions from the operators mentioned in
Sections \ref{non-univ} and \ref{nuR-couplings} will modify the predictions
for the observables from which the $S$ parameter should be extracted.

Another oblique parameter is $T$: it represents a direct tree contribution to
the $Z^0$ mass not affecting the $W^{\pm}$ mass. In our LEET, the size of such
custodial-breaking effects are parameterized by the spurion $\mathcal{Y}$
(i.e. the parameter $\eta$), via the $\mathcal{O} \left( p^2 \eta^4 \right)$
operator

\begin{eqnarray}
  \Lambda^2  \left\langle \Sigma^{\dag} D_{\mu} \Sigma \left( \mathcal{Y}_u
  \mathcal{Y}^{\dag}_u -\mathcal{Y}_d \mathcal{Y}^{\dag}_d \right)
  \right\rangle ^2 & \overset{\text{s.g.}}{=} & - \Lambda^2 \eta^4 
  \frac{e^2}{c^2 s^2} \nonumber\\
&& \times Z_{\mu} Z^{\mu} .  \label{T}
\end{eqnarray}

\subsection{Prefactors and low-energy constants}

So far, we have merely concentrated on the ordering of operators in the
effective lagrangian according to powers of momenta and spurions. This is
needed for the formal definition of a consistent quantum theory in a
systematic low-energy expansion. When expressed in the standard gauge, each
operator already contains a definite power of $g$, $g'$ and of spurion factors
$\xi$, $\eta$ {\emdash}c.f. the NLO operators (\ref{6.11}) and (\ref{6.15}), or the NNLO oblique
operators (\ref{S1}-\ref{Sl}). Actually, each of these operators will enter the lagrangian
multiplied by a LEC, which is not determined by symmetry
arguments. Here, we would like to comment about the possible order of magnitude estimate of such
LECs, keeping in mind that, though constrained by consistency of the
order-by-order renormalization, such estimates can only be validated by a
detailed quantitative analysis.

There are in fact three distinct classes of operators in the effective theory:
those that involve only elementary fields, those that involve only composite
fields, and finally those that mix elementary and composite fields. In the
first class, prefactors can be estimated since the elementary sector is
weakly-interacting.

For the second class, the precedent of $\chi$PT is of great help to estimate
factors of $1 / \left( 16 \mathpi^2 \right)$ or $f^2 / \Lambda^2$. In
particular, the constant in front of the operator (\ref{Sl}) corresponds to
$L_{10}$ in $\chi$PT \cite{Holdom:1990tc}. In both effective theories
($\chi$PT and the present one), the formal counting puts this operator at the
same level as one-loop graphs. This is indeed realized numerically if the LEC
in front of (\ref{Sl}) is of order the loop factor $1 / 16 \mathpi^2$, as observed
experimentally for the case of $\chi$PT (for a reasonable renormalization
scale). Such a QCD-like contribution to the $S$ parameter has the right
magnitude, but the wrong sign, to compensate for the absence of diagrams with
a Higgs \cite{Peskin:1992sw}. Still, this is not the end of the story here,
since operators of the third class also contribute to the $S$ parameter.

As for the operators belonging to the third class, which mix the elementary
and composite sector and therefore involve spurions, their prefactors can be estimated only by assuming
consistency with the power counting of the effective theory. Consider for
instance the operators (\ref{S1}, \ref{S2}), which have not been taken into account in
the literature: they have no equivalent in $\chi$PT (they contain the spurions
$\xi$ and $\eta$). One can fix a normalization of the spurions by assuming
that the two operators (\ref{S1}, \ref{S2}) enter the lagrangian with order one
prefactors. Comparing the contributions to $S$ of the operators (\ref{S1}, \ref{S2})
on one side, and of (\ref{Sl}) on the other side, and assuming that they are not
only of the same formal order, but of the same numerical order, we obtain
\begin{eqnarray}
  \xi^2 \sim \frac{g}{4 \mathpi}, &  & \eta^2 \sim \frac{g'}{4 \mathpi} . 
  \label{norm-spu}
\end{eqnarray}
Similarly, one expects that the operators (\ref{S3}-\ref{Sl-1}) appear with a prefactor
of order $1 / (4 \mathpi)$. Note that, in this case, the total tree
contribution to $S$ is still expected to be of the right order of magnitude to
compete with loop contributions.

Since we have arbitrarily normalized the spurions by assuming order one
prefactors in (\ref{S1}, \ref{S2}), slightly smaller/larger prefactors might appear in
front of other operators, for instance the fermion mass term (\ref{topmass}) and the
non-standard fermion-gauge interactions (\ref{6.11}, \ref{6.15}). In the case of the top mass, this
can be equivalently phrased as $\Lambda_Q$ being different from
$\Lambda_{\text{w}} \sim 3.1$ TeV. However, we estimate $\Lambda_Q \sim 4.7$
TeV, not a large relative factor. In the non-standard interactions (\ref{6.11}, \ref{6.15}), we see
that a prefactor of order $0.1$ would be required to satisfy electroweak
tests. Inside the effective theory, we do not know what the origin of this
additional suppression could be: the possibility cannot be ruled out
by logic alone
(one cannot expect the effective theory itself to exclude Higgs-less
scenario), and must be investigated in a detailed comparison with experiment 
.

\section{Lepton-number violation at the NLO} \label{LNV-NLO}

The left-handed and right-handed Majorana neutrino masses are the only
manifestation of LNV and of the spurion $\zeta_{\Iota}$ considered so far.
Here we concentrate on other LNV effects. The LEET approach, which presumes a
separation of low and high mass scales should allow to consider such effects
in full generality without commitment to a particular scenario for physics at
the high scale. For definiteness, we stick to the process
\begin{eqnarray}
  W^- + W^- & \longrightarrow & e^- + e^-,  \label{7.1}
\end{eqnarray}
which can be either real (possibly observable at a high luminosity collider),
or virtual. In the latter case, the process (\ref{7.1}) appears as a
sub-diagram of various neutrino-less double beta decays ($0 \nu 2 \beta$) such
as $\left( A, Z \right) \longrightarrow \left( A, Z + 2 \right) + e^- + e^-$,
rare $K$ decays $K^- \longrightarrow \pi^+ + e^- + e^-$, or their lepton
flavor violating variants. There are two distinct problems: i) one must
estimate the hadronic matrix elements of the product of the two hadronic
currents coupled to the $W$'s of (\ref{7.1}). This problem is yet another
challenge to the non-perturbative QCD phenomenology, and will not be addressed
in this work. We are more particularly concerned with the second problem: ii)
How could an experimental information about the strength, electron spectrum
and polarizations of the elementary process (\ref{7.1}) be exploited to learn
about the mechanism and parameters of LNV. In particular, could it be used to
extract an experimental information about neutrino masses and mixings?

\subsection{Tree contributions to $e^- + e^- \longrightarrow W^- + W^-$}

\begin{figure*}
\includegraphics{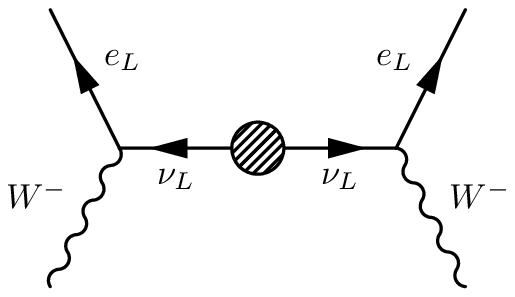}
\includegraphics{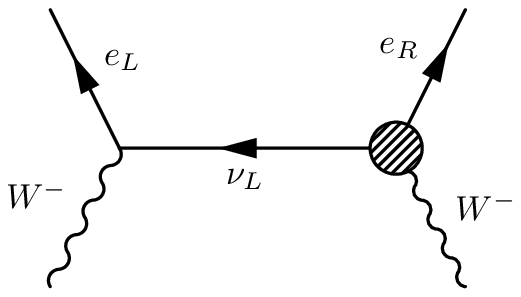}
\caption{\label{F1a}Majorana mass insertions and \label{F1b}odd chirality LNV vertex.}
\end{figure*}

\begin{figure*}
\includegraphics{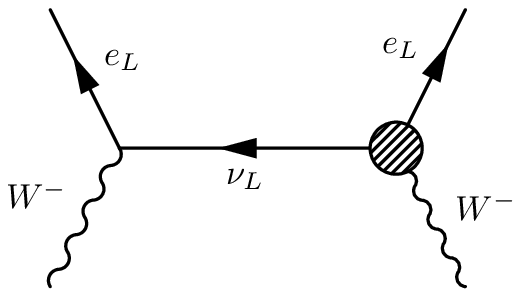}
\includegraphics{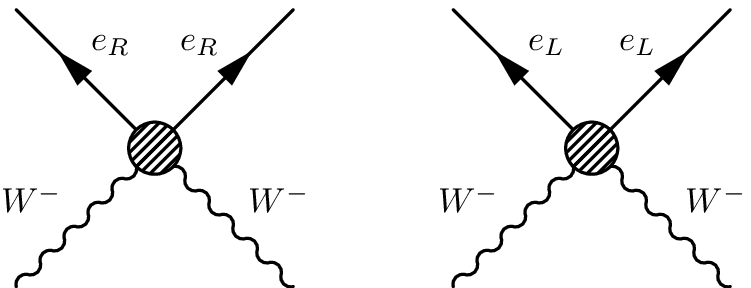}
\caption{\label{F1c}Magnetic type LNV vertex and \label{F1d}LNV contact terms.}
\end{figure*}

Four different types of contributions to the process (\ref{7.1}) are
classified in FIGs. \ref{F1a}-\ref{F1d}.

Let us first focus on the Majorana mass insertion in the $t$-channel neutrino
exchange shown in FIG. \ref{F1a}. In this contribution, the lepton number
violation is {\tmem{indirect}} since it does not involve any one-particle
irreducible LNV vertex. It stems from the square of the standard $\Delta L =
0$ interaction term
\begin{eqnarray}
  \mathcal{L}_{\text{CC}} & = & \frac{g}{\sqrt{2}}  \sum_{i, j}
  V_{\text{MNS}}^{i, j}  \left( \overline{\nu^i_L} \gamma^{\mu} e_L^j \right)
  W_{\mu}^+ + \text{h.c},  \label{7.2}
\end{eqnarray}
where $\nu_L^j$ ($j = 1, 2, 3$) is the $j$-th left-handed Majorana mass
eigenstate and $V_{\text{MNS}}$ stands for the MNS mixing matrix. Since the
asymptotic field $\nu^j_L$ obeys the chirally projected Majorana-Dirac
equation {\footnote{In the present case it is not particularly convenient to
use Majorana fields $\nu_M = \nu_M^c$ instead of the chiral ones $\gamma_5
\nu_L = - \nu_L$. These are just two equivalent sets of variables related by
the transformations $\nu_M = \nu_L + \nu_L^c$, $\nu_L = \frac{1 - \gamma_5}{2}
\nu_M$, describing the same lagrangian and the same physics.}}
\begin{eqnarray}
  \mathi \gamma^{\mu} \partial_{\mu} \nu_L^j & = & m_j  \left( \nu_L^j
  \right)^c,  \label{7.3}
\end{eqnarray}
there exists a $\Delta L = 2$ correlation leading to the propagator
\begin{eqnarray}
  \left\langle 0 | T \nu_L \left( x \right)  \overline{\left( \nu_L \right)^c}
  \left( y \right) | 0 \right\rangle & = &  m \frac{1 - \gamma_5}{2}\nonumber\\
&&\times \Delta_F
  \left( x - y, m \right),  \label{7.4}
\end{eqnarray}
which is represented in FIG. \ref{F1a} by the line with a shaded circle. The
corresponding indirect LNV contribution to $e^i + e^j \longrightarrow W + W$,
to $0 \nu 2 \beta$ decays and to other $\Delta L = 2$ processes is well-known
and described in the literature, see for instance {\cite{Atre:2005eb}}. It is
entirely given by the Majorana mass matrix element
\begin{eqnarray}
  \mu^{i, j} & = & \sum_k \left( V_{\text{MNS}}^{k, i} \right)^{\ast}  \left(
  V_{\text{MNS}}^{k, j} \right)^{\ast} m_k .  \label{7.5}
\end{eqnarray}
In the approximation $p^2 \gg m^2$, the contribution to the amplitude $e e
\longrightarrow W W$ becomes
\begin{eqnarray}
  M_1 & = & \frac{g^2}{2} \Delta_F \left( x - y \right)  \left[ \overline{e_L}
  \left( x \right) V^{\dag}_{\text{MNS}} mV_{\text{MNS}}^{\ast}  \left( e_L
  \left( y \right) \right)^c \right]\nonumber\\
&& \times W^{- \mu} \left( x \right) W_{\mu}^-
  \left( y \right) .  \label{7.9}
\end{eqnarray}

The diagrams shown in FIGs. \ref{F1b} and \ref{F1d} represent
different types of {\tmem{direct contributions}} from irreducible LNV
vertices. Within the present LEET framework, the existence and strength of the
latter depends on their chiral dimension and on their minimal spurion content
as dictated by the symmetry properties under $S_{\text{nat}}$. Before we
proceed to this analysis (Sections \ref{S7.2} and \ref{S7.3}), it is useful to
clarify the chirality structures of possible $\Delta L = 2$ vertices using
nothing but Lorentz invariance.

FIG. \ref{F1b} involves possible {\tmem{chirality-odd}} $\Delta L = 2$
vertex (indicated by a shaded circle) which, in the standard unitary gauge,
would be of the form
\begin{eqnarray}
  \mathcal{L}_{\Delta L = 2}^{\text{odd}} & = & g \left( \overline{e_R}
  \epsilon \gamma^{\mu}  \left( \nu_L \right)^c \right) W_{\mu}^- +
  \text{h.c}, \nonumber\\
  & = & - g \left( \overline{\nu_L} \epsilon^T \gamma^{\mu}  \left( e_R
  \right)^c \right) W^-_{\mu} + \text{h.c} .  \label{7.6}
\end{eqnarray}
Interference with the standard coupling (\ref{7.2}) contains the unsuppressed
chirality-odd part of the neutrino propagator
\begin{eqnarray}
  S_F \left( x - y, m \right) & \equiv & \left\langle 0 | T \nu_L \left( x
  \right)  \overline{\left( \nu_L \right)} \left( y \right) | 0 \right\rangle
\nonumber\\
&=& \frac{1 - \gamma_5}{2} \gamma^{\mu}
  \partial_{\mu} \Delta_F \left( x - y, m \right),  \label{7.7}
\end{eqnarray}
and the contribution to the amplitude $e e \longrightarrow W W$ shown in
FIG. \ref{F1b} becomes
\begin{eqnarray}
  M_2 & = & g^2  \left[ \overline{e_L} \left( x \right) \gamma^{\mu}
  V^{\dag}_{\text{MNS}} S_F \left( x - y, m \right) \epsilon^T \gamma^{\nu} 
  \left( e_R \left( y \right) \right)^c \right] \nonumber\\
&& \times W_{\mu}^- \left( x \right)
  W_{\nu}^- \left( y \right) .  \label{7.8}
\end{eqnarray}
For not too small neutrino momenta $p^2 \gg m^2$, the mass dependence of $S_F
\left( x - y, m \right)$ can be neglected and the coupling matrix in
(\ref{7.8}) reduces to $V^{\dag} \epsilon^T$, i.e. it is a priori
{\tmem{independent of neutrino masses}}. Comparing (\ref{7.9}) with
(\ref{7.8}), we note that the indirect contribution of FIG. \ref{F1a} and
the direct one of FIG. \ref{F1b} lead to different polarization spectra and
angular distribution of the two final leptons in $W W \longrightarrow e e$. In
principle, they could be distinguished experimentally.

FIG. \ref{F1c} describes the contribution to $W W \longrightarrow e e$
arising from direct LNV {\tmem{chirality-even}} vertices (shaded circle). Such
vertices are of magnetic type $\overline{e_L} \sigma^{\mu \nu}  \left( \nu_L
\right)^c \partial_{\mu} W_{\nu}^-$. Using equations of motion, they can be
reduced to
\begin{eqnarray}
  \mathcal{L}^{\text{even}}_{\Delta L = 2} & = & \frac{1}{\Lambda} W_{\mu}^- 
  \overline{e_L} \epsilon' \partial^{\mu} \left( \nu_L \right)^c + \text{h.c},
  \label{7.10}
\end{eqnarray}
with its own coupling matrix $\epsilon'$. The vertex (\ref{7.10}) contains one
more power of momentum as compared to the chirality-odd vertex (\ref{7.6}). It
is therefore dimensionally suppressed by the inverse power of an a priori
unknown scale $\Lambda$. The corresponding contribution to $W W
\longrightarrow e e$ (FIG. \ref{F1c}) reads
\begin{eqnarray}
  M_3 & = & \frac{g}{\Lambda}  \left[ \overline{e_L} \left( x \right)
  \gamma^{\mu} V^{\dag}_{\text{MNS}} \epsilon'^T S_F \left( x - y, m \right)
  \partial^{\nu}  \left( e_L \left( y \right) \right)^c \right] \nonumber\\ 
&& \times W_{\mu}^-
  \left( x \right) W_{\nu}^- \left( y \right) .  \label{7.11}
\end{eqnarray}
It has a similar helicity structure as the indirect contribution of FIG.
\ref{F1a}, but it differs by the distribution of final lepton momenta. Before
one concludes that this direct LNV term can be neglected on the basis of
momentum power counting, one has to check the spurion content of $\epsilon'$
compared to that of $\epsilon$, and to that of the Majorana mass $m$.

The direct LNV contact term of FIG. \ref{F1d} will be discussed in Section
\ref{contact}.

\subsection{The unique $\Delta L = 2$ chirality-odd vertex} \label{S7.2}

In order to infer the degree of suppression of the coupling $\epsilon$ and
$\epsilon'$, one has to go back to the formalism of Section \ref{4-spurions}.
The $\Delta L = 2$ vertices must be constructed out of lepton doublets
$\ell_L$ and $\ell_R$, as well as their conjugates $\left( \ell_{L, R}
\right)^c$, of the gauge connections, the GB matrix $\Sigma$ and the spurions
$\mathcal{X},\mathcal{Y}_{u, d}$ and $\mathcal{Z}$. The vertices must be
invariant under the whole group $S_{\text{nat}}$ and moreover, they have to
respect the $\mathbb{Z}_2$ $\nu_R$ {\tmem{sign-flip symmetry}} (\ref{5.148}).
It turns out that there is a single chirality-odd vertex with these properties
carrying the minimal chiral order and a minimal number of spurion insertions.
The unique result reads
\begin{eqnarray}
  \mathcal{L}^{\text{odd}}_{\Delta L = 2} & = & \sum_{i, j} c_{i j} 
  \overline{\ell^i_R} \mathcal{Y}_d^{\dag} \gamma^{\mu}  \left( \Sigma^{\dag}
  D_{\mu} \Sigma \right) \mathcal{Z} \Sigma^{\dag} \mathcal{X} \left( \ell^j_L
  \right)^c \nonumber\\
&& + \text{h.c} .  \label{7.16}
\end{eqnarray}
It is remarkable that the chain of factors in (\ref{7.16}) is uniquely
determined by the symmetry properties, including the $\nu_R$ sign-flip
symmetry which plays a crucial role in the above construction. It provides a
non-trivial illustration of the spurion formalism at work.

In the standard unitary gauge (\ref{7.16}) reduces to (\ref{7.6}) with the
coupling matrix $\epsilon$ given by
\begin{eqnarray}
  \epsilon_{i j} & = & \zeta_{\Iota}^2 \xi \eta c_{i j},  \label{7.17}
\end{eqnarray}
the factor $g$ being contained in $\Sigma^{\dag} D_{\mu} \Sigma$. Compared to
the left-handed Majorana mass matrix (\ref{i})
\begin{eqnarray}
  \left( M_L \right)_{i j} & = & \Lambda \zeta_{\Iota}^2 \xi^2 \mu^L_{i j}, 
  \label{7.18}
\end{eqnarray}
one observes that the degree of spurion suppression of the indirect and direct
LNV contribution is similar as long as $\xi \sim \eta$. A more detailed
comparison for the process $W W \longrightarrow e e$ should be made in terms
of the matrix elements $M_1$ and $M_2$ of (\ref{7.9}) and (\ref{7.8}). One
finds
\begin{eqnarray}
  \frac{M_2}{M_1} & \sim & \frac{\left\langle p \right\rangle}{\Lambda} 
  \frac{\eta}{\xi},  \label{7.19}
\end{eqnarray}
where $\left\langle p \right\rangle$ is the average momentum transfer by the
neutrino {\emdash}c.f. FIG. \ref{F1b}. $\Lambda$ is the value
which the masses of left-handed neutrinos would take in the absence of the
spurion suppression factors.

\subsection{Direct LNV chirality-even vertices} \label{S7.3}

We now consider $\Delta L = 2$ vertices of the form
\begin{eqnarray}
  \overline{\ell_L} T_L  \left( \ell_L \right)^c & \text{or} &
  \overline{\ell_R} T_R  \left( \ell_R \right)^c .  \label{7.20}
\end{eqnarray}
$T_{L, R}$ are matrices made up from gauge fields, GBs $\Sigma$ and from
spurions. They transform under $S_{\text{nat}}$  according to
\begin{eqnarray}
  T_{L, R} & \longmapsto & \mathe^{\mathi \alpha} G_{L, R} T_{L, R} G_{L,
  R}^{\dag} .  \label{7.21}
\end{eqnarray}
$T_{L, R}$ should contain an even number of $\gamma$-matrices. Furthermore, in
order to reconcile LNV with the symmetry $S_{\text{nat}}$, $T_{L, R}$ must
contain the spurions $\mathcal{Z}$.

Interactions of this type can either contribute to the $\Delta L = 2$
processes discussed before, such as $0 \nu 2 \beta$ decays, via diagrams shown
on FIGs. \ref{F1c} and \ref{F1d} or they can generate new independent
$\Delta L = 2$ processes, such as the decays $Z \longrightarrow \nu \nu$ or $Z
\longrightarrow \overline{\nu}  \overline{\nu}$, or magnetic-type transitions
\begin{eqnarray}
  \nu_L \hspace{0.2em} \longleftrightarrow \hspace{0.2em} \overline{\nu_L} +
  \gamma, &  & \nu_R \hspace{0.2em} \longleftrightarrow \hspace{0.2em}
  \overline{\nu_R} + \gamma,  \label{7.22}
\end{eqnarray}
between different Majorana mass eigenstates of the left-handed or right-handed
neutrinos. In the following, we give an essentially complete list of such
chirality-even $\Delta L = 2$ vertices and clarify their degree of suppression
by spurions.

\subsubsection{Right-handed Majorana magnetic coupling}

As a consequence of the presumably exact $\nu_R$ sign-flip symmetry
(\ref{5.148}), there are no Pauli-Dirac magnetic couplings of neutrinos.
Furthermore, due to the Fermi statistic, the diagonal part of Majorana
magnetic couplings vanishes, leaving no place for a neutrino magnetic moment
to all orders of our LEET. The Majorana {\tmem{magnetic transition moments}}
can be constructed in a full analogy with Majorana mass terms of Section
\ref{Maj-masses}. As in the latter case, the result, in particular the degree
of spurion suppression, could be different in the left-handed and right-handed
cases.

The lowest order $S_{\text{nat}}$-invariant right-handed magnetic vertex
merely involves the insertion of the spurion $\mathcal{Z}$
\begin{eqnarray}
  \mathcal{L}^R_{\text{mag}} & = & \frac{1}{\Lambda}  \overline{\ell^i_R}
  \mathcal{Y}_u^{\dag} \mathcal{Z} \mathcal{Y}_d \sigma^{\mu \nu}  \left(
  \ell_R^j \right)^c B_{\mu \nu} + \text{h.c} .  \label{7.23}
\end{eqnarray}
It is of the order $\mathcal{O} \left( p^2 \zeta_{\Iota}^2 \eta^2 \right)$.
$B_{\mu \nu}$ is the $\mathrm{U} \left( 1 \right)_{B - L}$ field strength and
$\Lambda$ is a scale reflecting the dimensional suppression. The vertex
(\ref{7.23}) is antisymmetric in the lepton flavor indices $\{ i, j \}$.
Furthermore, (\ref{7.23}) exhibits the $\nu_R$ sign-flip symmetry
automatically: in the standard unitary gauge, it reduces to a single term
quadratic in $\nu_R$
\begin{eqnarray}
  \mathcal{L}^R_{\text{mag}} & \overset{\text{s.g.}}{=} &
  \frac{\zeta_{\Iota}^2 \eta^2}{\Lambda}  \overline{\nu^i_R} \sigma^{\mu \nu} 
  \left( \nu_R^j \right)^c  \left( cA_{\mu \nu} - sZ_{\mu \nu} \right) \nonumber\\
&& +
  \text{h.c},  \label{7.24}
\end{eqnarray}
where $A_{\mu \nu}$ is the electromagnetic field and similarly, $Z_{\mu \nu}
\equiv \partial_{\mu} Z_{\nu} - \partial_{\nu} Z_{\mu}$.

The magnetic vertex (\ref{7.24}), together with the LN-conserving interaction
$Z \longrightarrow \overline{\nu_R} \nu_R$ (\ref{6.82}) completes the picture
of residual interactions of quasi-sterile right-handed neutrinos. They are
induced by spurions, $\mathcal{O} \left( p^2 \eta^2 \right)$ in the case of
(\ref{6.82}) and $\mathcal{O} \left( p^2 \zeta_{\Iota}^2 \eta^2 \right)$ in
the LNV case (\ref{7.24}). Note that, with three families, decays such as
(\ref{7.22}) imply that only the lightest $\nu_R$ and $\nu_L$ are stable.
However, the spurion factors in (\ref{7.24}) are such that, for masses below
the expected maximum of order $\tmop{keV}$, as allowed for $m_{\nu_R}$ in case
$\tmop{III}$ of Section \ref{alternatives-2}, the lifetime of all neutrino
species is longer than the age of the universe {\footnote{We thank Émilie
Passemar for an explicit check of this estimate.}}. Hence, we shall consider
that all neutrinos are stable for the purpose of the discussion of their
cosmological impact in Section \ref{6-nuR}.

\subsubsection{Left-handed LNV magnetic couplings}

We now turn to the magnetic couplings in the left-handed sector. The minimal
set of spurions which is needed to restore the invariance under
$S_{\text{nat}}$ is the same as in the case of left-handed Majorana masses
(\ref{i}). The extension of (\ref{7.23}) reads
\begin{eqnarray}
  \mathcal{L}^{L, 1}_{\text{mag}} & = & \frac{1}{\Lambda}  \overline{\ell^i_L}
  \mathcal{X}^{\dag} \Sigma \mathcal{Z} \Sigma^{\dag} \mathcal{X} \sigma^{\mu
  \nu}  \left( \ell_L^j \right)^c B_{\mu \nu} + \text{h.c} .  \label{7.25}
\end{eqnarray}
which is $\mathcal{O} \left( p^2 \xi^2 \zeta_{\Iota}^2 \right)$. In the
standard unitary gauge, it reduces to
\begin{eqnarray}
  \mathcal{L}^{L, 1}_{\text{mag}} & \overset{\text{s.g.}}{=} &
  \frac{\zeta_{\Iota}^2 \xi^2}{\Lambda}  \overline{\nu^i_L} \sigma^{\mu \nu} 
  \left( \nu_L^j \right)^c  \left( cA_{\mu \nu} - sZ_{\mu \nu} \right)  \nonumber\\
&& +
  \text{h.c} .  \label{7.26}
\end{eqnarray}
We see that the spurion factors suppressing the magnetic transition $\nu_L
\longrightarrow \overline{\nu_L} + \gamma$ and $\nu_R \longrightarrow
\overline{\nu_R} + \gamma$ are the same as for the respective Majorana mass
terms. In the left-handed sector, additional LNV magnetic couplings of the
order $\mathcal{O} \left( p^2 \xi^2 \zeta_{\Iota}^2 \right)$ are possible.
They are obtained inserting into appropriate places of the spurion chain
(\ref{7.25}) the field strength $G_{L \mu \nu} \longmapsto G_L G_{L \mu \nu}
G_L^{\dag}$ instead of the invariant $B_{\mu \nu}$
\begin{eqnarray}
  \mathcal{L}^{L, 2}_{\text{mag}} & = & \frac{1}{\Lambda}  \overline{\ell^i_L}
  G_{L \mu \nu} \mathcal{X}^{\dag} \Sigma \mathcal{Z} \Sigma^{\dag}
  \mathcal{X} \sigma^{\mu \nu}  \left( \ell_L^j \right)^c + \text{h.c} . 
  \label{7.27}
\end{eqnarray}
In the standard unitary gauge, this vertex reads
\begin{eqnarray}
  \mathcal{L}^{L, 2}_{\text{mag}} & \overset{\text{s.g.}}{=} &
  \frac{\zeta_{\Iota}^2 \xi^2}{\Lambda}  \left\{ \overline{\nu^i_L}
  \sigma^{\mu \nu}  \left( \nu_L^j \right)^c  \left( sA_{\mu \nu} + cZ_{\mu
  \nu} \right) \right. \nonumber\\
&&  \left. + \sqrt{2} c \overline{e^i_L} \sigma^{\mu \nu}  \left( \nu_L^j
  \right)^c W_{\mu \nu} + \text{h.c} \right\},  \label{7.28}
\end{eqnarray}
where $W_{\mu \nu} \equiv \nabla_{\mu} W_{\nu} - \nabla_{\nu} W_{\mu} +
\ldots$ is the charged component of the field strength $G_{L \mu \nu}$
expressed in the unitary gauge. The new element with respect to the
right-handed case, is the appearance of the charged current vertex $W^-_{\mu}
\longrightarrow e_L^- + \nu_L$. This was not possible in the right-handed
case, due to the $\nu_R$ sign-flip symmetry. This represents the
{\tmem{chirality-even direct}} contribution to the process $W W
\longrightarrow e_L e_L$ via the diagram represented in FIG. \ref{F1c}. The
order of magnitude estimate of this contribution suggests
\begin{eqnarray}
  M_3 & \sim & g \frac{m_{\nu_L}}{\Lambda^2}  \overline{e_L}  \left( e_L
  \right)^c W^-_{\mu} W^{- \mu} .  \label{7.29}
\end{eqnarray}
It appears that this contribution contains one power of $g$ less than the
indirect LNV contribution, equation (\ref{7.9}). Nevertheless, one should have
$M_3 \ll M_1$ due to the relative suppression factor $1 / g \left\langle p^2
\right\rangle / \Lambda^2$ (where $\left\langle p^2 \right\rangle$ is the
average momentum transfered by the neutrino).

\subsubsection{Contact contribution to $W W \longrightarrow e e$}
\label{contact}

We finally consider chirality-even LNV vertices containing two covariant
derivatives acting on GB fields $\Sigma$ {\footnote{Terms with derivatives of
fermion fields are related by equations of motion to the magnetic-type
vertices considered previously.}}. In the right-handed sector such vertices
are of the type
\begin{eqnarray}
  N_{R R}^{( a )} & = & \frac{1}{\Lambda}  \overline{\ell_R}
  \mathcal{Y}_a^{\dag}  \left( \Sigma^{\dag} D_{\mu} \Sigma \right)
  \mathcal{Z} \left( \Sigma^{\dag} D^{\mu} \Sigma \right) \mathcal{Y}_a^c 
  \left( \ell_R \right)^c  \nonumber\\
&& + \text{h.c},  \label{7.30}
\end{eqnarray}
where $a = u$ or $d$. In equation (\ref{7.30}), the occurrence (twice) of the
$\mathcal{Y}_a$ spurion guarantees that the the operator is invariant under
the $\nu_R$ sign-flip symmetry. Since in the standard unitary gauge we have
\begin{eqnarray}
  \mathi \Sigma^{\dag} D_{\mu} \Sigma & \overset{\text{s.g.}}{=} & \frac{e}{2
  cs}  \left\{ Z_{\mu} \tau^3 \right. \nonumber\\
&& + \left.\sqrt{2} c \left( W_{\mu}^+ \tau^+ + W_{\mu}^-
  \tau^- \right) \right\},  \label{7.31}
\end{eqnarray}
we obtain
\begin{eqnarray}
  N_{R R}^{( u )} & \overset{\text{s.g.}}{=} & - \frac{\zeta_{\Iota}^2
  \eta^2}{\Lambda}  \left( \frac{e}{2 cs} \right)^2  \overline{\nu_R}  \left(
  \nu_R \right)^c Z_{\mu} Z^{\mu} + \text{h.c},  \label{7.32}
\end{eqnarray}
whereas for $a = d$, the vertex (\ref{7.30}) reduces to
\begin{eqnarray}
  N_{R R}^{( d )} & \overset{\text{s.g.}}{=} & - \frac{\zeta_{\Iota}^2
  \eta^2}{2 \Lambda}  \left( \frac{e}{s} \right)^2  \overline{e_R}  \left( e_R
  \right)^c W^-_{\mu} W^{- \mu} + \text{h.c} .  \label{7.33}
\end{eqnarray}
Equation (\ref{7.33}) represents a direct LNV contact contribution to $W W
\longrightarrow e e$, as represented in FIG. \ref{F1d}. It is of order
$\mathcal{O} \left( p^2 \eta^2 \zeta_{\Iota}^2 \right)$, i.e. carries the
index $\kappa = 5$, and should normally be suppressed compared to the indirect
LNV diagram of FIG. \ref{F1a}. We indeed find
\begin{eqnarray}
  \frac{M_4^{R R}}{M_1} & \sim & \left( \frac{p}{\Lambda}  \frac{\eta}{\xi}
  \right)^2 .  \label{7.34}
\end{eqnarray}
In the left-handed sector, the two-derivative vertex similar to (\ref{7.30})
depicted in FIG. \ref{F1d} reads
\begin{eqnarray}
  N_{L L}^{( a )} & = & \frac{1}{\Lambda}  \overline{\ell_L}
  \mathcal{X}^{\dag}  \left( D_{\mu} \Sigma \right) \mathcal{Z} \left( D^{\mu}
  \Sigma^{\dag} \right) \mathcal{X} \left( \ell_L \right)^c  \nonumber\\
&& + \text{h.c}, 
  \label{7.35}
\end{eqnarray}
and in the standard unitary gauge it becomes
\begin{eqnarray}
  N_{L L} & \overset{\text{s.g.}}{=} & \frac{\zeta_{\Iota}^2 \xi^2}{\Lambda} 
  \left( \frac{e}{2 cs} \right)^2  \left\{ \overline{\nu_L}  \left( \nu_L  \right)^c Z_{\mu} Z^{\mu} \right. \nonumber\\
  &  & + 2 \sqrt{2} c \overline{e_L}  \left( \nu_L \right)^c W_{\mu}^-
  Z^{\mu}  + \text{h.c.} \nonumber\\
&& + \left. \frac{c^2}{2} W_{\mu}^- W^{- \mu}  \overline{e_L}  \left( e_L\right)^c \right\} .  \label{7.36}
\end{eqnarray}
It again contains the contact contribution to $W W \longrightarrow e_L e_L$ of
the order $\mathcal{O} \left( p^2 \xi^2 \eta^2 \zeta_{\Iota}^2 \right)$, i.e.
$\kappa = 5$. The latter is suppressed with respect to the indirect LNV
contribution of FIG. \ref{F1a} by a factor
\begin{eqnarray}
  \frac{M_4^{L L}}{M_1} & \sim & \frac{p^2}{\Lambda^2},  \label{7.37}
\end{eqnarray}
to be compared with (\ref{7.34}) and (\ref{7.19}).

The result of our analysis of the relative importance of different
contributions to the generic $\Delta L = 2$ process $W W \longrightarrow e e$
underlying various $0 \nu 2 \beta$-type processes may be summarized as
follows: if the $\mathcal{X}$ and $\mathcal{Y}$ spurions are of comparable
strength $\xi \sim \eta$, the LEET counting guarantees the dominance of the
indirect LNV contribution arising merely from the neutrino Majorana mass
terms. In this case, both final electrons are left-handed and the rate of LNV
is essentially determined by the Majorana mass matrix. On the other hand, it
is not possible to a priori exclude a hierarchy $\xi \sim \eta p / \Lambda$.

We know that $\xi \eta$ controls the scale of the top quark mass. A separate
information on $\xi$ and $\eta$ can be obtained studying non-standard
couplings of left-handed and right-handed fermions respectively, see Section
\ref{7-list}. A hierarchy $\xi \sim \eta p / \Lambda$ would imply that
indirect and direct LNV contributions could be of a comparable size, making
problematic the use of processes such as $0 \nu 2 \beta$ to measure the
absolute size of neutrino masses. Notice however, that even in this case, the
indirect Majorana mass term of FIG. \ref{F1a} would dominate the emission of
two left-handed charged leptons: the competing direct LNV contribution leads
to the emission of one (FIG. \ref{F1b}) or two (FIG. \ref{F1d})
right-handed leptons. The answer might come from the detailed quantitative
analysis of NLO in the LNV conserving sector. To conclude, we mention that,
although we have performed the comparison of respective contributions to the
LNV process $W W \longrightarrow e e$ in the hypothesis that ($B -
L$)-breaking is described by the case labelled $\Iota$ earlier, the relative
importance of the various terms will be unmodified if one were to analyze
cases $\tmop{II}$ and $\tmop{III}$.

\section{Quasi-sterile light right-handed neutrinos and cosmology}
\label{6-nuR}

This section is to a large extent self-contained. We describe the implications
of light quasi-sterile $\nu_R$ as a dark matter (DM) component, i.e. examine
whether the $\nu_R$ fit in the standard cosmology, without affecting
observables. We also consider the case where the $\nu_R$ would give the bulk
of the DM contribution, only to conclude that this seems to be excluded if one
includes the latest analyses. Before proceeding, we can only repeat the
warning that our colleagues cosmologists themselves issue: when interpreting
limits derived from cosmology, one should keep in mind that the analyses are
carried assuming the standard cosmological model with given priors on the
parameter ranges, whereas there could be more than one non-standard effect
compensating each other. In Section \ref{background}, we give a succinct
background necessary in order to follow the subsequent discussion. As a
primary source for more detailed information, we hasten to recommend the
reviews of {\cite{PDG}}, and the references therein. We will only quote
additional references for specific points.

The $\nu_R$ we have introduced in the Higgs-less effective theory are quite
different from the $N_R$ of the see-saw extension of the SM: they are not only
light, but also stable. Indeed, the only mass terms for neutrinos are of the
Majorana type, and our $\nu_R$ are characterized by the following properties:
i) they are neutral, ii) their main interactions are through neutral currents
with the $Z^0$, and are suppressed with respect to weak interactions
{\emdash}as parameterized by the factor $\eta^2$ in (\ref{6.82}){\emdash} and,
iii) they are lighter than the $\tmop{TeV}$ since they are part of the LEET,
and presumably, from the estimates of Sections \ref{alternatives-2} and
\ref{nuR-couplings}, lighter than $10 \tmop{keV}$.

Note that point iii) is valid for all three possible assumptions about ($B -
L$)-breaking, denoted $\Iota, \tmop{II}, \tmop{III}$ in Section
\ref{alternatives}. Note also, regarding point ii), that, from (\ref{6.82})
and bounds on the $Z^0$ invisible width, we deduced in Section
\ref{nuR-couplings} that $\eta^2$ should be smaller than about $10^{- 2}$.
Such a bound applies only for $m_{\nu_R} \lesssim M_Z / 2$, which we expect to
hold, given the remark after point iii) above.

\subsection{Relevant cosmological observations} \label{background}

The presence of light $\nu_R$ may impact cosmological observations, which are
influenced by the composition of the universe, i.e. the respective
contributions of baryonic matter, dark matter of various types, and dark
energy. The first input is the Hubble constant {\emdash}or its reduced version
$h${\emdash} from correlations of luminosity and redshift for standard
candles such as cepheids and type-Ia supernov\ae. The latter also constrain
one combination of parameters pertaining to the composition of the universe.
Other measurements give constraints on various combinations of such
parameters, and must be combined to resolve the degeneracies between them. One
can use the power spectrum of galaxy distribution, which gives information
about large-scale structure (LSS) formation, itself strongly influenced by the
dark matter density. Looking further back towards the past, anisotropies of
the cosmic microwave background radiation (CMB) recently helped to improve many
constraints. Considering the even earlier history of the universe, we recall
that the duration of the primordial nucleosynthesis (Big-Bang nucleosynthesis
or BBN) is influenced by the expansion rate of the universe, itself dependent
on the energy density.  If the latter is modified, the observed abundance of
the various elements in the galactic and intergalactic medium can be difficult
to reproduce. Intergalactic gas clouds can be used not only to determine the
relative abundance of different nuclides after BBN; one can also infer a
density spectrum along the line of sight in this medium {\emdash}from
absorption by neutral hydrogen at various redshifts{\emdash} which pertains to
the more recent history of structure formation. This is the so-called
Lyman-$\alpha$ forest observation. Since the density perturbations are smaller
than in the case of galaxy surveys, one can use the assumption of linear
growth down to smaller scales.

With the present accuracy, all these observations are well reproduced by a
flat $\Lambda$CDM model in which the largest contribution to the present day
density is provided in the form of a cosmological constant, representing a
fraction $\Omega_{\Lambda} \simeq 0.7$ of the critical density. Only a
fraction $\Omega_{\text{M}} \simeq 0.3$ is in the form of matter, yielding a
spatially flat universe {\emdash}as experimentally observed with the present
accuracy. Since the contribution of experimentally observed particles is known
to be marginal, the $\Lambda$CDM model assumes dark matter of the cold type
(CDM) to account for $\Omega_{\text{M}}$. However, admixtures of other types
of dark matter, or even a main contribution from warm dark matter (WDM)
particles with masses of order $1 \tmop{keV}$ can yield a good fit for
structure formation {\cite{Colin:2000dn,Bode:2000gq}}.

Since our light $\nu_R$'s do not oscillate or decay to $\nu_L$, they provide
an additional source of dark matter not usually considered. In the remainder
of this Section, we therefore explore the bounds on their masses and couplings
that can be deduced from cosmology. This reasoning assumes that the universe
was once hot enough for the $\nu_R$ to be in chemical equilibrium with the
other particles via their weaker-than-weak interactions (\ref{6.82}).

We also assume that the relative contributions of both the cosmological
constant and the dark matter (DM) are unchanged {\emdash}respectively
$\Omega_{\Lambda} \simeq 0.7$ and $\Omega_{\text{DM}} \simeq 0.3${\emdash},
but use different hypotheses as to the composition of the DM component.
Depending on the strength of their coupling to the $Z^0$ (\ref{6.82})
{\emdash}i.e. on the value of the spurion parameter $\eta${\emdash}, and on
their masses {\emdash}i.e. depending on whether we assume the ($B -
L$)-breaking option $\Iota, \tmop{II}$ or $\tmop{III}$ of Section
\ref{alternatives-2}{\emdash}, the $\nu_R$ contribution will generally not
yield the measured DM density. Clearly, we can exclude scenarios in which the
predicted density would be too high. In the other case where it is too low, it
simply means that the $\nu_R$ on their own cannot explain the whole DM
density: we then need to assume the presence of a CDM complement to obtain
$\Omega_{\text{DM}} \simeq 0.3$. We can then obtain stronger constraints by
using the list of observations mentioned above.

\subsection{Constraints on $\nu_R$ as dark matter}

The $\nu_R$'s couple mainly to the $Z^0$, via the operator (\ref{6.82}),  with
a coupling constant of order $\eta^2$ times the one of left-handed neutrinos.
Their freeze-out temperature $T_D$ will then be greater than that of $\nu_L$
(which is equal to a few $\tmop{MeV}$) by a factor
\begin{eqnarray}
  \frac{T_D}{T_D^{\nu_L}} & \sim & \eta^{- 4 / 3} .  \label{Tfr0}
\end{eqnarray}
For reasonable values of $\eta$ (see Section \ref{nuR-couplings}), this yields
the following range of variation for $T_D$
\begin{eqnarray}
  10 \tmop{MeV} \hspace{0.2em} \lesssim & T_D & \lesssim \hspace{0.2em} 1
  \tmop{GeV} .  \label{8.2}
\end{eqnarray}
This temperature is obtained by equating the annihilation rate, given at
temperatures below $M_Z$ by $n_{\nu_R} \times \left\langle \sigma v
\right\rangle_T \sim g^4 \eta^4 T^5 / M_{Z^{}}^4$, with the expansion rate $H
\sim T^2 / M_{\tmop{Planck}}$ of the universe at that same temperature. The
thermal density $n_{\nu_R}$ used assumes that the $\nu_R$ are relativistic at
that temperature, i.e. $m_{\nu_R} < T_D$ which we find is always true for the
range (\ref{8.2}) as long as $m_{\nu_R} \lesssim 1 \tmop{MeV}$, which we
expect to always hold. Hence our $\nu_R$ indeed decouple while they are still
relativistic, and will never yield a CDM candidate, whose mass should be above
a $\tmop{GeV}$ {\cite{Lee:1977ua,Kolb:1986nn}}. Note also that in the EWSB
scenario we consider, there are no new particles below a $\tmop{TeV}$.
Therefore, the range (\ref{8.2}) translates into a maximum number of degrees
of freedom at decoupling of $247 / 4$ {\cite{PDG}}, as in the SM at the same
scale.

\subsubsection{Relic density} \label{density}

Viable scenarios must first meet the condition that each particle contributing
to DM produces a relic density that is smaller or equal to $\Omega_{\text{DM}}
\simeq 0.3$. Remember that the contribution of a particle decreases as its
decoupling temperature increases {\cite{Olive:1982ak}}. Given the upper limit
of $247 / 4$ on the number of degrees of freedom at decoupling, the sum of the
masses of the $\nu_R$'s has to be below about $100 \tmop{eV}$ in order to
respect the bound. This upper limit would correspond to a situation where the
bulk of DM is provided by the $\nu_R$, i.e. a $\Lambda$WDM scenario. In that
case we would not need any additional source of DM. However seducing such a
scenario may appear, we will see in Section \ref{Lyman} that it now seems to
be excluded by more recent measurements of smaller scale structure formation
from Lyman-$\alpha$ forests, if we adopt such analyses.

\subsubsection{Combining CMB and LSS} \label{CMB}

If the $\nu_R$ would decouple as late as active $\nu_L$ neutrinos do
{\cite{Hannestad:2003ye,Crotty:2004gm}}, i.e. if they constituted hot hark
matter (HDM), the upper limit on their mass as deduced from CMB and LSS would
be of order the $\tmop{eV}$ in order for the density fluctuations not to be
erased by $\nu_R$ free-streaming. In practice this remains true if the
decoupling temperature is lower than the QCD transition temperature. If
$\eta^2$ is small enough that $T_D$ is above the QCD transition, the early
decoupling of the $\nu_R$ means that the decrease of the number density and
temperature due to the expansion of the universe is not compensated by the
subsequent annihilation of other species. Hence, the contribution of the
$\nu_R$ to the energy density after the QCD transition is small with respect
to that of other relativistic species. The limit from CMB and LSS on the mass
of such warm dark matter (WDM) is then almost inexistent~{\cite{Hannestad:2003ye}}.

Note that data on structure formation inferred from galaxy surveys must be
excluded from this analysis if they correspond to wave-numbers larger than
about $0.2 h \tmop{Mpc}^{- 1}$. The reason is that the growth of such
perturbations cannot be consistently considered to occur within the linear
regime. To extend the analysis and strengthen the constraints from structure
formation, we will see in Section \ref{Lyman} that the data set may be
extended by considering intergalactic gas clouds, where the density is lower.

\subsubsection{Influence on BBN} \label{BBN}

If the $\nu_R$'s decouple before the QCD transition, their influence on BBN
via the speed-up parameter (i.e. their contributions to the energy density at
the epoch $0.06 \tmop{MeV} < T_{\gamma} < 1 \tmop{MeV}$) is reduced, by the
same argument as above in Section \ref{CMB} {\cite{Steigman:1979xp}}.
Alternatively, one may say that the equivalent number of ``full-strength''
neutrino species $N_{\nu}$ is reduced.

If the $\nu_R$ decoupled at $T_D \simeq 1 \tmop{GeV}$, each of them would only
add $0.1$ to the effective number of neutrino species. In this situation,
three species of $\nu_R$ would modify the $N_{\nu}$ extracted from BBN within
current uncertainties {\cite{Olive:1999ij}} {\footnote{The situation is at
present unclear: according to some studies, even the standard BBN might have
difficulties, due to tension between the abundances of various elements. See
for instance the references in {\cite{Pastor:2003jx,PDG}}.}}. This would not
be possible if they decoupled after the QCD transition. Considering the main
interactions of these $\nu_R$ (\ref{6.82}), we deduce that the suppression
factor $\eta$ in the interaction (\ref{6.82}) might need to be slightly
smaller than the $0.1$ mentioned earlier. However, we are talking about
factors smaller than an order of magnitude here, which would be beyond the
predictive power of such naive estimates from operators in the LEET: there are
always unknown order-one coefficients.

\subsubsection{Constraints from Lyman-$\alpha$ forests} \label{Lyman}

Scenarios that satisfy constraints from LSS formation have to pass yet another
test pertaining to structure formation on scales in the range $\left( 1 \div
40 \right) h^{- 1} \tmop{Mpc}$. This  constrains particles with
a long free-streaming length, i.e. hot or warm dark matter. Such constraints
come from the observation of Lyman-$\alpha$ forests, whose interpretation has
been controversial in the past, but is becoming more and more widely accepted
these days.

One can guess that there should be a lower bound on the mass if the properties
for structure formation have to mimic those of CDM {\cite{Colombi:1995ze}},
and avoid the destructive effect of free-streaming. This is true for the case
where a large proportion of the DM comes from this WDM. The bound coming from
the study of Lyman-$\alpha$ forests is in this case that the mass of the WDM
particle should be higher than about $500 \tmop{eV}$
{\cite{Narayanan:2000tp,Viel:2005qj}} {\footnote{There are also claims of a
conflict with the early reionization deduced from CMB measurements by the WMAP
satellite, even for masses of order $10 \tmop{keV}$ {\cite{Yoshida:2003rm}}.
However, the relation between the measured optical depth and the inferred
redshift of reionization is not so direct
{\cite{Hansen:2003yj,Gnedin:2004nj}}. At any rate, this concern is irrelevant
for the present analysis: even though we started with a range of possible
masses for $\nu_R$ extending up to $10 \tmop{keV}$, we have seen in Section
$\ref{density}$ that, in our particular case, we could exclude $m_{\nu_R}
\gtrsim 100 \tmop{eV}$ since otherwise the decoupling could not occur early
enough as to yield $\Omega_{\nu_R} \lesssim 0.3$.}}. There would then be no
overlap with the upper bound of about $100 \tmop{eV}$ described in Section
\ref{density}.

There is however another alternative, where most of the DM is in the form of
unknown CDM, and the $\nu_R$ only provide a marginal fraction of the total
density {\emdash}as the $\nu_L$'s do. Remember that for the case of $\nu_L$,
the upper limit from CMB and LSS formation was of order $1 \tmop{eV}$. This
limit for $\nu_L$ is not much modified by the adjunction of Lyman-$\alpha$
forest results {\cite{Hannestad:2004bu}}.

The difference with the $\nu_L$ is that our $\nu_R$ a priori decouple earlier,
viz. (\ref{Tfr0}): Section \ref{BBN} indicated decoupling of $\nu_R$ before
the QCD transition {\emdash}in this case the bound from CMB and LSS (Section
\ref{CMB}) was almost inexistent. The inclusion of Lyman-$\alpha$ forest data,
if we adopt it, changes this picture drastically, and yields an upper bound
for our $\nu_R$ of about $10 \tmop{eV}$ {\cite{Hannestad:2004bu,Viel:2005qj}}.
This  disfavors the choice $\tmop{III}$ for the ($B - L$)-breaking
spurion $\mathcal{Z}$ (\ref{ZIII}) because of relation (\ref{136}), but leaves
room for the two other options.

\subsection{Conclusions on the different scenarios $\Iota, \tmop{II},
\tmop{III}$ for ($B - L$)-breaking}

The allowed range for the parameters are thus: i) suppression factor $\eta^2$
with respect to weak interactions of about $10^{- 2}$ or maybe smaller and,
ii) $\nu_R$ masses possibly up to $10 \tmop{eV}$, or smaller. Of the different
scenarios for ($B - L$)-breaking, only the choices labeled as $\Iota$ and
$\tmop{II}$ seem compatible with this limit on the masses. In particular, the
choice $\Iota$, which we have adopted for our main line of discussion, seems
safe with respect to current cosmological constraints.

Let us also mention the possibility studied in {\cite{Hansen:2001zv}}, and
references therein, in which $\tmop{keV}$ sterile neutrinos are populated via
oscillations with the active ones: such a situation cannot occur here if we
assume the $\nu_R$ sign-flip {\emdash}the $\mathbb{Z}_2$ symmetry of
(\ref{5.148}){\emdash} to be unbroken.

\section{Summary and conclusions} \label{8-concl}

i) Low-energy effective theories of EWSB without a Higgs in which deviations
from the SM would occur at leading order are untenable from the
phenomenological standpoint. A Higgs-less LEET based on the symmetry
$\tmop{SU} \left( 2 \right)_L \times \mathrm{U} \left( 1 \right)_Y$ and
operating with naturally light gauge bosons and chiral fermions unavoidably
suffers from this disease: at the leading order $\mathcal{O} \left( p^2
\right)$, the $\tmop{SU} \left( 2 \right)_L \times \mathrm{U} \left( 1
\right)_Y$ symmetry allows for non-standard oblique corrections, non-standard
fermion gauge boson vertices and {\emdash}last but not least{\emdash}
unsuppressed $\Delta L = 2$ Majorana masses for left-handed neutrinos. In the
SM all these operators would correspond to dimensionally suppressed operators
containing at least two powers of the elementary Higgs field. Consequently,
any Higgs-less scenario must first of all address the question of natural
suppression of non-standard operators at the leading order.

ii) We have shown that non-standard $\mathcal{O} \left( p^2 \right)$ vertices
are suppressed if the LEET is based from the onset on a non-linearly realized
higher symmetry $S_{\text{nat}} \supset S_{\text{red}} = \tmop{SU} \left( 2
\right)_L \times \mathrm{U} \left( 1 \right)_Y$, generalizing the concept of
custodial symmetry. $S_{\text{nat}}$ turns out to represent a hidden symmetry
of the Higgs-less vertices of the SM Lagrangian itself in the limit of
vanishing fermion masses. The manifestation of $S_{\text{nat}}$ is precisely
the absence of non-standard $\mathcal{O} \left( p^2 \right)$ vertices which
would be allowed by $S_{\text{red}}$ alone. $S_{\text{nat}}$ corresponds to
the maximal linear local symmetry the theory could have if its symmetry
breaking sector (Goldstone bosons) would be decoupled from the gauge/fermion
sector.

iii) In the LEET with a minimal light particle content one has $S_{\text{nat}}
= \left[ \tmop{SU} \left( 2 \right) \times \tmop{SU} \left( 2 \right)
\right]^2 \times \mathrm{U} \left( 1 \right)_{B - L}$. The set of variables on
which $S_{\text{nat}}$ acts linearly would thus involve thirteen gauge
connections. The gauge and symmetry breaking sectors are coupled and the
redundant fields are eliminated through constraints invariant under
$S_{\text{nat}}$. These constraints leave us with the four gauge fields of the
SM on which only the subgroup $S_{\text{red}} \subset S_{\text{nat}}$ acts
linearly, and with a set of three non-propagating spurion fields transforming
non trivially under $S_{\text{nat}}$. The very existence and properties of
spurions are not a matter of choice but follow from the covariant reduction of
the symmetry $S_{\text{nat}}$ to its subgroup $S_{\text{red}}$.

iv) Spurions are an integral part of the theory. They are crucial to maintain
the covariance under $S_{\text{nat}}$ even if the latter is explicitly broken.
There exists a gauge in which spurions reduce to three real constants $\xi,
\eta, \zeta$, which are used as small expansion parameters introducing a
technically natural hierarchy of symmetry breaking effects. Among the
spurions, there is necessarily one carrying two units of the $B - L$ charge:
it arises from the reduction of the right handed isospin and $\mathrm{U}
\left( 1 \right)_{B - L}$ subgroup of $S_{\text{nat}}$ to $\mathrm{U} \left( 1
\right)_Y$ symmetry of the SM. Consequently, the spurion formalism unavoidably
predicts $\Delta L = 2$ vertices. Their order of magnitude is controlled by a
single spurion parameter $\zeta^2$. The latter is postulated to be much
smaller than $\xi$ and $\eta$ which parametrize the lepton-number conserving
symmetry-breaking effects.

v) The effective Lagrangian is constructed and renormalized order by order in
a double expansion: according to the infrared dimensions {\emdash}c.f.
equation (\ref{3.16a}){\emdash} and according to powers of spurions. At each
order all operators invariant under $S_{\text{nat}}$ have to be included. At
the leading order $\mathcal{O} \left( p^2 \right)$ with no spurion insertion,
one recovers the SM interaction vertices between massive $W^{\pm}, Z^0$ and
massless fermions. As in the SM, $W^{\pm}$ and $Z^0$ acquire their masses by
the Higgs mechanism involving the three GBs. On the other hand, Dirac fermion
masses are necessarily suppressed by (at least) the spurion factor $\xi \eta$.
This fact allows to extend the Weinberg's power counting to fermions as
discussed in Section \ref{3.1.3}: it suggests the counting rule $\xi \eta
=\mathcal{O} \left( p \right)$.

vi) At higher spurion orders one finds all the non-standard couplings
encountered before (Sect \ref{3.2-irr-H-less}), but now suppressed by definite
spurion factors dictated by symmetry properties. The spurion formalism thus
suggests a hierarchy between possible effects beyond the SM. This might
indicate new directions in the search of the latter. For instance, in the
lepton-number conserving sector, we have concentrated on $\mathcal{O} \left(
p^2 \right)$ vertices containing at most two powers of spurions $\xi$ or
$\eta$. Such non-standard vertices represent universal modifications of both
left-handed and right-handed fermion couplings. They are suppressed with
respect to the leading-order SM contributions but, according to the power
counting, they should be more important than the loop contributions. The
complete list of these vertices is displayed in Section \ref{7-list}. They
exhaust all non-standard lepton-number conserving effects which arise in our
LEET at the NLO. Let us stress that the oblique corrections only arise at the
NNLO and cannot be disentangled from the loops. This is consistent with the
absence of observation of oblique corrections beyond the SM.

vii) As already stressed, the LNV sector is to a large extent determined by
the framework and by the symmetry properties of the LEET. First, in order to
allow the hidden custodial symmetry to be at work at energies at which the
LEET is relevant, the light rightshanded neutrinos must be present as doublet
partners of right handed electrons. At the leading order $\nu_R$'s decouple
from all gauge fields (as in the SM), but have a residual interaction with
$Z^0$, suppressed with respect to the SM coupling by the spurion factor
$\eta^2$ . This allows one to assume that the discrete sign flip symmetry
$\nu_R \longmapsto - \nu_R$ (not spoiled by anomalies) remains exact to all
orders of the LEET. This forbids to all orders any neutrino Dirac mass term as
well as charged leptonic right-handed currents. Both left-handed and
right-handed neutrinos (Majorana) masses $m_{\nu_L}$ and $m_{\nu_R}$ are
suppressed by the spurion factor $\zeta^2$.  They nevertheless differ, due to
different contributions of spurions $\xi$ and $\eta$. This represents an
alternative to the see-saw mechanism. With the existing constraints on the
mass of active neutrinos and naive estimates of spurion factors, one infers
that the heaviest $\nu_R$ can hardly be heavier than $10 \tmop{keV}$.

viii) We have reviewed pertinent cosmological constraints on super-weakly
interacting particles, coming from the observations of the CMB, of structure
formation and BBN. This leaves open the possibility of $m_{\nu_R} \lesssim 10
\tmop{eV}$, which satisfies all constraints in the standard $\Lambda
\tmop{CDM}$ cosmology. Either of the two scenarios $\Iota$ and $\tmop{II}$ for
$B - L$ breaking fall in this parameter space, and are therefore allowed. The
situation would be different for scenario $\tmop{III}$, with a heavier $\nu_R$
around the $\tmop{keV}$ which would provide the bulk of the DM in a $\Lambda
\tmop{WDM}$ cosmology. The recent interpretation of Lyman-$\alpha$ data for
smaller-scale structure formation, if taken at face value, would exclude this
theoretically well-motivated scenario.

ix) Finally we have analyzed the LO and NLO direct LNV vertices which may
compete with the indirect contribution of Majorana masses to the process $W^-
W^- \longrightarrow e^- e^-$, a building block for $0 \nu 2 \beta$ decay. We
classify, independently of the LEET, the different types of contributions
{\emdash}usually omitted in the literature{\emdash} besides the indirect LNV.
Note that this could a priori invalidate the extraction of the absolute
neutrino mass scale from $0 \nu 2 \beta$ decay. We then show that the LEET
provides estimates of their respective magnitudes: we find  that the
connection between Majorana masses and $0 \nu 2 \beta$ decay holds, up to
corrections in powers of spurions. Of course, such corrections might still
turn out to be large.

\begin{acknowledgments}

We are indebted to Fawzi Boudjema for pointing out the importance of
non-oblique corrections. We also profited from discussions with Christophe
Grojean, Thomas Hambye, Marc Knecht and Bachir Moussallam. JH would like to
thank Steen Hannestad and Shaaban Khalil for guidance, Verónica Sanz and Hagop
Sazdjian for helpful comments, and particularly Julien Lesgourgues and Sergio
Pastor for their help with the cosmology section.

This work was supported in part by the European Union EURIDICE network
under contract HPRN-CT-2002-00311.

\end{acknowledgments}

\appendix

\section{Proof of the generalized Weinberg power-counting formula}
\label{A-WPC}

Here we present a  derivation of the power-counting formula (\ref{3.28})
in the presence of gauge fields and chiral fermions, as described in Section
\ref{3-WPCF}. We start with the identity that simply counts the powers of $t$
coming from the rescalings (\ref{20}-\ref{22}) in a loop integral
\begin{eqnarray}
  d_{\text{IR}} \left[ \Gamma \right] & = & N_{\partial} + N_g + \frac{1}{2}
  N_f^{\text{ext}} - 2 I_b - I_f + 4 L .  \label{A.1}
\end{eqnarray}
Here, $N_{\partial}$ and $N_g$ are respectively the total number of
derivatives and of coupling constants that are contained in the vertices $1,
\cdots, V$ of the diagram $\Gamma$. $N_f^{\text{ext}}$ is the total number of
external (non-contracted) fermion fields. $I_b$ and $I_f$ denote the number of
internal boson and fermion lines. The last term $4 L$ accounts for the
$L$-loop momentum integral. One first observes that
\begin{eqnarray}
  I_f & = & \frac{1}{2}  \sum^V_{v = 1} n^{\text{int}}_f \left[ \mathcal{O}_v
  \right],  \label{A.2}
\end{eqnarray}
where $n^{\text{int}}_f \left[ \mathcal{O}_v \right]$ is the number of
internal fermion lines ending up in the vertex $\mathcal{O}$. Writing
\begin{eqnarray}
  N_f^{\text{ext}} & = & \sum^V_{v = 1} n_f^{\text{ext}} \left[ \mathcal{O}_v
  \right], 
\end{eqnarray}
where $n_f^{\text{ext}} \left[ \mathcal{O}_v \right]$ stands for the number of
external fermion lines ending up in the vertex $\mathcal{O}$, one finds that
\begin{eqnarray}
  2 I_f + N_f^{\text{ext}} & = & \sum^V_{v = 1} n_f \left[ \mathcal{O}_v
  \right],  \label{A.4}
\end{eqnarray}
where $n_f \left[ \mathcal{O}_v \right] = n^{\text{int}}_f \left[
\mathcal{O}_v \right] + n^{\text{ext}}_f \left[ \mathcal{O}_v \right]$
represents the number of fermion fields in the vertex $\mathcal{O}$.
Consequently, (\ref{A.1}) can be written as
\begin{eqnarray}
  d_{\text{IR}} \left[ \Gamma \right] & = & 4 L - 2 \left( I_b + I_f \right)  \nonumber\\
&& +
  \sum^V_{v = 1} \left( n_{\partial} \left[ \mathcal{O}_v \right] + n_g \left[
  \mathcal{O}_v \right] + \frac{1}{2} n_f \left[ \mathcal{O}_v \right]
  \right),  \label{A.5}
\end{eqnarray}
where the infrared dimension $d_{\text{IR}} \left[ \mathcal{O}_v \right]$
(\ref{3.16a}) of each vertex $\mathcal{O}_v$ appears. It remains to use the
identity
\begin{eqnarray}
  I_b + I_f & = & V + L - 1, 
\end{eqnarray}
and rewrite (\ref{A.5}) as the original Weinberg formula (\ref{3.28})
\begin{eqnarray}
  d_{\text{IR}} \left[ \Gamma \right] & = & 2 + 2 L + \sum^V_{v = 1} \left(
  d_{\text{IR}} \left[ \mathcal{O}_v \right] - 2 \right),  \label{A.8a}
\end{eqnarray}
which now holds beyond the framework of $\chi$PT. It is worth stressing that
the validity of (\ref{A.8a}) in the presence of light fermions is tied to the
assignment of infrared dimension $d_{\text{IR}} \left[ f \right] = 1 / 2$ to
chiral fermion fields. For a general $d_{\text{IR}} \left[ f \right]$, the
third term in the right-hand side of (\ref{A.1}) would be modified to
$d_{\text{IR}} \left[ f \right] N_f^{\text{ext}}$, whereas (\ref{A.2}) would
remain unchanged. Consequently, unless $d_{\text{IR}} \left[ f \right] = 1 /
2$, a suitably modified (\ref{A.4}) does not help in eliminating $I_f$ and
$N_f^{\text{ext}}$ in terms of $L$ and the total number of fermion fields in
the vertices $\sum n_f \left[ \mathcal{O}_v \right]$.

A last remark about the power-counting formula in a space-time of dimension
$D$ could be of interest. Note first that the mass-dimension of fields,
coupling constants $g$, and the GB coupling $f$ depends on $D$, whereas the
infrared dimension of all these quantities as introduced in section
\ref{3-WPCF} does {\tmem{not}}. The only modification of the power counting
formula (\ref{A.8}) thus arises from the loop integrals: the last term in
(\ref{A.1}) becomes $DL$ instead of $4 L$. Therefore with the definition
\begin{eqnarray}
  \Delta \left[ \Gamma \right] & \equiv & \sum^V_{v = 1} \left( d_{\text{IR}}
  \left[ \mathcal{O}_v \right] - 2 \right) \hspace{0.2em} \geqslant \hspace{0.2em}
  0,  \label{Delta}
\end{eqnarray}
the final formula reads
\begin{eqnarray}
  d_{\text{IR}} \left[ \Gamma \right] & = & 2 + \left( D - 2 \right) L +
  \Delta \left[ \Gamma \right],  \label{A.8}
\end{eqnarray}
showing explicitly the improved low-energy suppression of
loops as $D$ increases. Whereas higher dimensional (gauge) theories are too UV
singular to admit an expansion in powers of coupling constant only, they might
still be a privileged arena for LEETs formulated as a systematic low-energy
expansion.

\section{Unitarity order by order} \label{Unitarity}

The Weinberg power-counting formula (\ref{A.8}) allows one to compare how the
exact unitary $S$-matrix is successively approximated in a renormalized
expansion in powers of the coupling constant, and in the low-energy expansion.
To this end, we consider a transition matrix element $T_{i \rightarrow f}$
where the initial (final) state contains a set of $i$ ($f$) particles. We
first recall the unitary condition, which reads
\begin{eqnarray}
  \tmop{Im} T_{i \rightarrow f} & = & \sum_{n \geqslant 2} \int \mathd \left\{
  n \right\} \bignone  \left( T_{f \rightarrow n} \right)^{\ast} T_{i
  \rightarrow n},  \label{B.3}
\end{eqnarray}
where $\int \bignone \mathd \{ n \}$ stands for the integral over the Lorentz
invariant phase space of an $n$-particle intermediate state
\begin{eqnarray}
  \int \bignone \mathd \{ n \} & \propto & \int \left( \prod_{i = 1}^n
  \bignone \mathd^D p_i \delta \left( p_i^2 - m_i^2 \right) \theta \left(
  p_i^0 \right) \right)  \nonumber\\
&& \times \delta^D \left( P - \sum_{j = 1}^n p_j \right) . 
  \label{B.4}
\end{eqnarray}

We first deal with a renormalizable theory with trilinear $\sim g$ and
quadratic $\sim g^2$ couplings. The amplitude is approximated by a series in
$g$, where the power of $g$ appearing in a diagram with a given $i$ and $f$,
only depends on the number of loops $L$
\begin{eqnarray}
  T_{i \rightarrow f} & = & \sum_{L \geqslant 0} T^L_{i \rightarrow f}, 
  \label{B.1}
\end{eqnarray}
where $T_{i \rightarrow f}^L$ is of degree $2 L + i + f - 2$ in $g$. At a
given order in $g$, the condition (\ref{B.3}) becomes an identity between a
finite number of Feynman diagrams
\begin{eqnarray}
  \tmop{Im} T_{i \rightarrow f}^L & = & \sum^{L + 1}_{n \geqslant 2} \sum_{l,
  l' \geqslant 0} \delta_{l + l' + n, L + 1}  \nonumber\\
&& \times \int \mathd \left\{ n \right\}
  \bignone  \left( T^l_{f \rightarrow n} \right)^{\ast} T^{l'}_{i \rightarrow
  n} .  \label{B.5}
\end{eqnarray}
A few comments are in order:

i) As long as the theory is properly renormalized, the expected identity
(\ref{B.5}) automatically follows from the cutting rules of Feynman diagrams.
In principle, it holds at all energies. This however, does not give any
information on the size of violations of the unitary condition (\ref{B.3}) for
a finite energy and a given truncation of the expansion (\ref{B.1}).

ii) Equation (\ref{B.5}) is a recurrence relation. If one knows all amplitudes
$T^l$ for $l = 1, \cdots, L - 1$, it is possible to compute $\tmop{Im} T^L_{i
\rightarrow f}$ in all channels. $\tmop{Re} T^L$ is then, in principle,
determined by analyticity, up to a real subtraction polynomial. Requiring
renormalizability order by order is a restriction on the degree of these
polynomials, and thereby on the high-energy behavior of $\tmop{Im} T^L_{i
\rightarrow f}$.

iii) $\tmop{Im} T^0 = 0$, hence tree amplitudes violate unitarity as long as
$\tmop{Re} T^0 \neq 0$. This fact often motivates searches for additional
light tree-level exchanges (scalars, KK excited states) which would keep
$\tmop{Re} T^0$ small for energies as large as possible. Experiments will tell
us whether such a situation, allowing for a perturbative expansion, is
realized in the real world.

We now turn to the low-energy expansion, in which one expands the amplitudes
as
\begin{eqnarray}
  T_{i \rightarrow f} & = & \sum_{d_{\text{IR}} \geqslant 2} t_{i \rightarrow
  f}^{d_{\text{IR}}},   \label{B.6}
\end{eqnarray}
where $d_{\text{IR}}$ is the homogeneous degree of $t^{d_{\text{IR}}}_{i
\rightarrow f}$. The analysis of the unitary condition (\ref{B.3}) order by
order in the low-energy expansion (\ref{B.6}) requires the knowledge of the
infrared dimension of $\int \bignone \mathd \{ n \}$ (\ref{B.4}). Assuming
that all particles in the intermediate states are naturally light, i.e. their
masses are suppressed as at least $\mathcal{O} \left( p \right)$, one infers
the following {\tmem{suppression at low-energies}} of their Lorentz-invariant
phase-space
\begin{eqnarray}
  \int \mathd \left\{ n \right\} & = & \mathcal{O} \left( p^{\left( D - 2
  \right)  \left( n - 1 \right) - 2} \right) . 
\end{eqnarray}
It is crucial for the remainder that the creation of many-particle states is
suppressed {\footnote{As in Appendix \ref{A-WPC}, the suppression gets more
effective as the space-time dimension $D$ grows.}}. One then isolates
contributions of infrared dimension $d_{\text{IR}}$ in (\ref{B.3})
\begin{eqnarray}
  \tmop{Im} t_{i \rightarrow f}^{d_{\text{IR}}} & = & \sum^{\left[
  \frac{d_{\text{IR}} - 2}{D - 2} \right] + 1}_{n \geqslant 2} \sum_{d, d'
  \geqslant 2} \delta_{d + d' + \left( D - 2 \right)  \left( n - 1 \right),
  d_{\text{IR}} + 2}  \nonumber\\
&& \times \int \mathd \left\{ n \right\} \bignone  \left( t^d_{f
  \rightarrow n} \right)^{\ast} t^{d'}_{i \rightarrow n} .  \label{B.9}
\end{eqnarray}
In this low-energy expansion, the connection between the degree of a diagram
and the number of loops is more involved than following (\ref{B.1}). It is
therefore useful to introduce Feynman amplitudes $M^L_{i \rightarrow f}
\left\{ \Delta \right\}$ collecting $L$-loop diagrams and to use the sum
(\ref{Delta}) $\Delta = \sum_v \left( d_{\text{IR}} \left[ \mathcal{O}_v
\right] - 2 \right)$ over all vertices $\mathcal{O}$ involved, such that
\begin{eqnarray}
  M^L_{i \rightarrow f} \left\{ \Delta \right\} & = & \mathcal{O} \left( p^{2
  + \left( D - 2 \right) L + \Delta} \right) . 
\end{eqnarray}
We decompose $t^{d_{\text{IR}}}_{i \rightarrow f}$ as a sum of such
collections of Feynman diagrams, writing
\begin{widetext}
\begin{eqnarray}
  t_{i \rightarrow f}^{d_{\text{IR}}} & = & \sum_{L = 0}^{\left[
  \frac{d_{\text{IR}} - 2}{D - 2} \right]} M^L_{i \rightarrow f} \left\{
  \Delta = d_{\text{IR}} - \left( D - 2 \right) L - 2 \right\},  \label{B.12}
\end{eqnarray}
\end{widetext}
where we have taken into account that, for each term $M^L_{i \rightarrow f}
\left\{ \Delta \right\}$ to contribute to the order $d_{\text{IR}}$, the value
of $\Delta$ is fixed by the number of loops to be $\Delta = d_{\text{IR}} -
\left( D - 2 \right) L - 2$. Also, the the upper limit $L \leqslant \left[
\frac{d_{\text{IR}} - 2}{D - 2} \right]$ comes from the fact that $\Delta
\geqslant 0$.

In a LEET, renormalization is based on a cancellation of divergences between
terms of the finite sum (\ref{B.12}) with different $L$ but which carry the
same power $d_{\text{IR}}$. In other words, divergences of multi-loop diagrams
containing low infrared dimension vertices are absorbed by diagrams with fewer
loops, but higher dimension vertices {\cite{Bijnens:1999hw}}. This procedure
is repeated for every fixed $d_{\text{IR}} \geqslant 2$. For $d_{\text{IR}} =
2$ and $d_{\text{IR}} = 3$, $t^{d_{\text{IR}}}$ does not involve loops. It is
merely given by (finite) tree diagrams with $\Delta = 0$ and $\Delta = 1$.
Accordingly, $\tmop{Im} t^{2, 3} = 0$, in agreement with the unitarity
condition (\ref{B.9}).

It is instructive to compare the SM $L$-loop amplitude $T^L_{i \rightarrow f}$
with the Feynman amplitudes $M^L_{i \rightarrow f} \left\{ \Delta \right\}$ of
the minimal Higgs-less LEET. Since the $\mathcal{O} \left( p^2 \right)$
vertices of the LEET coincide with those of the SM without the Higgs fields,
the term with $\Delta = 0$ in $t^{d_{\text{IR}}}_{i \rightarrow f}$
(\ref{B.12}) is identical to the collection of $L$-loop SM diagrams that do
not contain internal Higgs lines. Symbolically
\begin{eqnarray}
  T^L_{i \rightarrow f} & = & T^L_{i \rightarrow f} |_{\text{Higgs exchange}}
 \nonumber\\
&& + M^L_{i \rightarrow f} \left\{ \Delta = 0 \right\} . 
\end{eqnarray}
Similarly, the LEET unitarity relation (\ref{B.9}) does not include physical
Higgs particles in the intermediate states $\left\{ n \right\}$, whereas the
SM unitarity relation (\ref{B.5}) does. The role of (virtual) Higgs
contributions in the renormalization and unitarization of the SM is replaced
in the Higgs-less LEET by effects of higher dimension vertices $\Delta > 0$ in
(\ref{B.12}). After renormalization, equation (\ref{B.12}) can be rewritten in
terms of renormalized amplitudes $\hat{M}^L_{i \rightarrow f} \left\{ \Delta,
\mu \right\}$
\begin{widetext}
\begin{eqnarray}
  t_{i \rightarrow f}^{d_{\text{IR}}} & = & \sum_{L = 0}^{\left[
  \frac{d_{\text{IR}} - 2}{D - 2} \right]} \hat{M}^L_{i \rightarrow f} \left\{
  \Delta = d_{\text{IR}} - \left( D - 2 \right) L - 2, \mu \right\} . 
  \label{B.14}
\end{eqnarray}
\end{widetext}
In this equation, the individual terms $\hat{M}^L_{i \rightarrow f} \left\{
\Delta, \mu \right\}$ depend on a renormalization scale $\mu$ (introduced e.g.
via dimensional regularization and renormalization) whereas the whole sum
(\ref{B.14}), i.e. the physical quantity $t^{d_{\text{IR}}}_{i \rightarrow
f}$, is both finite and $\mu$-independent {\cite{Bijnens:1999hw}}. The
recurrence unitarity relation (\ref{B.9}) then follows from the cutting
formula for the renormalized one-loop amplitudes $\hat{M}^L_{i \rightarrow f}
\left\{ \Delta, \mu \right\}$
\begin{widetext}
\begin{eqnarray}
  \tmop{Im} M_{i \rightarrow f}^L \left\{ \Delta, \mu \right\} & = & \sum^{L +
  1}_{n \geqslant 2} \sum_{\delta, \delta' \geqslant 0} \delta_{\delta +
  \delta', \Delta}  \sum_{l, l' \geqslant 0} \delta_{l + l' + n, L + 1}
 \int \mathd \left\{ n \right\} \bignone  \left( \hat{M}^l_{f
  \rightarrow n} \left\{ \delta, \mu \right\} \right)^{\ast}  \hat{M}^{l'}_{i
  \rightarrow n} \left\{ \delta', \mu \right\} .  \label{B.15}
\end{eqnarray}
\end{widetext}

This equation may be viewed as an extension of the cutting identity
(\ref{B.5}) for the case where one has to take into account vertices with
different dimensions. It should hold independently of the renormalization
scale $\mu$.

This analysis of course does not say much about the actual size of violations
of unitarity if the LEET (\ref{B.6}) is truncated at a given order
$d_{\text{IR}}$. By the very definition of the LEET, we can expect that for a
given $d_{\text{IR}}$, the violation of unitarity will increase with energy.
On the other hand, as demonstrated for the case of $\chi$PT in $\pi \pi$
scattering comparing $\mathcal{O} \left( p^4 \right)$ and $\mathcal{O} \left(
p^6 \right)$ {\emdash}see FIG. 1 of {\cite{Knecht:1995tr}} {\emdash}, one
may expect that the energy up to which unitarity is satisfied with a given
precision increases with $d_{\text{IR}}$.{\nocite{H,V,S,Gupta:1999du}}

\bibliography{bib-biblio.bib}

\end{document}